\documentclass[fleqn,usenatbib]{mnras}

\usepackage[T1]{fontenc}

\usepackage{graphicx}	
\usepackage{amsmath}	
\usepackage[export]{adjustbox}    
\usepackage[version=4]{mhchem}    
\usepackage{pdflscape}  
\usepackage{siunitx}    
\usepackage{xstring}    
\usepackage[nomessages]{fp}    
\usepackage[normalem]{ulem}
\usepackage{newtxtext,newtxmath}

\newcommand{\NqubricsI}[1]{
\IfEqCase{#1}{
{candidates}{1476}
{flagA}{327}
{QSO}{313}
{QSOzGT25}{203}}
{surveycovered}{22\%}
{successrate}{62\%}
}

\newcommand{\NqubricsII}[1]{
\IfEqCase{#1}{
{mainsample}{1014875}
{Selected}{1412}
{Candidates}{818}
{SelObserved}{266}
{KnownSel}{594}
{Completeness}{95\%}
{SuccessRate}{68\%}
{tot}{569}
{dupl}{58}          {dupl_F1}{10.2\%}
{dupl_n}{8}         {dupl_n_F2}{13.8\%}           {dupl_n_F1}{1.4\%}
{dupl_d}{29}        {dupl_d_F2}{50.0\%}           {dupl_d_F1}{5.1\%}
{dupl_g}{21}        {dupl_g_F2}{36.2\%}           {dupl_g_F1}{3.7\%}
{unique}{511}       {unique_F1}{89.8\%}
{flagA}{432}        {flagA_F2}{84.5\%}            {flagA_F1}{75.9\%}
{A_QSO}{390}        {A_QSO_F3}{90.3\%}            {A_QSO_F2}{76.3\%}            {A_QSO_F1}{68.5\%}
{A_QSO_zl25}{166}   {A_QSO_zl25_F4}{42.6\%}       {A_QSO_zl25_F3}{38.4\%}       {A_QSO_zl25_F2}{32.5\%}       {A_QSO_zl25_F1}{29.2\%}
{A_QSO_zg25}{224}   {A_QSO_zg25_F4}{57.4\%}       {A_QSO_zg25_F3}{51.9\%}       {A_QSO_zg25_F2}{43.8\%}       {A_QSO_zg25_F1}{39.4\%}
{A_QSO_zg35}{54}    {A_QSO_zg35_F4}{13.8\%}       {A_QSO_zg35_F3}{12.5\%}       {A_QSO_zg35_F2}{10.6\%}       {A_QSO_zg35_F1}{9.5\%}
{A_QSO_zg40}{15}    {A_QSO_zg40_F4}{3.8\%}        {A_QSO_zg40_F3}{3.5\%}        {A_QSO_zg40_F2}{2.9\%}        {A_QSO_zg40_F1}{2.6\%}
{A_GAL}{11}         {A_GAL_F3}{2.5\%}             {A_GAL_F2}{2.2\%}             {A_GAL_F1}{1.9\%}
{A_STAR}{31}        {A_STAR_F3}{7.2\%}            {A_STAR_F2}{6.1\%}            {A_STAR_F1}{5.4\%}
{flagB}{79}         {flagB_F2}{15.5\%}            {flagB_F1}{13.9\%}
{B_QSO}{18}         {B_QSO_F3}{22.8\%}            {B_QSO_F2}{3.5\%}             {B_QSO_F1}{3.2\%}
{B_QSO_zl25}{13}    {B_QSO_zl25_F4}{72.2\%}       {B_QSO_zl25_F3}{16.5\%}       {B_QSO_zl25_F2}{2.5\%}        {B_QSO_zl25_F1}{2.3\%}
{B_QSO_zg25}{5}     {B_QSO_zg25_F4}{27.8\%}       {B_QSO_zg25_F3}{6.3\%}        {B_QSO_zg25_F2}{1.0\%}        {B_QSO_zg25_F1}{0.9\%}
{B_QSO_zg35}{1}     {B_QSO_zg35_F4}{5.6\%}        {B_QSO_zg35_F3}{1.3\%}        {B_QSO_zg35_F2}{0.2\%}        {B_QSO_zg35_F1}{0.2\%}
{B_QSO_zg40}{1}     {B_QSO_zg40_F4}{5.6\%}        {B_QSO_zg40_F3}{1.3\%}        {B_QSO_zg40_F2}{0.2\%}        {B_QSO_zg40_F1}{0.2\%}
{B_MOSTRO}{31}      {B_MOSTRO_F3}{39.2\%}         {B_MOSTRO_F2}{6.1\%}          {B_MOSTRO_F1}{5.4\%}
{B_STAR}{7}         {B_STAR_F3}{8.9\%}            {B_STAR_F2}{1.4\%}            {B_STAR_F1}{1.2\%}
{B_???}{18}         {B_???_F3}{22.8\%}            {B_???_F2}{3.5\%}             {B_???_F1}{3.2\%}
{B_GAL}{5}          {B_GAL_F3}{6.3\%}             {B_GAL_F2}{1.0\%}             {B_GAL_F1}{0.9\%}
}[\PackageError{NqubricsII}{Undefined input: #1}{}]}

\newcommand{\NqubricsIII}[1]{
\IfEqCase{#1}{
{tot}{18}
{tot_NIR}{18}
{tot_QUIP}{15} {tot_QUIP_f}{83.3\%}
{tot_nonQUIP}{3} {tot_QUIP_f}{16.7\%}
{tot_BAL}{14} {tot_BAL_f}{93.3\%} {tot_BAL_ftot}{77.7\%}
{tot_LoBAL}{9} {tot_LoBAL_f}{64.3\%}
{tot_FeLoBAL}{8} {tot_FeLoBAL_f}{88.9\%} {tot_FeLoBAL_ftot}{44.4\%}
{tot_HiBAL}{5} {tot_HiBAL_f}{35.7\%}
{tot_other}{1} {tot_other_f}{6.7\%}
{A_QSO}{2} {A_QSO_f}{11.1\%}
{A_QSO_QUIP}{2} {A_QSO_QUIP_f}{13.3\%}
{A_QSO_BAL}{2} {A_QSO_BAL_f}{100.0\%}
{A_QSO_LoBAL}{1} {A_QSO_LoBAL_f}{50.0\%}
{A_QSO_FeLoBAL}{1} {A_QSO_FeLoBAL}{100.0\%} {A_QSO_FeLoBAL_qII_f}{0.3\%}
{A_QSO_HiBAL}{1} {A_QSO_HiBAL_f}{50.0\%}
{A_QSO_other}{0} {A_QSO_other}{0.0\%}
{B_any}{16} {B_any_f}{88.9\%}
{B_any_QUIP}{13} {B_any_QUIP_f}{86.7\%}
{B_any_BAL}{12} {B_any_BAL_f}{92.3\%}
{B_any_LoBAL}{8} {B_any_LoBAL_f}{66.7\%}
{B_any_FeLoBAL}{7} {B_any_FeLoBAL_f}{87.5\%} {B_any_FeLoBAL_qII_f}{8.9\%}
{B_any_HiBAL}{4} {B_any_HiBAL_f}{33.3\%}
{B_any_other}{1} {B_any_other}{7.7\%}
{B_QSO}{7}
{B_QSO_NIR}{4}
{B_QSO_QUIP}{4}
{B_MOSTRO}{8}
{B_MOSTRO_NIR}{8}
{B_MOSTRO_QUIP}{8}
{B_???}{1}
{B_???_NIR}{1}
{B_???_QUIP}{1}
{halpha}{13}
{hbeta}{11}
{hgamma}{8}
{efosc2}{7}
{ldss3}{7}
{wfccd}{6}
{mage}{5}
{lrs}{1}
}[\PackageError{NqubricsIII}{Undefined input: #1}{}]}

\defcitealias{2019ApJ...887..268C}{Paper~I}
\defcitealias{2020arXiv200803865B}{Paper~II}

\newcommand{\pI}{\citetalias{2019ApJ...887..268C}}
\newcommand{\pII}{\citetalias{2020arXiv200803865B}}

\newcommand{\AlIII}{\ion{Al}{iii}}
\newcommand{\CII}{\ion{C}{ii}}
\newcommand{\CIII}{\ion{C}{iii}}
\newcommand{\CIV}{\ion{C}{iv}}
\newcommand{\FeII}{\ion{Fe}{ii}}
\newcommand{\FeIII}{\ion{Fe}{iii}}
\newcommand{\Ha}{\ion{H}{$\alpha$}}
\newcommand{\Hb}{\ion{H}{$\beta$}}
\newcommand{\Hg}{\ion{H}{$\gamma$}}
\newcommand{\HI}{\ion{H}{i}}
\newcommand{\Lya}{\ion{Ly}{$\alpha$}}
\newcommand{\Lyb}{\ion{Ly}{$\beta$}}
\newcommand{\Lyg}{\ion{Ly}{$\gamma$}}
\newcommand{\MgII}{\ion{Mg}{ii}}

\newcommand{\NV}{\ion{N}{v}}

\newcommand{\OII}{\ion{O}{ii}}
\newcommand{\OIII}{[\ion{O}{iii}]}
\newcommand{\OVI}{\ion{O}{vi}}
\newcommand{\Pag}{\ion{Pa}{$\gamma$}}
\newcommand{\SIV}{\ion{S}{iv}}
\newcommand{\SiII}{\ion{Si}{ii}}
\newcommand{\SiIII}{\ion{Si}{iii}}
\newcommand{\SiIV}{\ion{Si}{iv}}

\newcommand{\lambdarf}{\lambda_\textrm{rf}}
\newcommand{\zem}{z_\textrm{em}}

\DeclareSIUnit\erg{erg}

\title[NIR spectroscopy of extreme QUBRICS BALQs]{Near-infrared spectroscopy of extreme BAL QSOs from the QUBRICS bright quasar survey}

\author[G. Cupani et al.]{Guido Cupani,$^{1,2}$\thanks{E-mail: guido.cupani@inaf.it},
Giorgio Calderone$^{1}$,
Pierluigi Selvelli$^{1}$,
Stefano Cristiani$^{1,2,5}$,
Konstantina Boutsia$^{3}$,
\newauthor
Andrea Grazian$^{4}$,
Fabio Fontanot$^{1,2}$,
Francesco Guarneri$^{1,6}$,
Valentina D'Odorico$^{1,2,7}$,
\newauthor
Emanuele Giallongo$^{8}$,
and Nicola Menci$^{8}$\\
\\
$^{1}$INAF--Osservatorio Astronomico di Trieste, Via Tiepolo 11, I-34143 Trieste, Italy\\
$^{2}$ IFPU--Institute for Fundamental Physics of the Universe, via Beirut 2, I-34151 Trieste, Italy \\
$^{3}$ Las Campanas Observatory, Carnegie Observatories, Colina El Pino, Casilla 601, La Serena, Chile\\
$^{4}$ INAF--Osservatorio Astronomico di Padova, Vicolo dell'Osservatorio 5, I-35122 Padova, Italy \\
$^{5}$ INFN--National Institute for Nuclear Physics, via Valerio 2, I-34127 Trieste, Italy \\
$^{6}$ Dipartimento di Fisica, Sezione di Astronomia, Università di Trieste, via Tiepolo 11, I-34131 Trieste, Italy \\
$^{7}$ Scuola Normale Superiore, piazza dei Cavalieri, I-56126 Pisa, Italy \\
$^{8}$ INAF--Osservatorio Astronomico di Roma, via Frascati 33, I-00078 Monte Porzio Catone, Italy \\
}

\date{Accepted XXX. Received YYY; in original form ZZZ}

\pubyear{2021}

\begin{document}
\label{firstpage}
\pagerange{\pageref{firstpage}--\pageref{lastpage}}
\maketitle
\sisetup{range-phrase=\textrm{--}}

\begin{abstract}
We report on the spectral confirmation of $\NqubricsIII{tot}$ QSO candidates from the ``QUasars as BRIght beacons for Cosmology in the Southern hemisphere'' survey  (QUBRICS), previously observed in the optical band, for which we acquired new spectroscopic data in the near-infrared band with the Folded-port InfraRed Echellette spectrograph (FIRE) at the Magellan Baade telescope. In most cases, further observations were prompted by the peculiar nature of the targets, whose optical spectra displayed unexpected absorption features. All candidates have been confirmed as bona fide QSOs, with average emission redshift $z\simeq 2.1$. The analysis of the emission and absorption features in the spectra, performed with {\sc Astrocook} and {\sc QSFit}, reveals that the large majority of these objects are broad-absorption line (BAL) QSOs, with almost half of them displaying strong \FeII{} absorption (typical of the so-called FeLoBAL QSOs). The detection of such a large fraction of rare objects (which are estimated to account for less than one percent of the general QSO population) is interpreted as an unexpected (yet favourable) consequence of the particular candidate selection procedure adopted within the QUBRICS survey. The measured properties of FeLoBAL QSOs observed so far provide no evidence that they are a manifestation of a particular stage in AGN evolution. In this paper we present an explorative analysis of the individual QSOs, to serve as a basis for a further, more detailed investigation.
\end{abstract}





\section{Introduction}

The study of luminous quasars (QSOs) at medium to high redshift is pivotal for a wide range of science cases in astrophysics, observational cosmology, and even fundamental physics. QSOs are among the best faraway beacons in the Universe, literally shedding light in the otherwise virtually invisible inter-galactic medium along their line of sight, but are also interesting in themselves, as the most apparent manifestation of the super-massive black hole growth in the early stage of galaxy evolution. The analysis of absorption and emission features in the optical and near-infrared (NIR) spectra of QSOs plays a key role in understanding the physical conditions of the primordial universe, the interplay between galaxies and large-scale structure, the mechanism of reionization, and related issues; it also helps constraining the primordial abundance of elements, the power spectrum of dark matter, the possible variation of fundamental constants across time, and the validity of General Relativity. All these undertakings rely on the availability of bright QSOs at $z\gtrsim 2.5$ across the whole sky.

The QUBRICS survey (``QUasars as BRIght beacons for Cosmology in the Southern hemisphere'') was
started in 2018 to even up a significant lack of identified QSOs in the Southern Hemisphere. The project entails the selection of QSO candidates from public databases, using innovative machine-learning techniques \citep{2019ApJ...887..268C,2021MNRAS.506.2471G} and their spectral confirmation through direct observation in the optical band \citep{2019ApJ...887..268C,2020arXiv200803865B}. The result is a growing catalogue of some 400 newly discovered bright QSOs, which will significantly enhance the feasibility of a redshift drift measurement with future facilities \citep[the so called Sandage Test;][]{2020arXiv200803865B} and were already used to put stronger constraints on the bright end of the QSO luminosity function \citep{2021ApJ...912..111B}.

Quite understandably, optical spectroscopy alone was not always sufficient to confirm or reject candidates as bona-fide QSOs. In particular, a number of candidates revealed notable absorption features that could not be explained as arising from structures either associated with the emitting AGN or located along the line of sight, given the scarce information provided by the emission component. We refer to these candidates as the ``QUBRICS Irregular and Peculiar'' (QUIP) targets. In other cases, a relative featureless optical spectra prevented a secure determination of the emission redshift (these unidentified objects are dubbed non-QUIP). To properly assess all these objects (QUIP and non-QUIP), we started a NIR observational campaign using the Folded-port InfraRed Echellette spectrograph (FIRE) at the Magellan Baade telescope. All the $\NqubricsIII{tot_NIR}$ targets observed so far in the NIR have been securely confirmed as QSOs through the observation of their Balmer series and/or other emission lines, with luminosities at \SI{5100}{\angstrom} ranging from about \SI{1.5e46}{\erg\per\second} to \SI{4.4e47}{\erg\per\second}. 

A relevant outcome of this observational campaign is that a large fraction of QUIPs appear to be broad-absorption line QSOs (BALQs), and in particular show strong intrinsic absorption from either high-ionization species  like \CIV{} and \SiIV{} (HiBALQs), or low-ionization species like \MgII{} and \AlIII{} (LoBALQs), and occasionally \FeII{} (FeLoBALQs). BALQs form an inherently interesting, yet not fully understood class of objects. Their characteristically strong absorption features, conventionally wider than \SI{2000}{\km\per\second} and at least 10 percent below the continuum level \citep{1991ApJ...373...23W}, are often blueshifted to velocities up to tens of thousands kilometers per second with respect to the QSO emission redshift, and are recognized as a signature of energetic AGN outflows \citep[e.g.][]{1983PASP...95..341F,1991ApJ...373...23W}, which are assumed to play an important role in quenching star formation in the host galaxies and self-regulating the growth of super-massive black holes \citep[e.g][]{1998A&A...331L...1S,2005Natur.433..604D,2012ARA&A..50..455F,2013ARA&A..51..511K}. A proper understanding of such feedback mechanism is fundamental to explain the observed properties of QSO host galaxies and to constrain the AGN-galaxy co-evolution. It is still unclear whether the strong outflows observed in BALQs represent a specific stage in the AGN evolution, or an ubiquitous feature that becomes apparent only when the QSO is observed at the right orientation \citep[see e.g.][and references therein]{Schulze_2017}. This is particularly true for the LoBALQ and FeLoBALQ subclasses, as different pieces of evidence from these objects support either the evolution scenario or the orientation scenario. LoBALQs appear to be more reddened than other QSOs \citep[e.g.][]{1992ApJ...390...39S,2003AJ....126.2594R,2007ApJ...662L..59F}, and this is consistent with the idea that they are young merger-induced QSOs caught in the process of blowing off their dust envelope, quenching star formation as a result \citep{2012ApJ...745..178F,2012MNRAS.420.1347F}. However, the star formation rate in LoBALQ host galaxies does not appear significantly different \citep{2012ApJ...755...29L,2016MNRAS.457.1371V} and there is no consensus about LoBALQs actually exhibiting larger Eddington ratios, as their younger age would imply \citep[compare e.g.][]{Schulze_2017,2012ApJ...757..125U}.

The identification of 7 previously unknown FeLoBALQs among the QUIPs represents a noteworthy addition to a class of objects estimated to account for only a tiny fraction of the whole QSO population \citep{2006ApJS..165....1T,2012ApJ...757..180D}. We present in this paper a preliminary analysis of all the NIR spectra acquired so far within the QUBRICS survey, with a specific focus on explaining the reasons for the relatively high (Fe)LoBALQ detection rate (which has been recently observed also in the SkyMapper survey, \citealt{10.1093/mnras/stz2955}). Despite the limitations set by the data sample, we are able to provide values of black hole mass and Eddington ratio for most of the targets, which give no indication that the FeLoBAL subsample is accreting at a higher ratio than other QSOs (as an evolutionary scenario would suggest). The analysis, performed with the packages {\sc Astrocook} \citep{10.1117/12.2561343} and {\sc QSFit} \citep{2017MNRAS.472.4051C} is aimed at providing the basis for a further, more detailed analysis of the individual targets.

The paper is organized as follows: in \autoref{sec:data} we describe how data were acquired and treated; in \autoref{sec:overview} we present a qualitative assessment of the spectra; in \autoref{sec:discussion} we discuss the statistics of the sample, with reference to the QUBRICS catalogue as a whole and to the general QSO population; finally, in \autoref{sec:conclusion} we draw the conclusions of the campaign. Magnitudes are expressed in the AB systems. Atomic transitions are denoted by their ionization state (\textsc{i} for neutral species, \textsc{ii} for singly ionized species, etc.) and, when required to avoid ambiguity, by their rest-frame vacuum wavelength in \SI{}{\angstrom} preceded by $\lambda$ (e.g.~\Ha{} ${\lambda}{6563}$).

\section{Data treatment}\label{sec:data}

\subsection{Acquisition}
In \citet[hereafter \pI]{2019ApJ...887..268C} we presented the first results of our survey, aimed at identifying previously unknown QSOs with $i\leq 18$ at $z\geq 2$ in the Southern Hemisphere using the photometric information from existing databases (Skymapper, Gaia DR2, WISE, and 2MASS). We implemented a new method based on the Canonical Correlation Analysis and extracted a preliminary list of about 1500 QSO candidates, 54 of which were spectroscopically confirmed as QSOs at $z>2.5$. The survey was further extended as documented in \citet[hereafter \pII]{2020arXiv200803865B}, bringing the number of confirmed QSOs at $z>2.5$ to $\NqubricsII{A_QSO_zg25}$ and the total number of confirmed QSOs to $\NqubricsII{A_QSO}$.

Alongside these sources, which were securely confirmed by spectroscopic observations in the optical band (``flag A''), the survey produced a remaining $\NqubricsII{flagB}$ candidates (``flag B'') whose exact nature was still uncertain, for a variety of reasons: i) spectra were noisy and could not provide a clear redshift indication; ii) spectra showed broad and numerous absorption features that could not be unambiguously associated to a single redshift solution; 3) only a single emission line was visible, preventing a secure identification.

The observational campaign discussed in this paper was aimed at assessing the nature of $\NqubricsIII{B_any}$ of the brightest flag-B candidates, including $\NqubricsIII{B_any_QUIP}$ targets with peculiar absorption features (QUIPs) and $\NqubricsIII{tot_nonQUIP}$ other bright targets. The details of the observations, including both optical and NIR observations, are given in \autoref{tab:summary}. Two flag-A candidates exhibiting QUIP features were added to the sample, bringing the total to $\NqubricsIII{tot_NIR}$. We acquired a total of 26 optical spectra (taking into account targets that were observed more than once) and 18 NIR spectra.

The acquisition of the optical spectra is described in \pI{} and \pII{}; here we summarize it as follows:
\begin{enumerate}
    \item $\NqubricsIII{efosc2}$ targets were observed with EFOSC2 at ESO-NTT during ESO period P103 (April–September 2019, PI A.~Grazian, proposal 0103.A-0746). Grism \#13 was used, with a wavelength range $\lambda\simeq\SIrange{3700}{9300}{\angstrom}$ and a central-wavelength FWHM of $\sim$\SI{21}{\angstrom} or $\sim$\SI{1000}{km/s} ($R\simeq 300$), with a $\ang{;;1.5}$ slit. The range of exposure times was $\SIrange{360}{600}{s}$.
    \item $\NqubricsIII{ldss3}$ targets were observed with LDSS3 at the Magellan Clay Telescope between September and November 2019. Grism VPH-ALL was used, with a wavelength range $\lambda\simeq\SIrange{4250}{10000}{\angstrom}$ and a resolution of $\sim$\SI{860}{} (FWHM $\simeq\SI{8}{\angstrom}$ or $\sim$\SI{350}{km/s}), with a $\ang{;;1.0}$ slit. The range of exposure times was $\SIrange{400}{600}{s}$.
    \item $\NqubricsIII{wfccd}$ targets were observed with WFCCD at the du Pont Telescope between August and September 2019. The blue grism was used, with a wavelength range $\lambda\simeq\SIrange{3600}{7600}{\angstrom}$ and a dispersion of $\SI{2}{\angstrom/pixel}$, together with a $\ang{;;1.6}$ slit, achieving a resolution of $\sim$\SI{800}{} (FWHM $\simeq\SI{6}{\angstrom}$ or $\sim$\SI{375}{km/s}). The range of exposure times was $\SIrange{400}{900}{s}$.
    \item $\NqubricsIII{mage}$ targets were observed with MagE at the Magellan Baade Telescope between October 2018 and November 2019, with a wavelength range $\lambda\simeq\SIrange{3300}{10000}{\angstrom}$ and a slit of $\ang{;;0.85}$ or $\ang{;;1.0}$, achieving a resolution from $\numrange{4100}{4800}$ (FWHM $\simeq\SIrange{1.4}{1.6}{\angstrom}$ or $\SIrange{60}{70}{km/s}$). The range of exposure times was $\SIrange{1200}{1800}{s}$.
    \item $\NqubricsIII{lrs}$ target was observed with DOLORES (LRS) at the Telescopio Nazionale Galileo in June 2021, with a wavelength range $\lambda\simeq\SIrange{3500}{8000}{\angstrom}$ and a slit of $\ang{;;1.5}$, achieving a resolution of $\sim$ \SI{390}{} (FWHM $\simeq\SI{15}{\angstrom}$ or $\sim$\SI{770}{km/s}). The total exposure time was $\SI{580}{s}$.
\end{enumerate}

The acquisition of NIR spectra was carried out with the FIRE spectrograph at the Magellan Baade Telescope between December 2018 and November 2019. The high throughput prism mode was used, with slits of $\ang{;;0.6}$ or $\ang{;;1.0}$; nominal resolution ranged from $R\simeq\numrange{400}{500}$ in the J band to $R\simeq\numrange{200}{300}$ in the K band. Four to six exposures of $t_\textrm{exp}\simeq\SI{126}{s}$ were taken for each target, depending on the observing conditions, for a total exposure time ranging from \SI{510}{s} to \SI{760}{s}.

\begin{table*}
    \setlength{\tabcolsep}{4pt}
	\centering
	\caption{Summary of the observations of QUIP targets, with approximate exposure times, sorted by ascending RA. SkyMapper ID from Data Release 1 \citep{2018PASA...35...10W} are provided for reference.}
	\label{tab:summary}
	\begin{tabular}{cccccccccccccc}
		\hline
		Name        &QUBRICS&SkyMapper&Flag&           &              &       & EFOSC2         & LDSS3          & WFCCD          & MagE           & LRS            & \multicolumn{2}{c}{FIRE}       \\
		            &ID     &ID (DR1)     &    &RA         &Dec           &mag$_i$&$t_\textrm{exp}$&$t_\textrm{exp}$&$t_\textrm{exp}$&$t_\textrm{exp}$&$t_\textrm{exp}$& $t_\textrm{exp}$& slit\\
		\hline
        J0008$-$5058&963183 &317765879&B   &00:08:11.96&$-$50:58:44.95&17.591 & $\SI{420}{s}$  &                &                & $\SI{1800}{s}$ &                & $\SI{760}{s}$   & $\ang{;;1.0}$\\
        J0010$-$3201&1030576&6425629  &B   &00:10:40.66&$-$32:01:11.14&16.770 &                & $\SI{600}{s}$  & $\SI{600}{s}$  &                &                & $\SI{510}{s}$   & $\ang{;;1.0}$\\
        J0140$-$2531&1035925&6814119  &B   &01:40:30.83&$-$25:31:37.48&17.291 & $\SI{360}{s}$  &                &                &                &                & $\SI{510}{s}$   & $\ang{;;1.0}$\\
        J0407$-$6245&921925 &314058510&B   &04:07:36.82&$-$62:45:49.28&16.346 &                & $\SI{400}{s}$  &                &                &                & $\SI{510}{s}$   & $\ang{;;1.0}$\\
        J0514$-$3854&999243 &11288048 &B   &05:14:21.33&$-$38:54:42.57&16.888 &                & $\SI{600}{s}$  & $\SI{500}{s}$  &                &                & $\SI{510}{s}$   & $\ang{;;1.0}$\\
        J1215$-$2129&814912 &64056228 &B   &12:15:02.40&$-$21:29:14.13&16.741 & $\SI{600}{s}$  &                &                &                &                & $\SI{760}{s}$   & $\ang{;;0.6}$\\
        J1318$-$0245&823202 &65974091 &B   &13:18:33.31&$-$02:45:36.22&17.876 &                &                &                &                & $\SI{580}{s}$  & $\SI{1010}{s}$  & $\ang{;;0.6}$\\
        J1503$-$0451&882537 &100099370&B   &15:03:50.13&$-$04:51:45.09&16.896 & $\SI{600}{s}$  &                &                &                &                & $\SI{510}{s}$   & $\ang{;;0.6}$\\
        J2012$-$1802&1052318&170868221&B   &20:12:25.59&$-$18:02:46.77&17.092 & $\SI{360}{s}$  &                &                &                &                & $\SI{760}{s}$   & $\ang{;;1.0}$\\
        J2018$-$4546&1089108&306376125&A   &20:18:47.29&$-$45:46:48.42&16.462 &                & $\SI{600}{s}$  &                & $\SI{1800}{s}$ &                & $\SI{760}{s}$   & $\ang{;;0.6}$\\
        J2105$-$4104&891578 &307248146&B   &21:05:26.94&$-$41:04:52.58&17.099 & $\SI{600}{s}$  &                & $\SI{600}{s}$  &                &                & $\SI{760}{s}$   & $\ang{;;1.0}$\\
        J2134$-$7243&990244 &305553291&B   &21:34:58.78&$-$72:43:11.83&16.889 &                & $\SI{600}{s}$  &                & $\SI{1800}{s}$ &                & $\SI{760}{s}$   & $\ang{;;0.6}$\\
        J2154$-$0514&862715 &4025749  &B   &21:54:56.69&$-$05:14:50.37&17.778 &                & $\SI{600}{s}$  &                &                &                & $\SI{760}{s}$   & $\ang{;;1.0}$\\
        J2157$-$3602&875768 &397340   &A   &21:57:28.21&$-$36:02:15.11&17.367 &                &                & $\SI{900}{s}$  &                &                & $\SI{760}{s}$   & $\ang{;;0.6}$\\
        J2222$-$4146&917913 &1098401  &B   &22:22:26.09&$-$41:46:29.99&16.228 &                & $\SI{600}{s}$  &                & $\SI{1200}{s}$ &                & $\SI{760}{s}$   & $\ang{;;0.6}$\\
        J2255$-$5404&892403 &308459017&B   &22:55:08.37&$-$54:04:14.01&16.975 &                &                & $\SI{400}{s}$  &                &                & $\SI{760}{s}$   & $\ang{;;1.0}$\\
        J2319$-$7322&846931 &305771865&B   &23:19:31.05&$-$73:22:56.45&17.263 &                &                & $\SI{400}{s}$  &                &                & $\SI{510}{s}$   & $\ang{;;1.0}$\\
        J2355$-$5253&962517 &308944978&B   &23:55:52.05&$-$52:53:50.37&17.665 & $\SI{420}{s}$  &                &                & $\SI{1800}{s}$ &                & $\SI{760}{s}$   & $\ang{;;0.6}$\\
		\hline
	\end{tabular}
\end{table*}

\subsection{Reduction and analysis}\label{sec:red_an}

The reduction of optical spectra taken with WFCCD, EFOSC2, and LDSS-3 has been described in \pI{} and \pII{}. MagE spectra were reduced using the CarPy pipeline\footnote{https://code.obs.carnegiescience.edu/mage-pipeline} \citep{2003PASP..115..688K,2000ApJ...531..159K}. The CarPy product is a sky subtracted, wavelength calibrated spectrum for each separate order. Every night a target was observed, an associated spectro-photometric standard star was also observed and used for relative flux calibration. Flux calibration was performed with {\sc iraf} \citep{1993ASPC...52..173T} routines. The task \textit{standard} was used to calibrate the flux of a standard star based on tabulated calibration data included in the {\sc iraf} database, consisting of wavelengths, calibration magnitudes and bandpass widths. The output was then used by the task \textit{sensfunc} in order to obtain the system sensitivity as a function of wavelength. This task also produces a revised extinction function based on the residuals to the input extinction table. In this case the {\sc ctio} extinction table has been used. Finally, the task \textit{calibrate} applies the sensitivity function to the source spectra that are now corrected for extinction and calibrated to the correct flux scale. The flux calibrated spectrum of each order has then been combined using the {\sc iraf} task \textit{scombine}, to obtain the 1D spectrum over the full wavelength range.

NIR spectra taken with the FIRE spectrograph were reduced with the FireHose IDL pipeline \citep{jonathan_gagne_2015_18775}, and in particular
with the procedure designed for prism spectra, which includes flat-fielding, manual wavelength calibration, and optimal extraction. A set of reference stars were also reduced with the same procedure, to be used for telluric absorption removal and flux calibration. Each star was individually associated with a given QSO and observed immediately before or after the QSO itself.

The set of reduced spectra were post-processed with the {\sc Astrocook} package\footnote{\url{https://github.com/DAS-OATs/astrocook}} for QSO spectral analysis \citep{2018SPIE10707E..23C,2020ASPC..522..187C,10.1117/12.2561343}. The post-processing included five steps:
\begin{enumerate}
\item\emph{Coaddition of the NIR spectra.} The extracted NIR exposures were combined into a single spectrum and rescaled to matching count values. {\sc Astrocook} creates a combined spectrum by retaining all the information from the contributing exposures, including the wavelengths and sizes of individual pixels. This combined spectrum was then rebinned into a log-wavelength grid with step $\Delta v=c\Delta\log\lambda=\SI{50}{\km\per\second}$; $\Delta\lambda\simeq\SI{1.67}{\angstrom}\times(\lambda/\SI{10000}{\angstrom})$, optimizing the wavelength range to the instrument setup adopted for each night. The reference star spectra were rebinned to the same wavelength grid of the associated QSO spectra, to maintain the compatibility.
\item\emph{Telluric absorption removal and flux calibration of the NIR spectra.} Reference star spectra were normalized to a black-body spectrum with proper effective temperature. The spectra were ``cleaned'' by visually identifying their most prominent absorption features and clipping them out. The resulting spectra are ideally free from features intrinsic to the stars and contain only the signatures of the telluric absorption and the instrument response. QSO spectra were divided by these calibrators, respecting the association between each QSO and the star observed closest in time. It is worth remarking that the flux calibration performed in this way is only relative: the spectral shape of the QSO is reconstructed, but it is not rescaled to proper physical units and it differs from the actual flux density profile by a constant factor.
\item\emph{Coaddition of the optical spectra.} Optical spectra were flux calibrated (at least in a relative sense) during reduction, and in general they are much less affected by telluric absorption than NIR spectra. Reduced exposures from different instruments were thus directly combined into a flux calibrated spectrum, rescaling their flux to matching values in the superposition regions. Noisy regions at the ends of the spectra were cut out before coaddition. The combined spectrum was then rebinned into a log-wavelength grid with step $\Delta v=\SI{300}{\km\per\second}$; $\Delta\lambda\simeq\SI{10}{\angstrom}\times(\lambda/\SI{10000}{\angstrom})$, to accomodate for the different resolving power of the instrument and avoid oversampling.
\item\emph{Merging of the optical and NIR spectra.} The combination and rescaling procedure was adopted also to merge optical and NIR spectra of each target into a final spectrum. In some cases, the combination procedure highlighted a discrepancy between the wavelength calibration in the two bands. As the discrepancy was within the estimated accuracy of manual wavelength calibration as implemented by the FireHose pipeline, we decide to correct it by rigidly shifting the NIR spectrum along wavelengths to match the optical spectrum, using shared features as a reference. The typical shift was of $\sim$\SI{10}{\angstrom}, never exceeding \SI{40}{\angstrom}. The combined spectrum was further downsampled into a log-wavelength grid; the step was $\Delta v=\SI{500}{\km\per\second}\simeq\SI{16.7}{\angstrom}\times(\lambda/\SI{10000}{\angstrom})$ for all targets, except for J1318$-$0245, which was resampled at $\Delta v=\SI{800}{\km\per\second}\simeq\SI{26.7}{\angstrom}\times(\lambda/\SI{10000}{\angstrom})$ in account of the lower resolution of the LRS optical spectrum.
\item\emph{Photometric adjustment of flux calibration.} Merged spectra were further calibrated using photometry from the SkyMapper Data Release 1 \citep{2018PASA...35...10W}. Flux densities were integrated within the SkyMapper photometric bands and compared with the available magnitudes. We assumed that the relative flux calibration performed in the previous steps was good enough to reconstruct the shape of the spectrum up to a normalization factor, and we computed it as the average of the correction factors extracted from the available bands. The resulting spectra are flux-calibrated in an absolute sense and are suitable to infer the properties of the emitting sources from flux measurements.
\item\emph{Redshift estimation.} A first estimate of the emission redshifts was obtained by shifting an emission line mask along the spectra and visually finding the better alignment with the QSO emission features. The mask contained transitions from the Lyman (\Lya{} ${\lambda}{1216}$, \Lyb{} ${\lambda}{1026}$, \Lyg{} ${\lambda}{973}$) and Balmer series (\Ha{} ${\lambda}{6563}$, \Hb{} ${\lambda}{4861}$, \Hg{} ${\lambda}{4340}$) and metal transitions 
(\NV{} ${\lambda}{1241}$; \SiII{} ${\lambda}{1304}$; \SiIV{} ${\lambda}{1398}$; \CIV{} ${\lambda}{1549}$; \AlIII{} ${\lambda}{1859}$; \MgII{} ${\lambda}{2800}$); the list of transitions was adjusted to match the actual presence of features and the wavelength coverage of the spectrum. The redshift estimation was then improved by shifting each merged spectrum to rest frame and computing the cross-correlation with the above-mentioned QSO spectrum template in the region of the Balmer series emissions. The uncertainty of the estimation was assessed with a bootstrap method similar to the one described by \cite{Peterson1998}: we created an ensemble of 100 realizations for each merged spectrum using random sampling with replacement and determined the redshift and its error from the statistics of the cross-correlation maxima over the ensemble. All the errors were below $10^{-3}$; combined with the uncertainty on wavelength calibration, we conservatively assume a maximum redshift uncertainty of $\pm 0.01$ for all our targets.
\item\emph{Reddening correction.}
Flux calibrated spectra were cleaned from spikes and other spurious features and corrected for both Galactic and instrinsic extinction, using the parametrization by \citet{1994-odonnell-updateCCM} and
assumed a total selective extinction $A(V)/E(B-V)=3.1$. Intrinsic extinction was estimated by shifting each merged spectrum to rest frame and comparing it with a QSO spectrum template (combined from \citealt{2001AJ....122..549V} and \citealt{2006ApJ...640..579G}). We tested values of intrinsic colour excess $E(B-V)$ between 0.00 and 0.30, with a step of 0.05\footnote{Given the quality of the data, we refrained from determining the intrinsic $E(B-V)$ at a better precision, considering that its impact on the continuum normalization at wavelength larger than $\sim$\SI{4500}{\angstrom} rest-frame would have been nevertheless negligible; see \autoref{sec:cont}.}, and visually selected the value that yielded the best agreement with the template in the full spectral range. The $E(B-V)$ values used to de-redden the spectra (both Galactic and intrinsic) are listed in \autoref{tab:analysis}. We also created composite spectra for the two groups of targets described in \autoref{sec:overview}. These composite spectra were obtained from rest-frame spectra (de-reddened and non de-reddened) by rebinning them to a fixed log-wavelength grid with step $\Delta v=c\Delta\log\lambda=\SI{800}{\km\per\second}$; $\Delta\lambda\simeq\SI{16.7}{\angstrom}\times(\lambda/\SI{10000}{\angstrom})$, normalizing the flux at \SI{3800}{\Angstrom} rest-frame, and computing their geometric mean.
\end{enumerate}

Some of the target characteristics determined by this analysis are listed in \autoref{tab:analysis}. The whole analysis procedure was firstly run using the {\sc Astrocook} graphical user interface, to define the steps and the parameters. Once the procedure was frozen, it was translated into a set of JSON files and Bash scripts to allow for a cascade execution. Both the input data and the processing files are released together with the processed data as ancillary material to this paper (see ``Data availability'' below).

\subsection{Modeling the broad band QSO emission}\label{sec:cont}

An estimate of the broad band continuum emitted by the QSO would require simultaneous observations spanning several bands at optical and NIR wavelengths. Here we attempt to estimate the slope and luminosity of the QSO continuum using just the NIR spectrum and the approach described in \citet{2017MNRAS.472.4051C}. In particular, we used the {\sc QSFit}\footnote{\url{https://qsfit.inaf.it/}} package to constrain a global model spanning the whole wavelength range covered by the NIR spectrum, and simultaneously fitting all relevant spectral components. We excluded from the modeling the regions most affected by telluric absorption (\SIrange{13500}{14500}{}, \SIrange{18000}{19500}{}, and $>$\SI{25000}{\Angstrom} in the observed frame) and most contaminated by metal absorption ($<$\SI{2800}{\Angstrom} rest-frame). The main model components are: the QSO continuum itself (modelled as a simple power law); the blended \FeII{} and \FeIII{} emission lines at optical/ultraviolet (rest-frame) wavelengths; and the most prominent emission lines (\SiIV{}, \CIV{}, \CIII{} ${\lambda}{1909}$, \MgII{}, \Hg{}, \Hb{}, \OIII{} ${\lambda\lambda}{4959,5007}$ and \Ha{}).  

\begin{figure*}
    \centering
	\includegraphics[width=0.85\textwidth]{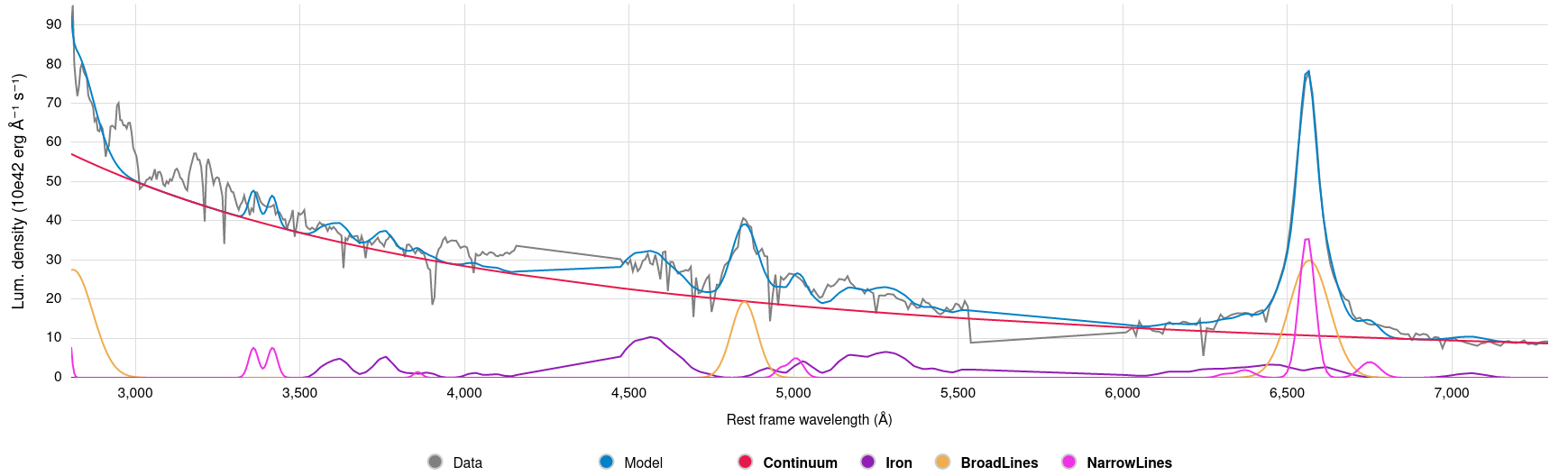}
    \caption{NIR spectrum of J2105$-$4104 and best fit model.  Although the model does not account for all the features in the observed spectra, it still provides a reliable estimate of the broad band QSO continuum (here modelled as a power law). The two emission lines at \SI{4861}{\angstrom} and \SI{6563}{\angstrom} rest-frame are $H_\beta$ and $H_\alpha$, respectively.}
    \label{fig:QSFit_J2105}
\end{figure*}

The best fit parameters and their 1-$\sigma$ uncertainties (as obtained with the Fisher matrix method, i.e. by considering the square root of the diagonal elements of the covariance matrix) are shown in \autoref{tab:analysis}. An example of fitted model is shown in \autoref{fig:QSFit_J2105}. The spectral range used to fit the model was different for each QSO, depending on redshift and presence of telluric absorption.

The continuum slope was successfully estimated in 17 cases out of \NqubricsIII{tot}: the continuum of J0140$-$2531 was not compatible with a power-law profile since it shows systematically negative residuals at wavelengths larger than $\sim$\SI{3300}{\Angstrom} (rest-frame), hence we decided to neglect the analysis for this target, rather than using a more complicated model, and maintain the uniformity of the procedure. The continuum of J2319$-$7322 is also uncertain, because it was constrained only for $\lambda>\SI{4000}{\angstrom}$; below this limit the shape was not compatible with a power-law profile, much like for J0140$-$2531 above.

The FWHMs and luminosities of the Balmer series lines were successfully estimated for at least one line in 16 cases out of 17, along with the corresponding luminosity at $\lambda=\SI{5100}{\angstrom}$. In the case of J2157$-$3602, the relatively high redshift prevented the observation of the Balmer series in the wavelength range covered by FIRE. In other cases, we could not obtain a reliable estimate of the parameters for all the emission lines since the profiles are affected by strong telluric absorptions, especially at wavelengths slightly shorter than the \MgII{} and \Hb{} lines. The end-of-scale FWHM value of \SI{15e3}{\km\per\second} denotes the cases where the fitting procedure was not able to converge to a value below this figure. We regard these values as mere upper limits for the real FWHMs of the lines. Additionally, the 1-$\sigma$ error of FWHM values does not take into account the uncertainty associated with the power-law continuum fitting and with other non-modeled emission features. Values of $L_{\textrm{H}\alpha}$, $L_{\textrm{H}\beta}$, and $\lambda L_{5100}$ are in principle affected by errors in the intrinsic $E(B-V)$ used in de-reddening; even an error as large as 0.05 would nevertheless produce only a $>$2\% variation in the flux level at $\lambda\gtrsim\SI{4500}{\angstrom}$ rest-frame, rendering it negligible when compared to the uncertainties resulting from fitting.

\begin{figure}
    \centering
	\includegraphics[width=0.49\textwidth]{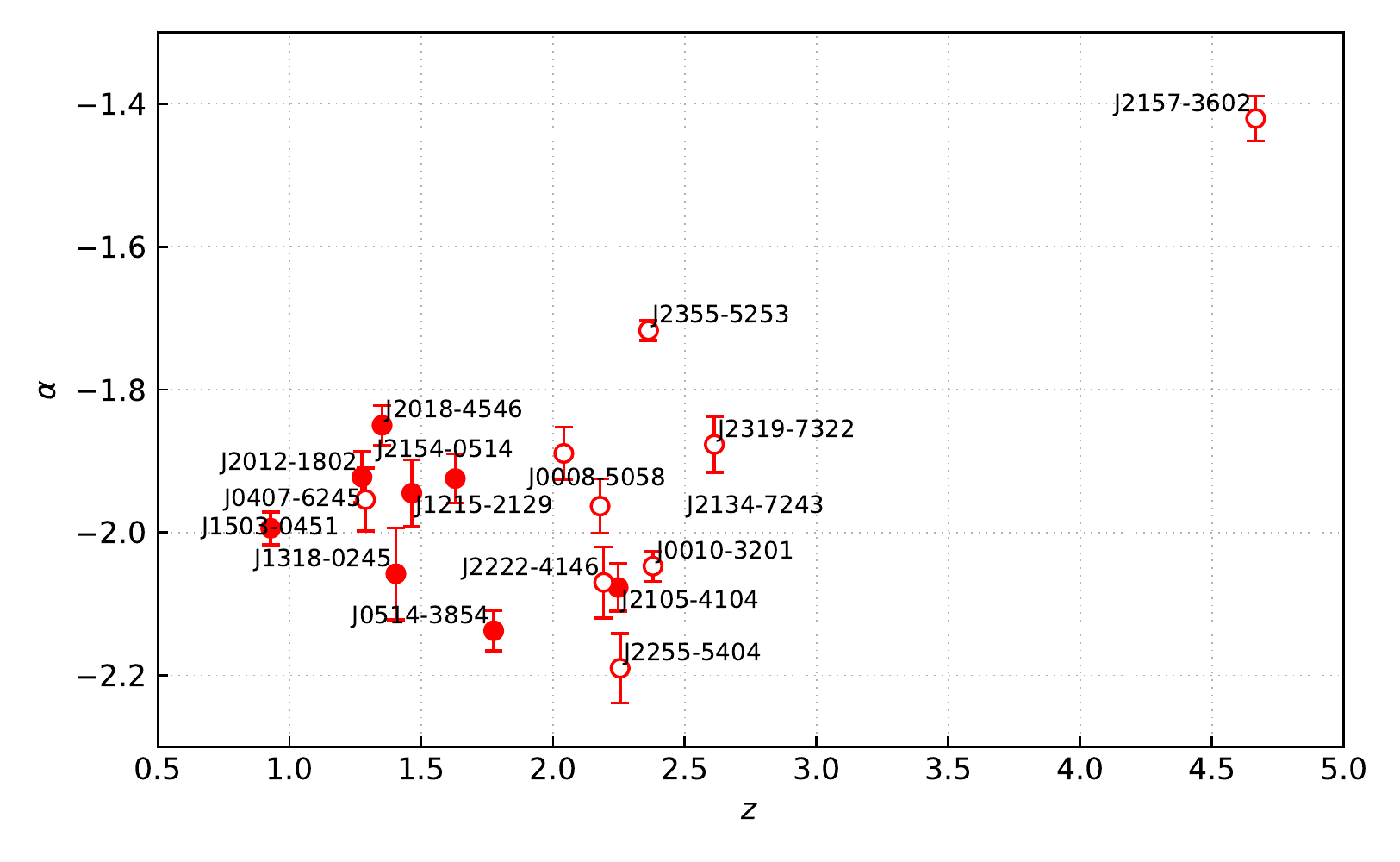}
    \caption{Broad band QSO continuum slope ($F_\lambda \propto \lambda^\alpha$) vs. source redshift for all the targets in our sample. FeLoBALQs (see \autoref{sec:felobal}) are denoted with filled circles, while other targets are denoted with empty circles. Error bars are the 1-$\sigma$ uncertainties, as derived from the Fisher matrix.}
    \label{fig:z_vs_slope}
\end{figure}

On the other hand, the Balmer decrement (ratio of integrated luminosities of H$\alpha$ and H$\beta$) is $2.9\pm0.9$ in agreement with, e.g., \citealt{2019-Lu-BalmerDecrement}. The distribution of continuum slopes is $-1.89\pm0.18$ ($L_\lambda \propto \lambda^\alpha$) and the values show no dependence on redshift (see \autoref{fig:z_vs_slope}).

\section{Overview of the spectra}\label{sec:overview}

\begin{table*}
	\centering
	\caption{Main characteristics of the observed targets after analysis. Q, H, L, and F are abbreviations for QUIP, HiBALQ, LoBALQ, and FeLoBALQ, respectively. No assigned class means that the target is neither a QUIP nor a (Hi/Lo/FeLo)BALQs.
	The determination of the emission redshift $\zem$ is described in \autoref{sec:red_an}. The fit of the power-law index $\alpha$ to the continuum emission is discussed in \autoref{sec:cont}.}
	\label{tab:analysis}
	\begin{tabular}{
	    c
	    c
	    c
	    c
	    c
	    r@{${}\pm{}$}l
	    r@{${}\pm{}$}l
	    r@{${}\pm{}$}l
	    r@{${}\pm{}$}l
	    r@{${}\pm{}$}l
	    r@{${}\pm{}$}l
	}
		\hline
		QUBRICS ID    & class & $\zem$ & \multicolumn{2}{c}{$E(B-V)$} & \multicolumn{2}{c}{$\alpha$} & \multicolumn{2}{c}{$\textrm{FWHM}_{\textrm{H}\alpha}$} & \multicolumn{2}{c}{$\textrm{FWHM}_{\textrm{H}\beta}$} & \multicolumn{2}{c}{$L_{\textrm{H}\alpha}$} & \multicolumn{2}{c}{$L_{\textrm{H}\beta}$} & \multicolumn{2}{c}{$\lambda L_{5100}$} \\
		& & & gal. & intr. & \multicolumn{2}{c}{} & \multicolumn{2}{c}{\SI{e3}{\km\per\second}} & \multicolumn{2}{c}{\SI{e3}{\km\per\second}} & \multicolumn{2}{c}{\SI{e42}{\erg\per\second}} & \multicolumn{2}{c}{\SI{e42}{\erg\per\second}} & \multicolumn{2}{c}{\SI{e44}{\erg\per\second}} \\
		\hline
J0008$-$5058 & Q,H & 2.041 & 0.014 & 0.10 & $-1.89$ & 0.04       & 7.5 & 1.2            &  6.8 & 1.1              & 2300 & 200           &  770 &  100.         &  490 & 20            \\
J0010$-$3201 & Q,H & 2.379 & 0.013 & 0.00 & $-2.05$ & 0.02       & 8.4 & 0.9            & \multicolumn{2}{c}{<15} & 2700 & 200           & 1300 &  150.         &  500 & 20            \\
J0140$-$2531 & Q   & 2.947 & 0.012 & 0.00 & \multicolumn{2}{c}{} & \multicolumn{2}{c}{} & \multicolumn{2}{c}{}    & \multicolumn{2}{c}{} & \multicolumn{2}{c}{} & \multicolumn{2}{c}{} \\
J0407$-$6245 & Q,L & 1.289 & 0.032 & 0.00 & $-1.95$ & 0.04       & 6.3 & 1.0            &  5.6 & 1.1              &  500 &  80           &  171 &   27          &  200 &  10           \\
J0514$-$3854 & Q,F & 1.775 & 0.035 & 0.00 & $-2.14$ & 0.03       & 7.6 & 0.8            & 14.7 & 3.1              & 1130 &  80           &  440 &  130          &  330 &  10           \\
J1215$-$2129 & Q,F & 1.464 & 0.051 & 0.20 & $-1.94$ & 0.05       & 9.7 & 1.3            & 13.8 & 4.0              & 1620 & 180           &  510 &  150          &  610 &  20           \\
J1318$-$0245 & Q,F & 1.404 & 0.023 & 0.15 & $-2.06$ & 0.06       & 7.0 & 2.2            &  4.7 & 1.0              & 260  &  70           &  180 &   20          &  150 &   6           \\
J1503$-$0451 & Q,F & 0.929 & 0.081 & 0.15 & $-1.99$ & 0.02       & 7.6 & 0.5            & 11.0 & 0.9              & 920  &  40           &  350 &   30          &  182 &   4           \\
J2012$-$1802 & Q,F & 1.275 & 0.104 & 0.25 & $-1.92$ & 0.03       & 6.8 & 0.7            &  6.9 & 0.9              & 1550 & 110           &  610 &   70          &  470 &  10           \\
J2018$-$4546 & Q,F & 1.352 & 0.027 & 0.15 & $-1.85$ & 0.03       & 6.9 & 0.8            & 10.2 & 1.4              & 1320 &  90           &  580 &   70          &  510 &  10           \\
J2105$-$4104 & Q,F & 2.247 & 0.029 & 0.10 & $-2.08$ & 0.03       & 6.5 & 0.5            &  5.0 & 0.5              & 5000 & 300           & 1780 &  140          &  940 &  30           \\
J2134$-$7243 &     & 2.178 & 0.037 & 0.00 & $-1.96$ & 0.04       & 5.9 & 1.1            & 10.3 & 2.3              &  880 & 140           &  620 &  110          &  400 &  20           \\
J2154$-$0514 & Q,F & 1.629 & 0.022 & 0.20 & $-1.92$ & 0.03       & 7.4 & 1.4            &  5.4 & 0.8              &  860 &  80           &  360 &   60          &  330 &  10           \\
J2157$-$3602 & Q,H & 4.665 & 0.013 & 0.05 & $-1.42$ & 0.03       & \multicolumn{2}{c}{} & \multicolumn{2}{c}{}    & \multicolumn{2}{c}{} & \multicolumn{2}{c}{} & 4400 & 200           \\
J2222$-$4146 &     & 2.192 & 0.013 & 0.00 & $-2.07$ & 0.05       & 9.3 & 3.0            & \multicolumn{2}{c}{<15} & 1440 & 290           & 1020 &  210          &  650 &  30           \\
J2255$-$5404 & Q,H & 2.255 & 0.013 & 0.15 & $-2.19$ & 0.05       & 6.7 & 0.7            &  8.5 & 1.3              & 3700 & 300           & 1860 &  220          &  810 &  40           \\
J2319$-$7322 & Q,H & 2.612 & 0.025 & 0.00 & $-1.88$ & 0.04       & 6.5 & 1.1            &\multicolumn{2}{c}{<15}  & 1390 & 190           & 1260 &  140          &  620 &  20           \\
J2355$-$5253 &     & 2.363 & 0.013 & 0.00 & $-1.72$ & 0.01       & 2.6 & 0.6            &  1.6 & 1.9              &  550 &  60           &  132 &   42          &  310 &  10           \\
\hline
	\end{tabular}
\end{table*}

\autoref{tab:analysis} lists the main characteristics of the QUIPs, resulting from the analysis with {\sc Astrocook} and QSFit. In $\NqubricsIII{halpha}$ out of $\NqubricsIII{tot_QUIP}$ cases the redshift is firmly established by the detection of a strong \Ha{} emission line, frequently complemented by a comparably clear detection of \Hb{} ($\NqubricsIII{hbeta}$ cases out of $\NqubricsIII{halpha}$) and \Hg{} ($\NqubricsIII{hgamma}$ cases out of $\NqubricsIII{halpha}$). Occasionally, the \Hb{} and \Hg{} lines fall within the strong telluric \ce{H2O} absorption band at $\sim$\SI{14000}{\angstrom}, but they are marginally detectable anyway.

The line mask described \autoref{sec:red_an} helped in visually identifying notable absorption systems. According to the classification given in \autoref{tab:analysis}, $\NqubricsIII{tot_BAL}$ out of $\NqubricsIII{tot_QUIP}$ ($\NqubricsIII{tot_BAL_f}$) QUIPs are BALQs, either HiBALQs ($\NqubricsIII{tot_HiBAL}$ out of $\NqubricsIII{tot_BAL}$, $\NqubricsIII{tot_HiBAL_f}$) or LoBALQs ($\NqubricsIII{tot_LoBAL}$ out of $\NqubricsIII{tot_BAL}$, $\NqubricsIII{tot_LoBAL_f}$). $\NqubricsIII{tot_FeLoBAL}$ out of $\NqubricsIII{tot_LoBAL}$ LoBALQs ($\NqubricsIII{tot_FeLoBAL_f}$) show evidence of strong \FeII{} absorption (FeLoBALQs).

The spectra of the individual targets are shown at the end of the paper in \autoref{fig:felobal} (FeLoBALQs) and \autoref{fig:non_felobal} (non-FeLoBALQs). A description of the same targets in the two groups is given in \autoref{sec:felobal} and \autoref{sec:other}, with targets sorted by their right ascension. Composite spectra for the two groups are displayed in \autoref{fig:composite} (blue and green for FeLoBALQs and non-FeLoBALQs, respectively), superimposed to the individual spectra in transparency. Identified emission and absorption features are highlighted in the plots with dotted bars and described in the text moving bluewards of the QSO emission redshift to the observer.

In the formulae, we will refer to FeLoBALQs and non-FeLoBALQs with subscripts $\textrm{F}$ and $\textrm{nF}$, respectively. We will also use $\textrm{H}$ and $\textrm{L}$ to refer to HiBALQs and LoBALQs specifically. Values of quantities labeled with these subscripts are the average and the standard deviation across the respective groups.

\subsection{FeLoBAL QSOs}\label{sec:felobal}

A significant fraction of QUIP targets ($\NqubricsIII{tot_FeLoBAL_f}$) shows BAL features from ions like \MgII{}, \AlIII{}, and  \FeII{}, sometimes alongside absorption from higher-ionization ions like \CIV{}, \SiIV{}, and \NV{}. Low-ionization BALQs with \FeII{} absorption systems, or FeLoBALQs, are a widely-investigated class of objects \citep[see e.g.][]{1993ApJS...88..357K,2007ApJ...662L..59F,2012MNRAS.420.1347F}. Low-ionization BAL features have large column densities, thick enough to extend beyond the hydrogen ionization front \citep{1987MNRAS.229..371H}. FeLoBALQs in particular reach the highest column densities among BALQs \citep{2014ApJ...783...58L}, in some cases resulting in a saturated trough bluewards of the \MgII{} ${\lambda}{2800}$ emission (these objects are also called ``overlapping trough'' QSOs or OFeLoBALQs; see e.g.~\citealt{2014ApJ...783...58L,2002ApJS..141..267H}). A detailed analysis of such complex systems is non-trivial and requires a dedicated approach \citep[e.g.][]{2020ApJ...891...53C} which is beyond the scope of this paper. Here we provide only a qualitative description of the most relevant features we detected on the QUIP FeLoBAL spectra, which are displayed in \autoref{fig:felobal}.

\paragraph*{J0514$-$3854 ($\boldsymbol{\zem=1.775}$, $\boldsymbol{\alpha=-2.14}$).} A QSO with a strong \MgII{} emission, which corroborates the redshift estimation, mainly based on \Ha{} (which falls within a telluric band and may appear distorted for this reason; significant residuals of telluric removal are observed at $\lambdarf\simeq\SI{4950}{\angstrom}$ and $\lambdarf\simeq\SI{6740}{\angstrom}$). Broad \SiIV{}, \AlIII{}, \FeII{}, and \MgII{} absorption lines are observed, confirming the identification as a FeLoBALQ. The absorption system has a complex velocity structure with at least three components; the strongest one has a redshift $z=1.678$. No saturated trough is present. An absorption system at $\lambdarf\simeq\SI{1970}{\angstrom}$ has no secure identification.

\paragraph*{J1215$-$2129 ($\boldsymbol{\zem=1.464}$, $\boldsymbol{\alpha=-1.94}$).} Possibly the most peculiar of all QUIPs. The emission redshift is firmly determined by a strong \Ha{} line, despite the fact that \Hb{} is weak and \Hg{} barely detected, and is compatible with an associated absorber at $z=1.463$, observed in \MgII{}, \FeII{}, \CIV{} (the identification of these species must be regarded as tentative). The main absorption line exhibits a wide and shallow profile, which probably emerges from an unresolved velocity structure. No line appears to be saturated.

\paragraph*{J1318$-$0245 ($\boldsymbol{\zem=1.404}$, $\boldsymbol{\alpha=-2.06}$).} A BALQ with a strong associated absorption system at $z=1.390$, observed in \MgII{} ad \FeII{}, at $\Delta v\simeq\SI{-1.8e3}{km/s}$ with respect to the emission redshift, securely determined from the Balmer series (\Ha{}, \Hb{}, and \Hg{}). A second saturated system is likely identified as \MgII{} at $z=1.344$, but lacks a clear \FeII{} counterpart. The region bluewards from the \FeII{} absorption is populated by several systems that could be identified only through accurate modeling, which is beyond the scope of this paper.

\paragraph*{J1503$-$0451 ($\boldsymbol{\zem=0.929}$, $\boldsymbol{\alpha=-1.99}$).} At $\zem=0.930$, this is the lowest-redshift QUIP in our list. The emission redshift is consistent with the possibile identification of \Pag{} ${\lambda}{1094}$; a possible \MgII{} emission is almost completely suppressed by an associated absorber at $z=0.923$, which is observed also in \FeII. This absorbtion system also includes a lower-redshift component  ($z=\numrange{0.888}{0.892}$) and a not clearly resolved \FeII{} component at slightly higher redshift. \FeII{} resonance is also observed at shorter wavelengths; accurate modeling is required for a detailed identification of all the absorption features.

\paragraph*{J2012$-$1802 ($\boldsymbol{\zem=1.275}$, $\boldsymbol{\alpha=-1.92}$).} Also this QSO show emission lines other than the Balmer series (\OII{} ${\lambda}{3729}$, \MgII); \MgII{} in particular appears to be partly obscured by an associated absorber at $z=1.261$, observed also in \FeII, and \AlIII. In fact, at least two other absorption systems with the same transitions are observed at $z=1.175$ and $z=1.036$. These three systems together explain most of the strongest feature observed along the line of sight; as in the other cases, the level of unabsorbed continuum is hard to determine and blended absorption may actually be responsible for a consistent decrement in the observed flux bluewards of the \MgII{} emission.

\paragraph*{J2018$-$4546 ($\boldsymbol{\zem=1.352}$, $\boldsymbol{\alpha=-1.85}$).} This QUIP is a quite clear example of OFeLoBALQ, with an almost completely saturated absorption trough bluewards of \MgII{} emission. Two associated \MgII{}/\FeII{} broad absorption lines are observed at the emission redshift and at $\Delta v\simeq\SI{-8.5e3}{km/s}$ with respect to the emission redshift ($z=1.286$; this one possibly including \FeII{} ${\lambda}{2344}$ alongside \FeII{} ${\lambda\lambda}{2382,2600}$). Partial coverage likely accounts for the fact that the lines do not reach the zero level, despite exhibiting a clearly saturated profile. The amount of blended absorption prevents the secure identification of additional systems along the the line of sight.

\paragraph*{J2105$-$4104 ($\boldsymbol{\zem=2.247}$, $\boldsymbol{\alpha=-2.08}$).} This object shows features typical of both LoBALQs (\MgII{}, \FeII{}, \AlIII{} absorption) and HiBALQs (\SiIV{}, \CIV{} absorption). Low-ionization absorbers are observed at $z=2.231,2.153$; a narrower system (possibly a mini-BAL) is observed along the line of sight at $z=1.568$ (at $\Delta v\simeq\SI{-6.3e4}{km/s}$). High-ionization absorbers, on the other hand, are observed at $z=2.215,2.085,2.026,1.926$. These identifications are not enough to explain all the absorption lines bluewards of the \MgII{} emission; most of these lines are narrow ($\textrm{FWHM}\sim\SI{e3}{km/s}$) and not too much affected by blending.

\paragraph*{J2154$-$0514 ($\boldsymbol{\zem=1.629}$, $\boldsymbol{\alpha=-1.92}$).} For this QSO, the Balmer series is complemented by marginally detected \MgII{} and possibly \AlIII{} emission lines. An almost saturated system is observed at $z=1.611$ ($\Delta v\simeq\SI{-2.1e3}{km/s}$) in \MgII{}, \FeII{}, and \AlIII{}; \FeII{} is particularly strong, corroborating the identification of this QUIP as a FeLoBALQ. The system may have an absorption ``tail'', observed for \MgII{} at $\lambdarf\simeq\SIrange{2630}{2760}{\angstrom}$ and for \FeII{} ${\lambda}{2600}$ at $\lambdarf\simeq\SIrange{2420}{2540}{\angstrom}$. A second weaker system at $z=1.563$ is observed only in \MgII{} and \FeII{} ${\lambda}{2382}$, and a third system at $z=1.069$ in \MgII{} and \FeII{} ${\lambda\lambda}{2382,2600}$. Also in this cases, a dense pattern of absorbers below $\lambdarf\simeq\SI{1800}{\angstrom}$ is lacking a secure identification.

\subsection{Other QSOs}\label{sec:other}

Some QUIP targets do not show significant \FeII{} absorption in their spectra, and derive their peculiar character from individual features (most notably, broad metal absorption complex at high ionization) which are described in more detail below. We include in this group also the non-QUIP targets that were selected for NIR observations, to secure their identification.

\paragraph*{J0008$-$5058 ($\boldsymbol{\zem=2.041}$, $\boldsymbol{\alpha=-1.89}$).} This QSOs shows an extended list of clearly detected emission lines at the emission redshift in addition to the Balmer series, including \MgII{}, \CIII{}, \AlIII{}, \CIV{}, \SiIV{}, \SiIII{}, \NV{}, and \Lya{}. An associated absorber at $z=2.0113$ ($\Delta v\simeq\SI{-3.0e3}{km/s}$) is observed in \MgII{}, \AlIII{}, \CIV{}, \SiIV{}, \NV{}, and \Lya{}, identifying this object as a HiBALQ. A second component of this absorber at $z=1.986$ ($\Delta v\simeq\SI{-5.4e3}{km/s}$) is possibly observed in \CIV{} and \Lya{}. Other absorption lines bluewards of the \MgII{} emission remain unexplained.

\paragraph*{J0010$-$3201 ($\boldsymbol{\zem=2.379}$, $\boldsymbol{\alpha=-2.05}$).} Another HiBALQ, exhibiting a particularly strong \CIV{}/\SiIV{} forest with at least two BAL features. Emission lines of \MgII{}, \AlIII{}, and \CIV{} at the redshift of the Balmer series appear significantly masked by an absorber at slightly higher redshift ($z=2.393$, $\Delta v\simeq\SI{1.2e3}{km/s}$). Two possible BAL systems, almost saturated in \CIV{}, \SiIV{}, and \Lya{}, are observed at $z=2.295$ and $z=2.126$ (with absorption also in \MgII{}, \AlIII{}, and \NV{}). Both systems display complex velocity patterns that we did not try to model, but give rise to comparable line profiles in \CIV{}, \SiIV{}, and \Lya{}.

\paragraph*{J0140$-$2531 ($\boldsymbol{\zem=2.947}$).} This QUIP has the second highest redshift in our sample, mainly anchored to the relative position of the \MgII{}, \CIV{}, and \NV{} emission lines; \Lya{} and \Lyb{} are marginally detected at the emission redshift. The object does not appear to be BALQ, showing only relatively weak absorptions in \CIV{} and \SiIV{} at $z=2.848$ and $z=2.746$ (the redder of the two absorbers, which is also the stronger, has also a clear \HI{} counterpart). The \Lya{} forest along the line of sight to this object also appears relatively under-absorbed. A strong absorption at $\sim$\SI{1050}{\Angstrom} rest-frame may be associated to a \SIV{} ${\lambda}{1063}$ resonance closer to the emission redshift.

\paragraph*{J0407$-$6245 ($\boldsymbol{\zem=1.289}$, $\boldsymbol{\alpha=-1.95}$).} A LoBALQ with almost no detectable \FeII{} absorption, this object displays \MgII{} and \AlIII{} emission lines complementing the Balmer series (which appears comparatively weaker than in other objects); the low-ionization metal emission is not entirely suppressed by an associated absorber slightly more redshifted than the QSO itself ($z=1.310$, corresponding to $\Delta v\simeq\SI{2.7e3}{km/s}$ with respect to the emission redshift). A strong \MgII{}/\AlIII{} absorption system with a complex velocity structure is observed bluewards of the emission, with at least two components at $z=1.227$ and $z=1.193$ ($\Delta v\simeq\SI{-8.2e3}{km/s}$ and $\SI{-1.29e4}{km/s}$ respectively). Sparse absorption bluewards of the \MgII{} emission may be possibly due to \FeII{}, an hypothesis that could only be confirmed by an accurate modeling.

\paragraph*{J2134$-$7243 ($\boldsymbol{\zem=2.178}$, $\boldsymbol{\alpha=-1.96}$).} A rather featureless, non-QUIP QSO, with no notable metal absorption complex at either high or low ionization. The redshift is determined mainly from the \Ha{} and \MgII{} emission lines; \Hb{} is very weak and \Hg{} is undetected (it is expected to fall in a region of strong telluric absorption). Enhanced emission is observed at the emission redsfhit for \AlIII{}, \CIV{}, and particularly \Lya{}; the only unequivocal absorption feature is in fact identified as a \Lya{}, precisely at the QSO emission redshift.

\paragraph*{J2157$-$3602 ($\boldsymbol{\zem=4.665}$, $\boldsymbol{\alpha=-1.42}$).} The relatively high redshift of this QSO is determined almost only on the basis of the \MgII{} emission line at $\lambda\simeq\SI{15860}{\angstrom}$, the Balmer series falling well outside the wavelength range covered by our observations; however, marginal detections of \CII{} ${\lambda}{1908}$, \AlIII{}, \CIV{}, \SiII{}, \NV{}, and possibly also \SiIV{}, \Lya{}, and \OVI{} ${\lambda}{1037}$ confirm the assessment. The QSO shows a strong associated metal system at $z=4.608$ ($\Delta v\simeq\SI{-3.0e3}{km/s}$), observed in \CIV{}, \SiIV{}, \NV{} and \OVI{}; the corresponding \Lya{} and \Lyb{} signatures are not too clear. A second \SiIV{}/\SiIII{} absorption at $z=4.438$ may have a correspondence in the \Lya{} (which is in general considerably opaque, as expected at this redshift); the whole region $\lambda\simeq\SIrange{7600}{7900}{\angstrom}$, bluewards of the \SiIV{} ${\lambda}{1398}$ emission is quite peculiar and possibly contaminated by uncorrected telluric absorption.

\paragraph*{J2222$-$4146 ($\boldsymbol{\zem=2.192}$, $\boldsymbol{\alpha=-2.07}$).} A non-QUIP object quite similar to J2134$-$7243, with weak emission lines (\Ha{} and \MgII{}, constraining the emission redshift; \Hb{}, \SiIV{}, \Lya{}) and no remarkable absorption feature outside the \Lya{} forest. Weak, extended absorption features bluewards of the \Lya{} emission lack a clear identification, as well as a strong absorption complex in the \Lya{} forest, at $\sim$\SI{1100}{\angstrom} rest-frame, which may be associated to \FeIII{} ${\lambda}{1123}$. One could be led to categorize this QUIP as a BALQ, based on the latter complex, but the identification is uncertain and not corroborated by high- or low-ionization metal absorption.

\paragraph*{J2255$-$5404 ($\boldsymbol{\zem=2.255}$, $\boldsymbol{\alpha=-2.19}$).} Another HiBALQ, with weak \AlIII{} and almost absent \MgII{} and \FeII{} absorption. Among the emission lines, \Hb{} is marginally detected and \Hg{} is partly contaminated by a telluric band, but the determination of the emission redshift is secured, in addition to \Ha{}, by \MgII{}, \CIV{}, \NV{}, \Lya{}, and possibly \AlIII{}, \FeII{} ${\lambda}{1608}$ (with a quite peculiar profile), \SiIV{}, and \SiII. Two associated absorbers, redwards and bluewards of the QSO ($z=2.283$ and $z=2.234$, corresponding to
$\Delta v\simeq\SI{2.6e3}{km/s}$ and $\Delta v\simeq\SI{-1.9e3}{km/s}$ respectively) are observed in \AlIII{}, \CIV{}, and \SiIV{}; the low-redshift one (which is also the stronger) is observed also in \MgII{}, \FeII{}, \SiII{}, \NV{}, and \Lya{}, and appears to be slightly bluer for the low-ionization components ($z=2.228$ instead of $z=2.234$). A third strong metal absorber at $z=2.160$ is also observed (\AlIII{}, \CIV{}, \SiIV{}, \SiII{}, possibly \Lya{}).

\paragraph*{J2319$-$7322 ($\boldsymbol{\zem=2.612}$, $\boldsymbol{\alpha=-1.88}$).} A HiBALQ with almost completely saturated \CIV{}/\SiIV{} forest. Emission lines are generally very weak (\Hb{}, \AlIII{}, \CIV{}), with the possible exception of \Ha{}, \MgII{}, and \Lya{} (which nevertheless appear to be strongly contaminated by nearby absorption). A relatively narrow associated system at $z=2.558$ ($\Delta v\simeq\SI{-4.5e3}{km/s}$) is observed in \MgII{}, \AlIII{}, \CIV{}, \NV{}, and also \SiIV{} and \Lya{}, with a slight shift in velocity. Hints of another associated absorber, slightly redshifted with respect to $\zem$ ($z=2.702$, $\Delta v\simeq\SI{7.4e3}{km/s}$) are possibly seen in \AlIII{} and \Lya{}. Most of the remaining absorption can be explained with (roughly) two low-ionization components (\MgII{}, \AlIII{}) at $z=2.457$ and $z=2.359$ and four high-ionization components (\CIV{}, \SiIV{}) at $z=2.486,2.413,2.359,2.291$, the latter being consistent also with an almost totally absorbed through in the \Lya{} forest, in the range $\lambdarf\simeq\SIrange{1170}{1210}{\angstrom}$.

\paragraph*{J2355$-$5253 ($\boldsymbol{\zem=2.363}$, $\boldsymbol{\alpha=-1.72}$).} Another rather featureless non-QUIP QSO, similar to J2134$-$7243 and J2255$-$5404. A relatively weak Balmer series (\Ha{}, \Hb{}, possibly \Hg{}) is complemented by \MgII{}, \Lya{}, and possibly extended \CIV{} emission in constraining $\zem$. Notable absorption is observed only in the \Lya{} forest. A correspondence between narrow absorption lines is found for a system at $z=2.360$, observed in \NV{}, \CIV{}, and possibly \SiIV{}. No evidence for a BALQ classification is found.

\section{Discussion}\label{sec:discussion}

\begin{figure*}
    \centering
	\includegraphics[width=0.9\textwidth]{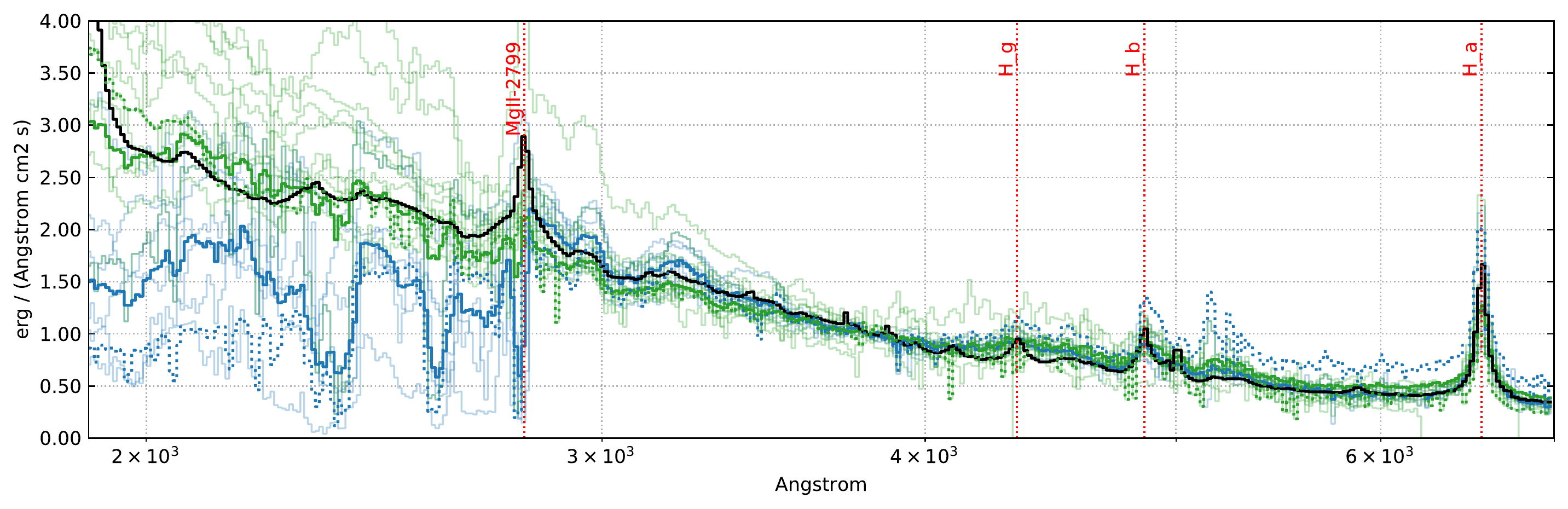}\\
    \caption{Composite rest-frame spectra of FeLoBALQs (blue) and non-FeLoBALQs (green) from our catalogue. De-reddened composites are shown with solid lines, while non-dereddened composites are shown with dashed lines. The de-reddened spectra of the individual QSOs are shown in transparency; spectra were normalized at \SI{3800}{\Angstrom} rest-frame to create the composites. The black solid line is the QSO template combined from \citealt{2001AJ....122..549V} and \citealt{2006ApJ...640..579G}.}
    \label{fig:composite}
\end{figure*}

\begin{figure}
    \centering
	\includegraphics[width=0.49\textwidth]{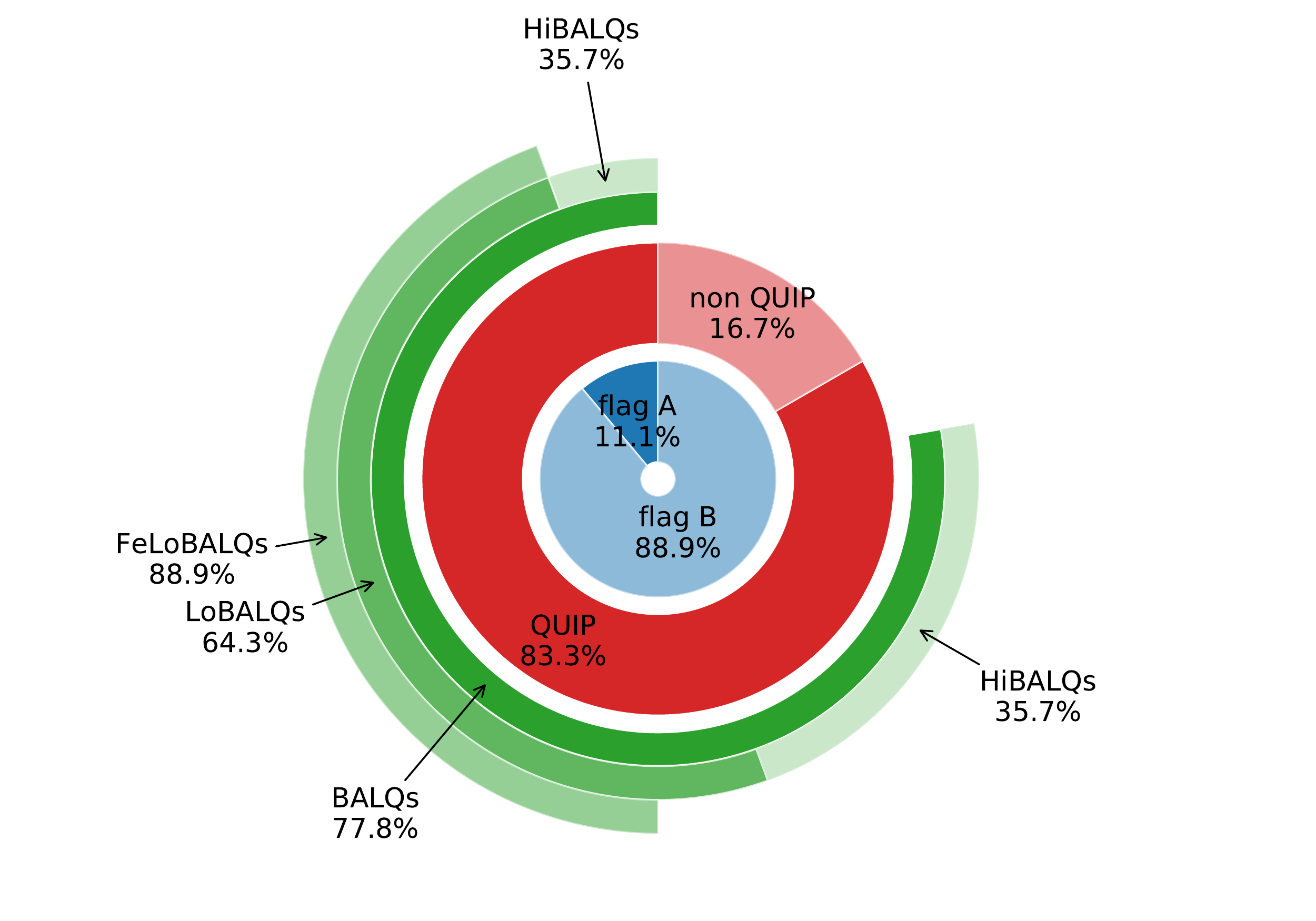}\\
    \caption{Classification of targets in our sample. Moving from the centre outwards, the rings display: 1. (blue) classification from \pI{}; 2. (red) classification from current paper; 3. (green) distribution of BALQs among the previous groups; 4. (green) distribution of HiBALQs and LoBALQs among BALQs; 5. (green) distribution of FeLoBALQs among LoBALQs. Percentages in ring 4 refer to ring 3, while percentages in ring 5 refer to ring 4.}
    \label{fig:fraction}
\end{figure}

\subsection{Fraction of BALQs}

The classification of targets in our sample based on their BAL features is graphically represented in \autoref{fig:fraction}. We observe that:
\begin{itemize}
    \item Both previously confirmed QSOs (flag A) are BALQs. Their inclusion among the QUIP targets was prompted by the peculiar absorption trough at the blue end of the spectra. We refrain from generalizing from this observation, though, due to the very low number of flag-A objects in our sample ($\NqubricsIII{A_QSO_QUIP}$ out of $\NqubricsIII{tot_QUIP}$ QUIPs). Despite their different emission redshift, these two QSOs (J2018$-$4546 and J2157$-$3602) display qualitatively similar spectra, with a significant drop in flux at $\lambda_\textrm{obs}\simeq6500$--$\SI{7000}{\angstrom}$, almost completely suppressing the emission continuum in the $u$, $g$, and $r$ bands. In the case of J2018$-$4546, the drop is due to metal absorbers (\MgII{} and \FeII{}) at $z\simeq1.3$, while in the case of J2157$-$3602 it is an effect of the \Lya{} forest at $z<4.6$.
    \item A relatively large fraction of newly confirmed QSOs (previously flag B) are BALQs ($\NqubricsIII{B_any_BAL}$ out of $\NqubricsIII{B_any}$ flag-B candidates, $\NqubricsIII{B_any_BAL_f}$), with a marginal prevalence of LoBALQs over HiBALQs ($\NqubricsIII{B_any_LoBAL}$ v. $\NqubricsIII{B_any_HiBAL}$). These objects failed to be identified as QSOs in our previous analysis (\pI, \pII) because of their pattern of strong absorption features (arising from either low- or high-ionization metal absorbers), which significantly altered the expected distribution of flux in the $u$, $g$, and $r$ bands. The peculiarity that motivated their inclusion in the QUIP sample is now totally reconciled with the QSO nature of these sources, thanks to the information provided by the NIR spectra.
    \item All but one non-BAL QSOs are not included in the QUIP sample. This is consistent with the interpretation that a peculiar spectral appearance arises as result of significant absorption associated with the emitting source or located along the line of sight. The only QUIP not identified as a BALQ is J0140$-$2531, the second highest-redshift object in our sample, whose QUIP nature is probably due to the combination of an unevenly absorbed \Lya{} forest and an excess emission around and redwards of \MgII{}.
\end{itemize}

The fraction of BALQs among the general QSO population, $F_\textrm{BAL}$, is typically assessed at $\sim$$\numrange{10}{15}$\% (e.g.~\citealt{2003AJ....125.1784H}, from pre-SDSS data; \citealt{2003AJ....126.2594R,2006ApJS..165....1T,2008MNRAS.386.1426K,2009ApJ...692..758G}, from different SDSS releases), based on \CIV{} absorption observed in the optical band, and is possibly increasing with redshift \citep{2011MNRAS.410..860A}. Higher fractions have been advocated by some authors from observations in other bands: \citet{2008ApJ...672..108D} measured $F_\textrm{BAL}\simeq\numrange{25}{40}$\% (depending on the classification criterion) on a sample 2MASS-selected QSOs, while  \citet{2019A&A...630A.111B} obtained $F_\textrm{BAL}\simeq24$\% from targets of the WISSH quasar project. The fraction of LoBALQs among BALQs, $F_\textrm{LoBAL/BAL}$, is similarly uncertain, being assessed at $\sim$15\% \citep[e.g.][]{1992ApJ...390...39S,2003AJ....126.2594R,2007ApJ...662L..59F} and possibly ranging between $\sim$$\numrange{5}{30}$\% \citep{2019A&A...630A.111B}. FeLoBALQs appear to be particularly rare, with estimates of $F_\textrm{FeLoBAL/BAL}$ as low as some percent \citep{2006ApJS..165....1T,2012ApJ...757..180D}.


The fraction of FeLoBALQs currently confirmed among QUBRICS QSOs is $\NqubricsIII{tot_FeLoBAL}$ out of $\NqubricsII{unique}$ ($\sim$2\%).
We interpret this evidence as a consequence of the criteria adopted by the QUBRICS survey to select QSOs at $z>2.5$ (\pI{}). The selection procedure was trained to interpret a relative dearth of flux in the $g$ band and in the bands bluewards as a signature of the \Lya{} forest in the relevant redshift range. A similar signature can nevertheless be produced by metal absorbers at lower redshift, provided they are strong enough to significantly impact the transmission in the band. QSOs at $z<2.5$ with strong associated metal absorption can thus be mistaken as QSOs at $z>2.5$, and at the same time be regarded as peculiar because their emission redshift cannot be properly assessed from optical spectra alone. The same occurrence was observed also in the SkyMapper survey, where a large fraction of low-redshift contaminants (16 out of 24) were identified as FeLoBALQs \citep{10.1093/mnras/stz2955}. In particular:
\begin{itemize}
    \item In the case of HiBALQs, \CIV{} and \SiIV{} absorption can mimic the appearance of the \Lya{} in the $g$ band for  $1.8\lesssim z\lesssim 2.6$. All the HiBALQs in our sample are consistent with this range, with the exception of J2157$-$3602  (which shows a significantly absorbed \Lya{} forest in the $g$ and $r$ bands, and is regarded as a QUIP due to a peculiar absorption feature at $\lambda\simeq 7600$--$\SI{7900}{\angstrom}$); if we neglect the outlier, the remaining targets have $z_\textrm{H}=2.32\pm 0.24$.
    \item In the case of LoBALQs, \MgII{} and \FeII{} complexes can similarly mimic the appearance of the \Lya{} in the $g$ band for $0.6\lesssim z\lesssim 1.8$. This explains both the redshift distribution of LoBALQs and FeLoBALQs in our sample ($z_\textrm{L}=1.48\pm 0.37$; $z_\textrm{F}=1.50\pm 0.39$) and the high fraction of detected FeLoBALQs: only LoBALQs with significantly strong \FeII{} absorption are likely to be mistaken for QSOs at higher redshift, due to the superficial similarity between the \Lya{} forest and the \FeII{} complexes.
\end{itemize}

The serendipitous discovery of $\NqubricsIII{tot_FeLoBAL}$ FeLoBALQs among the $\NqubricsIII{tot_NIR}$ QSOs discussed in this paper presents a noteworthy addition to the overall census of FeLoBALQs, not only in the Southern Hemisphere but in the whole sky.
We remark that the relatively high fraction of identified FeLoBALQs in our survey ($\sim$2\%, see above) is computed over a population of QSOs at higher redshift ($z>2.5$), which at magnitudes $i<18$ have a lower surface density with respect to QSOs in the same redshift range of our FeLoBALQs ($0.6\lesssim z\lesssim 1.8$). This fraction is therefore not directly comparable with the lower fraction from the literature \citep{2006ApJS..165....1T,2012ApJ...757..180D}, which is computed over matching redshift ranges for the FeLoBALQs and the general QSO population.

In other respects, the QSOs in our sample (both QUIPs and non-QUIPs) are not peculiar. The distributions of continuum slopes indexes (see \autoref{sec:cont}) give $\alpha_\textrm{F}=-1.98\pm 0.09$, $\alpha_\textrm{nF}=-1.83\pm 0.19$, with no dependence on redshift on either group (\autoref{fig:z_vs_slope}). This value is in agreement with the literature \citep[e.g.][]{2016MNRAS.462.2478C}. The small size of our sample, combined with the relatively high uncertainty associated with the flux calibration and continuum fitting procedure, prevents us from drawing stronger conclusions on this point.


\subsection{Reddening and Eddington ratios}\label{sec:mbh_lam}

Despite the low accuracy in the measurement of the intrinsic $E(B-V)$, FeLoBALQs in our sample are significantly more reddened than other QSOs, with $E(B-V)|_\textrm{F}=0.15\pm 0.08$, compared to $E(B-V)|_\textrm{nF}=0.03\pm 0.05$. This is consistent with higher levels of dust extinction \citep[e.g.][]{1992ApJ...390...39S,2003AJ....126.2594R,2009ApJ...692..758G}. As shown in \autoref{fig:composite} \FeII{} absorption bluewards of $\sim$\SI{2700}{\Angstrom} rest-frame is responsible for a decrease in the observed flux ranging from a factor of 2 to 3, most noticeable in correspondence of the absorbing features at $\sim$\SI{2350}{\Angstrom} and $\sim$\SI{2600}{\Angstrom}. Apart from these differences, the non-dereddened composite of FeLoBALQs and non-FeLoBALQs do not look excessively different redwards of the \MgII{} emission (dashed lines in \autoref{fig:composite}). De-reddening accounts only for a marginal increase in continuum steepness at $\lambda>\SI{3000}{\Angstrom}$. The overall shape of the emission lines alone provide no indication of a difference between the two groups.

\begin{table}
	\centering
	\caption{Black hole masses and Eddington ratios computed from line FWHMs and luminosities in \autoref{tab:analysis}, using formulae in \autoref{sec:mbh_lam}. The errors are propagated from FWHM and $L$ measurements and do not reflect the intrinsic scatter of the distribution. Estimates from \autoref{eq:Hb5100} and \ref{eq:lam5100} are in roman while estimates from \autoref{eq:Ha} and \ref{eq:lama} are in italic.}
	\label{tab:mbh_lam}
	\begin{tabular}{
	    c
	    c
	    r@{${}\pm{}$}l
	    r@{${}\pm{}$}l
	}
		\hline
		QUBRICS ID    & class &
		\multicolumn{2}{c}{$\log M_\textrm{BH}/M_\odot$} &
		\multicolumn{2}{c}{$\log\lambda_{\textrm{Edd}}$} \\
        \hline
J0008-5058 & Q,H & \textit{10.2} & \textit{0.1}  & $\mathit{-0.81}$ & \textit{0.13} \\
J0010-3201 & Q,H & \textit{10.3} & \textit{0.1}  & $\mathit{-0.87}$ & \textit{0.10} \\
J0140-2531 & Q   & \multicolumn{2}{c}{}          & \multicolumn{2}{c}{}             \\
J0407-6245 & Q,L &          9.6  &         0.1   & $        -0.38$  &         0.15  \\
J0514-3854 & Q,F & \textit{10.0} & \textit{0.1}  & $\mathit{-0.97}$ & \textit{0.1}  \\
J1215-2129 & Q,F & \textit{10.3} & \textit{0.1}  & $\mathit{-1.11}$ & \textit{0.12} \\
J1318-0245 & Q,F &          9.3  &         0.2   & $        -0.30$  &         0.15  \\
J1503-0451 & Q,F &         10.1  &         0.1   & $        -1.00$  &         0.07  \\
J2012-1802 & Q,F &          9.9  &         0.1   & $        -0.38$  &         0.10  \\
J2018-4546 & Q,F &         10.3  &         0.1   & $        -0.71$  &         0.11  \\
J2105-4104 & Q,F &          9.8  &         0.1   & $         0.05$  &         0.09  \\
J2134-7243 &     & \textit{9.75} & \textit{0.15} & $\mathit{-0.79}$ & \textit{0.16} \\
J2154-0514 & Q,F &          9.6  &         0.1   & $        -0.25$  &         0.11  \\
J2157-3602 & Q,H & \multicolumn{2}{c}{}          & \multicolumn{2}{c}{}             \\
J2222-4146 &     & \textit{10.3} & \textit{0.2}  & $\mathit{-1.10}$ & \textit{0.24} \\
J2255-5404 & Q,H & \textit{10.2} & \textit{0.1}  & $\mathit{-0.59}$ & \textit{0.1}  \\
J2319-7322 & Q,H &  \textit{9.9} & \textit{0.14} & $\mathit{-0.78}$ & \textit{0.15} \\
J2355-5253 &     &  \textit{8.9} & \textit{0.2}  & $\mathit{-0.13}$ & \textit{0.18} \\
        \hline
    \end{tabular}
\end{table}

The parameters of the \Ha{} and \Hb{} lines extracted by {\sc QSFit} as discussed in \autoref{sec:cont} can be used to estimate the mass $M_\textrm{BH}$ of the black holes powering the QSOs that we observed, albeit with limited accuracy.
We adopted two different $M_\textrm{BH}$ estimates:
\begin{itemize}
\item $M_\textrm{BH}^{5100}$, using the \Hb{} broad component FWHM and the optical continuum luminosity at \SI{5100}{\Angstrom} \citep[Equation 5]{2006ApJ...641..689V}:
\begin{equation}\label{eq:Hb5100}
    M_\textrm{BH}^{5100}=10^{5.91} \left[\frac{\textrm{FWHM}_{\textrm{H}\beta}}{\SI{1000}{\km\per\second}}\right]^2\left[\frac{\lambda L_{5100}}{\SI{e42}{\erg\per\second}}\right]^{0.50} M_\odot;
\end{equation}
\item $M_\textrm{BH}^{\textrm{H}\alpha}$, using the \Ha{} broad component FWHM and luminosity \citep[Equation 2]{Schulze_2017}:
\begin{equation}\label{eq:Ha}
    M_\textrm{BH}^{\textrm{H}\alpha}=10^{6.711} \left[\frac{\textrm{FWHM}_{\textrm{H}\alpha}}{\SI{1000}{\km\per\second}}\right]^{2.12}\left[\frac{\lambda L_{\textrm{H}\alpha}}{\SI{e42}{\erg\per\second}}\right]^{0.48} M_\odot;
\end{equation}
\end{itemize}
We also computed the Eddington ratios $\lambda_\textrm{Edd} =
L_\textrm{bol}/L_\textrm{Edd}$ using two different bolometric corrections: \begin{itemize}
\item \citet{2011ApJS..194...45S} for $L_{5100}$:
\begin{equation}\label{eq:lam5100}
\lambda_{\textrm{Edd},5100}\simeq
9.26L_{5100}\big/\SI{1.25e38}{}M_\textrm{BH}^{5100};
\end{equation}
\item\citet{2012MNRAS.423..600S} for $L_{\textrm{H}\alpha}$:
\begin{equation}\label{eq:lama}
\lambda_{\textrm{Edd},\textrm{H}\alpha} =
130L_{\textrm{H}\alpha}\big/\SI{1.25e38}{}M_\textrm{BH}^{\textrm{H}\alpha}.
\end{equation}
\end{itemize}
The virial mass estimators in the equations above are calibrated using relatively low-$z$ sources, whose $\textrm{H}\alpha$ and $\textrm{H}\beta$ emission lines are still observable in the optical wavebands. Here we are extrapolating their usage to $z \gtrsim 2$, and since the luminosities increase with redshift we expect to find larger values for the black hole mass with respect to the population of QSOs at $z<1$ \citep[see discussion in \S 5.1 of][]{2004ApJ...614..547S}.

\begin{figure}
    \centering
	\includegraphics[width=0.49\textwidth]{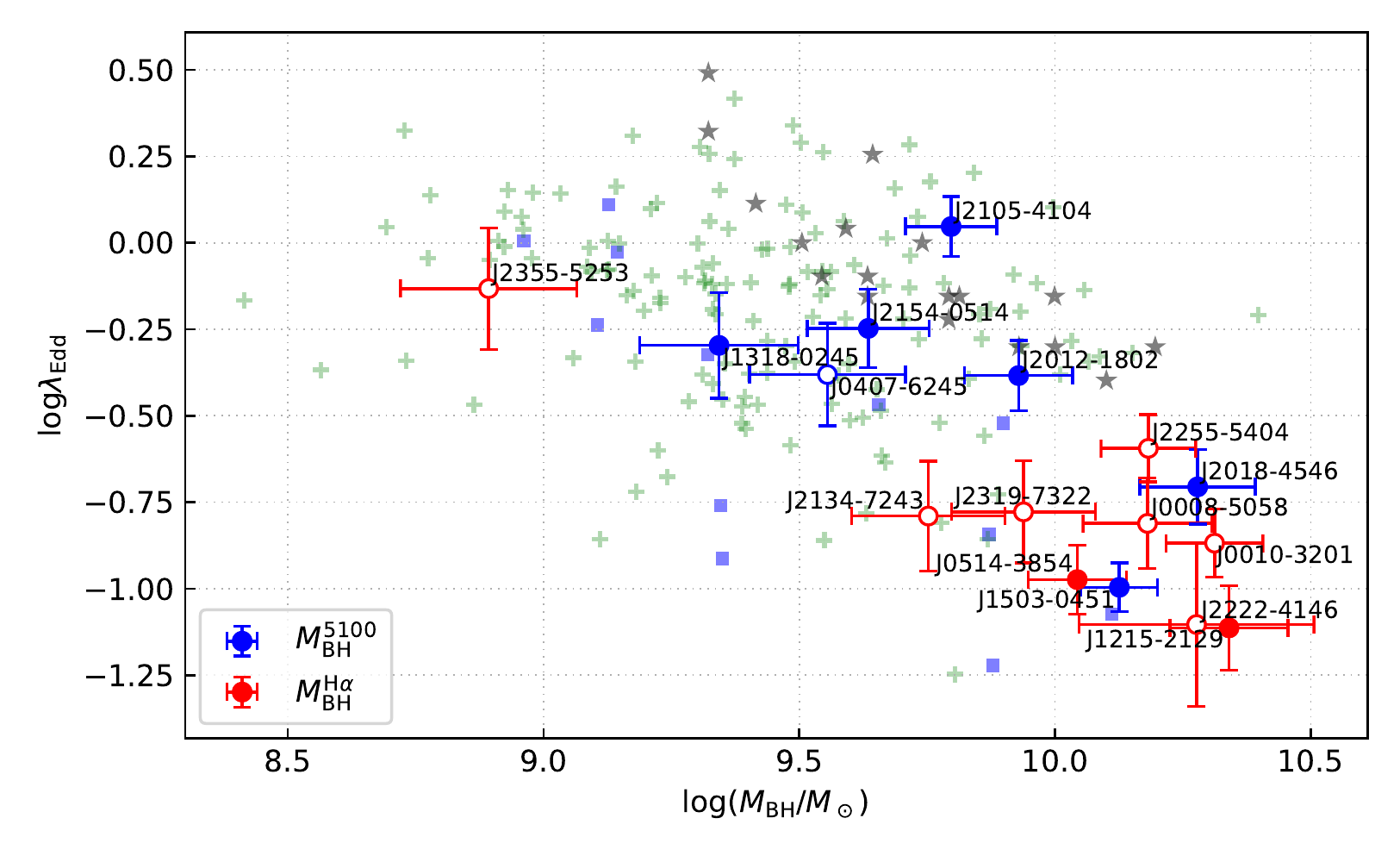}
    \caption{Eddington ratios estimated from {\sc QSFit} models of the targets in our sample. 
    Blue dots are computed from \autoref{eq:Hb5100} and \ref{eq:lam5100} while red dots from \autoref{eq:Ha} and \ref{eq:lama}. FeLoBALQs are denoted with filled circles, while other targets are denoted with empty circles. 
    Light-blue squares are from \citet{Schulze_2017}; green crosses from \citet{2017MNRAS.465.2120C}; grey stars from \citet{2018A&A...617A..81V}.
    }
    \label{fig:lam}
\end{figure}

The estimated values of $M_\textrm{BH}$ and $\lambda_\textrm{Edd}$ for the QSOs in our sample are listed in \autoref{tab:mbh_lam}. Only targets with a reliable line model (16 out of 18) were used to estimate the black hole mass.
$M_\textrm{BH}^{5100}$ (and consequently $\lambda_{\textrm{Edd},5100}$) was computed only for targets that allowed a proper modeling of the \Hb{} line. In some cases (J0010$-$3201, J2222$-$4146, and J2319$-$7322), {\sc QSFit} provided only an upper limit for $\textrm{FWHM}_{\textrm{H}\beta}$; in other cases (J0008$-$5058, J0514$-$3854, J1215$-$2129, J2134$-$7243, J2255$-$5404, and J2355$-$5253), the \Hb{} line appeared to be contaminated by telluric absorption or not prominent enough. For all these targets, $M_\textrm{BH}^{\textrm{H}\alpha}$ and $\lambda_{\textrm{Edd},\alpha}$ were used instead. We remark that our best fit model parameters should be considered rough estimates, and are provided as best effort values. Their limited reliability sums up with the significant uncertainties associated with single epoch virial mass estimates ($\sim$\SI{0.5}{dex}), providing no significant evidence for a tension between our mass values and the values from the literature (\autoref{fig:lam}).

Despite the limitations discussed in \autoref{sec:cont}, the continuum and emission line models obtained by {\sc QSFit} are reliable enough to map the distribution of $\lambda_\textrm{Edd}$ across our sample (\autoref{fig:lam}). A comparison with the distribution of 230 luminous QSOs at redshift $1.5<z<4.0$ \citep[][green crosses]{2017MNRAS.465.2120C} and with the 18 WISE/SDSS selected hyper-luminous (WISSH) QSOs at $z\simeq2$–$4$ \citep[][grey stars]{2018A&A...617A..81V} shows an overall agreement. No statistically significant difference between FeLoBALQs and non-FeLoBALQs (filled and empty circles, respectively) is observed: we measured $\log\lambda_\textrm{Edd,F}=-0.58\pm 0.41$ for the former and $\log\lambda_\textrm{Edd,nF}=-0.68\pm 0.32$ for the latter; these values are consistent with those obtained by \citet{Schulze_2017} and corroborate their conclusion that (Fe)LoBALQs do not appear to accrete at a higher rate compared to the general QSO population. Overall, no evidence to support an evolutionary scenario for (Fe)LoBALQs is observed.

\section{Conclusions}\label{sec:conclusion}

We have presented the combined optical-to-NIR spectra of $\NqubricsIII{tot}$ QSOs from the QUBRICS survey \citep{2019ApJ...887..268C,2020arXiv200803865B}, which were previously unconfirmed or lacking a secure redshift estimation. Redshift values ranging from $0.928$ to $4.665$ have been determined for all objects with a fiducial uncertainty of $0.001$, based on the identification of the Balmer series and/or \MgII{} emission lines made accessible by the NIR spectroscopy. Emission lines as well as several dozens of absorption systems, either associated with the emitting sources or intervening along the line of sight, have been detected using {\sc Astrocook} \citep{10.1117/12.2561343}. The continuum emission has been modeled with {\sc QSFit} \citep{2017MNRAS.472.4051C}, resulting in best-fit power-law slopes ranging from $-2.19$ to $-1.42$.

In most cases ($\NqubricsIII{tot_QUIP_f}$ of the observed targets), the acquisition of NIR spectra was prompted by peculiarities in the already available optical spectra (hence the designation of ``QUBRICS Irregular and Peculiar'' targets, or QUIPs). An unexpectedly high fraction of targets have been identified as broad-absorption line quasars or BALQs ($\NqubricsIII{tot_BAL_ftot}$ of the observed targets), and in particular BALQs with significant low-ionization \FeII{} absorption or FeLoBALQs ($\NqubricsIII{tot_FeLoBAL_ftot}$ of the observed targets), with a great degree of superposition between the original QUIP assessment and the successive BALQ confirmation ($\NqubricsIII{tot_BAL_f}$ BALQs among QUIPs). Such large detection rates arise as a serendipitous consequence of the selection criteria adopted by the QUBRICS survey: the procedure is optimized to identify QSOs at $z>2.5$ through their \Lya{} forest, and is therefore triggered by significant metal absorption in the $g$ band, leading to the identification of several BALQs (and especially FeLoBALQs) at $0.6\lesssim z\lesssim 1.8$. This is a convenient result for all science cases relying on the statistical and individual analysis of such rare objects.

QUBRICS FeLoBALQs appear significantly more reddened than other QSOs (with an average colour excess of 0.015), confirming what observed by other studies \citep{1992ApJ...390...39S,2003AJ....126.2594R,2009ApJ...692..758G}. However, the interpretation of the (Fe)LoBALQ phenomenon as an early stage in the QSO evolution is not supported by any evidence of increased accretion rate, as the mean Eddington ratio of FeLoBALQs is observed to be low (typically between $1$ and $8$ percent). The black hole masses measured across our whole sample (including both BALQs and non-BALQs) are consistent with those measured for luminous QSOs by other authors \citep{2017MNRAS.465.2120C,2018A&A...617A..81V}, indicating no difference in the mass distribution of BALQs (and in particular FeLoBALQs). A better understanding of the individual characteristics of the FeLoBALQs in our sample would require further observations at higher resolution, e.g. with VLT X-shooter.

\section*{Acknowledgements}

AG and FF acknowledge support from PRIN MIUR project `Black Hole winds and the Baryon Life Cycle of Galaxies: the stone-guest at the galaxy evolution supper', contract 2017-PH3WAT.

GCu would like to thank Manuela Bischetti and Andrea Travascio for insightful discussion.

This work is based on data products from observations made with (1) ESO Telescopes at La Silla Paranal Observatory, Chile (ESO programmes ID 103.A-0746(A), 0103.A-0746(B), and 0104.A-0754(A)), (2) the 6.5 meter Magellan Telescopes located at Las Campanas Observatory, Chile, and (3) the Italian Telescopio Nazionale Galileo (TNG) operated on the island of La Palma by the Fundaci\'on Galileo Galilei of the INAF (Istituto Nazionale di Astrofisica) at the Spanish Observatorio del Roque de los Muchachos of the Instituto de Astrofisica de Canarias.


This work has made use of data from the European Space Agency (ESA) mission {\it Gaia} (\url{https://www.cosmos.esa.int/gaia}), processed by the {\it Gaia} Data Processing and Analysis Consortium (DPAC, \url{https://www.cosmos.esa.int/web/gaia/dpac/consortium}). Funding for the DPAC has been provided by national institutions, in particular the institutions participating in the {\it Gaia} Multilateral Agreement.



\section*{Data Availability}

The spectra and the software tools described in this paper are made publicly available to ensure that the analysis is fully reproducible, and to foster a full exploitation of the collected data. The package has DOI \url{10.20371/INAF/DS/2021_00003} and is available at \url{https://www.ict.inaf.it/index.php/31-doi/137-ds-2021-03}. It contains:
\begin{itemize}
    \item Input spectra, i.e.~reduced optical spectra; 
    \item {\sc Astrocook} and {\sc QSFit} scripts, with instructions to run them;
    \item Output spectra and figures for reference.
\end{itemize}



\bibliographystyle{mnras}
\bibliography{main} 

\begin{thebibliography}{}
\makeatletter
\relax
\def\mn@urlcharsother{\let\do\@makeother \do\$\do\&\do\#\do\^\do\_\do\%\do\~}
\def\mn@doi{\begingroup\mn@urlcharsother \@ifnextchar [ {\mn@doi@}
  {\mn@doi@[]}}
\def\mn@doi@[#1]#2{\def\@tempa{#1}\ifx\@tempa\@empty \href
  {http://dx.doi.org/#2} {doi:#2}\else \href {http://dx.doi.org/#2} {#1}\fi
  \endgroup}
\def\mn@eprint#1#2{\mn@eprint@#1:#2::\@nil}
\def\mn@eprint@arXiv#1{\href {http://arxiv.org/abs/#1} {{\tt arXiv:#1}}}
\def\mn@eprint@dblp#1{\href {http://dblp.uni-trier.de/rec/bibtex/#1.xml}
  {dblp:#1}}
\def\mn@eprint@#1:#2:#3:#4\@nil{\def\@tempa {#1}\def\@tempb {#2}\def\@tempc
  {#3}\ifx \@tempc \@empty \let \@tempc \@tempb \let \@tempb \@tempa \fi \ifx
  \@tempb \@empty \def\@tempb {arXiv}\fi \@ifundefined
  {mn@eprint@\@tempb}{\@tempb:\@tempc}{\expandafter \expandafter \csname
  mn@eprint@\@tempb\endcsname \expandafter{\@tempc}}}

\bibitem[\protect\citeauthoryear{{Allen}, {Hewett}, {Maddox}, {Richards}  \&
  {Belokurov}}{{Allen} et~al.}{2011}]{2011MNRAS.410..860A}
{Allen} J.~T.,  {Hewett} P.~C.,  {Maddox} N.,  {Richards} G.~T.,   {Belokurov}
  V.,  2011, \mn@doi [\mnras] {10.1111/j.1365-2966.2010.17489.x}, \href
  {https://ui.adsabs.harvard.edu/abs/2011MNRAS.410..860A} {410, 860}

\bibitem[\protect\citeauthoryear{{Boutsia} et~al.,}{{Boutsia}
  et~al.}{2020}]{2020arXiv200803865B}
{Boutsia} K.,  et~al., 2020, arXiv e-prints, \href
  {https://ui.adsabs.harvard.edu/abs/2020arXiv200803865B} {p. arXiv:2008.03865}

\bibitem[\protect\citeauthoryear{{Boutsia} et~al.,}{{Boutsia}
  et~al.}{2021}]{2021ApJ...912..111B}
{Boutsia} K.,  et~al., 2021, \mn@doi [\apj] {10.3847/1538-4357/abedb5}, \href
  {https://ui.adsabs.harvard.edu/abs/2021ApJ...912..111B} {912, 111}

\bibitem[\protect\citeauthoryear{{Bruni} et~al.,}{{Bruni}
  et~al.}{2019}]{2019A&A...630A.111B}
{Bruni} G.,  et~al., 2019, \mn@doi [\aap] {10.1051/0004-6361/201834940}, \href
  {https://ui.adsabs.harvard.edu/abs/2019A&A...630A.111B} {630, A111}

\bibitem[\protect\citeauthoryear{{Calderone}, {Nicastro}, {Ghisellini},
  {Dotti}, {Sbarrato}, {Shankar}  \& {Colpi}}{{Calderone}
  et~al.}{2017}]{2017MNRAS.472.4051C}
{Calderone} G.,  {Nicastro} L.,  {Ghisellini} G.,  {Dotti} M.,  {Sbarrato} T.,
  {Shankar} F.,   {Colpi} M.,  2017, \mn@doi [\mnras] {10.1093/mnras/stx2239},
  \href {https://ui.adsabs.harvard.edu/abs/2017MNRAS.472.4051C} {472, 4051}

\bibitem[\protect\citeauthoryear{{Calderone} et~al.,}{{Calderone}
  et~al.}{2019}]{2019ApJ...887..268C}
{Calderone} G.,  et~al., 2019, \mn@doi [\apj] {10.3847/1538-4357/ab510a}, \href
  {https://ui.adsabs.harvard.edu/abs/2019ApJ...887..268C} {887, 268}

\bibitem[\protect\citeauthoryear{{Choi}, {Leighly}, {Terndrup}, {Gallagher}  \&
  {Richards}}{{Choi} et~al.}{2020}]{2020ApJ...891...53C}
{Choi} H.,  {Leighly} K.~M.,  {Terndrup} D.~M.,  {Gallagher} S.~C.,
  {Richards} G.~T.,  2020, \mn@doi [\apj] {10.3847/1538-4357/ab6f72}, \href
  {https://ui.adsabs.harvard.edu/abs/2020ApJ...891...53C} {891, 53}

\bibitem[\protect\citeauthoryear{{Coatman}, {Hewett}, {Banerji}, {Richards},
  {Hennawi}  \& {Prochaska}}{{Coatman} et~al.}{2017}]{2017MNRAS.465.2120C}
{Coatman} L.,  {Hewett} P.~C.,  {Banerji} M.,  {Richards} G.~T.,  {Hennawi}
  J.~F.,   {Prochaska} J.~X.,  2017, \mn@doi [\mnras] {10.1093/mnras/stw2797},
  \href {https://ui.adsabs.harvard.edu/abs/2017MNRAS.465.2120C} {465, 2120}

\bibitem[\protect\citeauthoryear{{Cristiani}, {Serrano}, {Fontanot}, {Vanzella}
   \& {Monaco}}{{Cristiani} et~al.}{2016}]{2016MNRAS.462.2478C}
{Cristiani} S.,  {Serrano} L.~M.,  {Fontanot} F.,  {Vanzella} E.,   {Monaco}
  P.,  2016, \mn@doi [\mnras] {10.1093/mnras/stw1810}, \href
  {https://ui.adsabs.harvard.edu/abs/2016MNRAS.462.2478C} {462, 2478}

\bibitem[\protect\citeauthoryear{{Cupani}, {Calderone}, {Cristiani}, {Di
  Marcantonio}, {D'Odorico}  \& {Taffoni}}{{Cupani}
  et~al.}{2018}]{2018SPIE10707E..23C}
{Cupani} G.,  {Calderone} G.,  {Cristiani} S.,  {Di Marcantonio} P.,
  {D'Odorico} V.,   {Taffoni} G.,  2018, \mn@doi [Proc. SPIE]
  {10.1117/12.2312093}, \href
  {https://ui.adsabs.harvard.edu/abs/2018SPIE10707E..23C} {10707, 1070723}

\bibitem[\protect\citeauthoryear{{Cupani}, {Calderone}, {Cristiani},
  {D'Odorico}  \& {Taffoni}}{{Cupani} et~al.}{2020a}]{2020ASPC..522..187C}
{Cupani} G.,  {Calderone} G.,  {Cristiani} S.,  {D'Odorico} V.,   {Taffoni} G.,
   2020a, Astronomical Society of the Pacific Conference Series, \href
  {https://ui.adsabs.harvard.edu/abs/2020ASPC..522..187C} {522, 187}

\bibitem[\protect\citeauthoryear{Cupani, D'Odorico, Cristiani, Russo, Calderone
   \& Taffoni}{Cupani et~al.}{2020b}]{10.1117/12.2561343}
Cupani G.,  D'Odorico V.,  Cristiani S.,  Russo S.~A.,  Calderone G.,   Taffoni
  G.,  2020b, \mn@doi [SPIE Conference Series] {10.1117/12.2561343}, 11452, 372

\bibitem[\protect\citeauthoryear{{Dai}, {Shankar}  \& {Sivakoff}}{{Dai}
  et~al.}{2008}]{2008ApJ...672..108D}
{Dai} X.,  {Shankar} F.,   {Sivakoff} G.~R.,  2008, \mn@doi [\apj]
  {10.1086/523688}, \href
  {https://ui.adsabs.harvard.edu/abs/2008ApJ...672..108D} {672, 108}

\bibitem[\protect\citeauthoryear{{Dai}, {Shankar}  \& {Sivakoff}}{{Dai}
  et~al.}{2012}]{2012ApJ...757..180D}
{Dai} X.,  {Shankar} F.,   {Sivakoff} G.~R.,  2012, \mn@doi [\apj]
  {10.1088/0004-637X/757/2/180}, \href
  {https://ui.adsabs.harvard.edu/abs/2012ApJ...757..180D} {757, 180}

\bibitem[\protect\citeauthoryear{{Di Matteo}, {Springel}  \& {Hernquist}}{{Di
  Matteo} et~al.}{2005}]{2005Natur.433..604D}
{Di Matteo} T.,  {Springel} V.,   {Hernquist} L.,  2005, \mn@doi [\nat]
  {10.1038/nature03335}, \href
  {https://ui.adsabs.harvard.edu/abs/2005Natur.433..604D} {433, 604}

\bibitem[\protect\citeauthoryear{{Fabian}}{{Fabian}}{2012}]{2012ARA&A..50..455F}
{Fabian} A.~C.,  2012, \mn@doi [\araa] {10.1146/annurev-astro-081811-125521},
  \href {https://ui.adsabs.harvard.edu/abs/2012ARA&A..50..455F} {50, 455}

\bibitem[\protect\citeauthoryear{{Farrah}, {Lacy}, {Priddey}, {Borys}  \&
  {Afonso}}{{Farrah} et~al.}{2007}]{2007ApJ...662L..59F}
{Farrah} D.,  {Lacy} M.,  {Priddey} R.,  {Borys} C.,   {Afonso} J.,  2007,
  \mn@doi [\apjl] {10.1086/519492}, \href
  {https://ui.adsabs.harvard.edu/abs/2007ApJ...662L..59F} {662, L59}

\bibitem[\protect\citeauthoryear{{Farrah} et~al.,}{{Farrah}
  et~al.}{2012}]{2012ApJ...745..178F}
{Farrah} D.,  et~al., 2012, \mn@doi [\apj] {10.1088/0004-637X/745/2/178}, \href
  {https://ui.adsabs.harvard.edu/abs/2012ApJ...745..178F} {745, 178}

\bibitem[\protect\citeauthoryear{{Faucher-Gigu{\`e}re}, {Quataert}  \&
  {Murray}}{{Faucher-Gigu{\`e}re} et~al.}{2012}]{2012MNRAS.420.1347F}
{Faucher-Gigu{\`e}re} C.-A.,  {Quataert} E.,   {Murray} N.,  2012, \mn@doi
  [\mnras] {10.1111/j.1365-2966.2011.20120.x}, \href
  {https://ui.adsabs.harvard.edu/abs/2012MNRAS.420.1347F} {420, 1347}

\bibitem[\protect\citeauthoryear{{Foltz}, {Wilkes}, {Weymann}  \&
  {Turnshek}}{{Foltz} et~al.}{1983}]{1983PASP...95..341F}
{Foltz} C.,  {Wilkes} B.,  {Weymann} R.,   {Turnshek} D.,  1983, \mn@doi
  [\pasp] {10.1086/131170}, \href
  {https://ui.adsabs.harvard.edu/abs/1983PASP...95..341F} {95, 341}

\bibitem[\protect\citeauthoryear{Gagné, Lambrides, Faherty  \& Simcoe}{Gagné
  et~al.}{2015}]{jonathan_gagne_2015_18775}
Gagné J.,  Lambrides E.,  Faherty J.~K.,   Simcoe R.,  2015, FireHose\_v2:
  Firehose v2.0, \mn@doi{10.5281/zenodo.18775}, \url
  {https://doi.org/10.5281/zenodo.18775}

\bibitem[\protect\citeauthoryear{{Gibson} et~al.,}{{Gibson}
  et~al.}{2009}]{2009ApJ...692..758G}
{Gibson} R.~R.,  et~al., 2009, \mn@doi [\apj] {10.1088/0004-637X/692/1/758},
  \href {https://ui.adsabs.harvard.edu/abs/2009ApJ...692..758G} {692, 758}

\bibitem[\protect\citeauthoryear{{Glikman}, {Helfand}  \& {White}}{{Glikman}
  et~al.}{2006}]{2006ApJ...640..579G}
{Glikman} E.,  {Helfand} D.~J.,   {White} R.~L.,  2006, \mn@doi [\apj]
  {10.1086/500098}, \href
  {https://ui.adsabs.harvard.edu/abs/2006ApJ...640..579G} {640, 579}

\bibitem[\protect\citeauthoryear{{Guarneri}, {Calderone}, {Cristiani},
  {Fontanot}, {Boutsia}, {Cupani}, {Grazian}  \& {D'Odorico}}{{Guarneri}
  et~al.}{2021}]{2021MNRAS.506.2471G}
{Guarneri} F.,  {Calderone} G.,  {Cristiani} S.,  {Fontanot} F.,  {Boutsia} K.,
   {Cupani} G.,  {Grazian} A.,   {D'Odorico} V.,  2021, \mn@doi [\mnras]
  {10.1093/mnras/stab1867}, \href
  {https://ui.adsabs.harvard.edu/abs/2021MNRAS.506.2471G} {506, 2471}

\bibitem[\protect\citeauthoryear{{Hall} et~al.,}{{Hall}
  et~al.}{2002}]{2002ApJS..141..267H}
{Hall} P.~B.,  et~al., 2002, \mn@doi [\apjs] {10.1086/340546}, \href
  {https://ui.adsabs.harvard.edu/abs/2002ApJS..141..267H} {141, 267}

\bibitem[\protect\citeauthoryear{{Hazard}, {McMahon}  \& {Morton}}{{Hazard}
  et~al.}{1987}]{1987MNRAS.229..371H}
{Hazard} C.,  {McMahon} R.~G.,   {Morton} D.~C.,  1987, \mn@doi [\mnras]
  {10.1093/mnras/229.3.371}, \href
  {https://ui.adsabs.harvard.edu/abs/1987MNRAS.229..371H} {229, 371}

\bibitem[\protect\citeauthoryear{{Hewett} \& {Foltz}}{{Hewett} \&
  {Foltz}}{2003}]{2003AJ....125.1784H}
{Hewett} P.~C.,  {Foltz} C.~B.,  2003, \mn@doi [\aj] {10.1086/368392}, \href
  {https://ui.adsabs.harvard.edu/abs/2003AJ....125.1784H} {125, 1784}

\bibitem[\protect\citeauthoryear{{Kelson}}{{Kelson}}{2003}]{2003PASP..115..688K}
{Kelson} D.~D.,  2003, \mn@doi [\pasp] {10.1086/375502}, \href
  {https://ui.adsabs.harvard.edu/abs/2003PASP..115..688K} {115, 688}

\bibitem[\protect\citeauthoryear{{Kelson}, {Illingworth}, {van Dokkum}  \&
  {Franx}}{{Kelson} et~al.}{2000}]{2000ApJ...531..159K}
{Kelson} D.~D.,  {Illingworth} G.~D.,  {van Dokkum} P.~G.,   {Franx} M.,  2000,
  \mn@doi [\apj] {10.1086/308445}, \href
  {https://ui.adsabs.harvard.edu/abs/2000ApJ...531..159K} {531, 159}

\bibitem[\protect\citeauthoryear{{Knigge}, {Scaringi}, {Goad}  \&
  {Cottis}}{{Knigge} et~al.}{2008}]{2008MNRAS.386.1426K}
{Knigge} C.,  {Scaringi} S.,  {Goad} M.~R.,   {Cottis} C.~E.,  2008, \mn@doi
  [\mnras] {10.1111/j.1365-2966.2008.13081.x}, \href
  {https://ui.adsabs.harvard.edu/abs/2008MNRAS.386.1426K} {386, 1426}

\bibitem[\protect\citeauthoryear{{Korista}, {Voit}, {Morris}  \&
  {Weymann}}{{Korista} et~al.}{1993}]{1993ApJS...88..357K}
{Korista} K.~T.,  {Voit} G.~M.,  {Morris} S.~L.,   {Weymann} R.~J.,  1993,
  \mn@doi [\apjs] {10.1086/191825}, \href
  {https://ui.adsabs.harvard.edu/abs/1993ApJS...88..357K} {88, 357}

\bibitem[\protect\citeauthoryear{{Kormendy} \& {Ho}}{{Kormendy} \&
  {Ho}}{2013}]{2013ARA&A..51..511K}
{Kormendy} J.,  {Ho} L.~C.,  2013, \mn@doi [\araa]
  {10.1146/annurev-astro-082708-101811}, \href
  {https://ui.adsabs.harvard.edu/abs/2013ARA&A..51..511K} {51, 511}

\bibitem[\protect\citeauthoryear{{Lazarova}, {Canalizo}, {Lacy}  \&
  {Sajina}}{{Lazarova} et~al.}{2012}]{2012ApJ...755...29L}
{Lazarova} M.~S.,  {Canalizo} G.,  {Lacy} M.,   {Sajina} A.,  2012, \mn@doi
  [\apj] {10.1088/0004-637X/755/1/29}, \href
  {https://ui.adsabs.harvard.edu/abs/2012ApJ...755...29L} {755, 29}

\bibitem[\protect\citeauthoryear{{Lu}, {Zhao}, {Bai}  \& {Fan}}{{Lu}
  et~al.}{2019}]{2019-Lu-BalmerDecrement}
{Lu} K.-X.,  {Zhao} Y.,  {Bai} J.-M.,   {Fan} X.-L.,  2019, \mn@doi [\mnras]
  {10.1093/mnras/sty3229}, \href
  {https://ui.adsabs.harvard.edu/abs/2019MNRAS.483.1722L} {483, 1722}

\bibitem[\protect\citeauthoryear{{Lucy}, {Leighly}, {Terndrup}, {Dietrich}  \&
  {Gallagher}}{{Lucy} et~al.}{2014}]{2014ApJ...783...58L}
{Lucy} A.~B.,  {Leighly} K.~M.,  {Terndrup} D.~M.,  {Dietrich} M.,
  {Gallagher} S.~C.,  2014, \mn@doi [\apj] {10.1088/0004-637X/783/1/58}, \href
  {https://ui.adsabs.harvard.edu/abs/2014ApJ...783...58L} {783, 58}

\bibitem[\protect\citeauthoryear{{O'Donnell}}{{O'Donnell}}{1994}]{1994-odonnell-updateCCM}
{O'Donnell} J.~E.,  1994, \mn@doi [\apj] {10.1086/173713}, \href
  {http://adsabs.harvard.edu/abs/1994ApJ...422..158O} {422, 158}

\bibitem[\protect\citeauthoryear{{Peterson}, {Wanders}, {Horne}, {Collier},
  {Alexander}, {Kaspi}  \& {Maoz}}{{Peterson} et~al.}{1998}]{Peterson1998}
{Peterson} B.~M.,  {Wanders} I.,  {Horne} K.,  {Collier} S.,  {Alexander} T.,
  {Kaspi} S.,   {Maoz} D.,  1998, \mn@doi [\pasp] {10.1086/316177}, \href
  {https://ui.adsabs.harvard.edu/abs/1998PASP..110..660P} {110, 660}

\bibitem[\protect\citeauthoryear{{Reichard} et~al.,}{{Reichard}
  et~al.}{2003}]{2003AJ....126.2594R}
{Reichard} T.~A.,  et~al., 2003, \mn@doi [\aj] {10.1086/379293}, \href
  {https://ui.adsabs.harvard.edu/abs/2003AJ....126.2594R} {126, 2594}

\bibitem[\protect\citeauthoryear{Schulze et~al.,}{Schulze
  et~al.}{2017}]{Schulze_2017}
Schulze A.,  et~al., 2017, \mn@doi [The Astrophysical Journal]
  {10.3847/1538-4357/aa8e4c}, 848, 104

\bibitem[\protect\citeauthoryear{{Shemmer}, {Netzer}, {Maiolino}, {Oliva},
  {Croom}, {Corbett}  \& {di Fabrizio}}{{Shemmer}
  et~al.}{2004}]{2004ApJ...614..547S}
{Shemmer} O.,  {Netzer} H.,  {Maiolino} R.,  {Oliva} E.,  {Croom} S.,
  {Corbett} E.,   {di Fabrizio} L.,  2004, \mn@doi [\apj] {10.1086/423607},
  \href {https://ui.adsabs.harvard.edu/abs/2004ApJ...614..547S} {614, 547}

\bibitem[\protect\citeauthoryear{{Shen} et~al.,}{{Shen}
  et~al.}{2011}]{2011ApJS..194...45S}
{Shen} Y.,  et~al., 2011, \mn@doi [\apjs] {10.1088/0067-0049/194/2/45}, \href
  {https://ui.adsabs.harvard.edu/abs/2011ApJS..194...45S} {194, 45}

\bibitem[\protect\citeauthoryear{{Silk} \& {Rees}}{{Silk} \&
  {Rees}}{1998}]{1998A&A...331L...1S}
{Silk} J.,  {Rees} M.~J.,  1998, \aap, \href
  {https://ui.adsabs.harvard.edu/abs/1998A&A...331L...1S} {331, L1}

\bibitem[\protect\citeauthoryear{{Sprayberry} \& {Foltz}}{{Sprayberry} \&
  {Foltz}}{1992}]{1992ApJ...390...39S}
{Sprayberry} D.,  {Foltz} C.~B.,  1992, \mn@doi [\apj] {10.1086/171257}, \href
  {https://ui.adsabs.harvard.edu/abs/1992ApJ...390...39S} {390, 39}

\bibitem[\protect\citeauthoryear{{Stern} \& {Laor}}{{Stern} \&
  {Laor}}{2012}]{2012MNRAS.423..600S}
{Stern} J.,  {Laor} A.,  2012, \mn@doi [\mnras]
  {10.1111/j.1365-2966.2012.20901.x}, \href
  {https://ui.adsabs.harvard.edu/abs/2012MNRAS.423..600S} {423, 600}

\bibitem[\protect\citeauthoryear{{Tody}}{{Tody}}{1993}]{1993ASPC...52..173T}
{Tody} D.,  1993, in {Hanisch} R.~J.,  {Brissenden} R.~J.~V.,   {Barnes} J.,
  eds,  Astronomical Society of the Pacific Conference Series Vol. 52,
  Astronomical Data Analysis Software and Systems II. p.~173

\bibitem[\protect\citeauthoryear{{Trump} et~al.,}{{Trump}
  et~al.}{2006}]{2006ApJS..165....1T}
{Trump} J.~R.,  et~al., 2006, \mn@doi [\apjs] {10.1086/503834}, \href
  {https://ui.adsabs.harvard.edu/abs/2006ApJS..165....1T} {165, 1}

\bibitem[\protect\citeauthoryear{{Urrutia}, {Lacy}, {Spoon}, {Glikman},
  {Petric}  \& {Schulz}}{{Urrutia} et~al.}{2012}]{2012ApJ...757..125U}
{Urrutia} T.,  {Lacy} M.,  {Spoon} H.,  {Glikman} E.,  {Petric} A.,   {Schulz}
  B.,  2012, \mn@doi [\apj] {10.1088/0004-637X/757/2/125}, \href
  {https://ui.adsabs.harvard.edu/abs/2012ApJ...757..125U} {757, 125}

\bibitem[\protect\citeauthoryear{{Vanden Berk} et~al.,}{{Vanden Berk}
  et~al.}{2001}]{2001AJ....122..549V}
{Vanden Berk} D.~E.,  et~al., 2001, \mn@doi [\aj] {10.1086/321167}, \href
  {https://ui.adsabs.harvard.edu/abs/2001AJ....122..549V} {122, 549}

\bibitem[\protect\citeauthoryear{{Vestergaard} \& {Peterson}}{{Vestergaard} \&
  {Peterson}}{2006}]{2006ApJ...641..689V}
{Vestergaard} M.,  {Peterson} B.~M.,  2006, \mn@doi [\apj] {10.1086/500572},
  \href {https://ui.adsabs.harvard.edu/abs/2006ApJ...641..689V} {641, 689}

\bibitem[\protect\citeauthoryear{{Vietri} et~al.,}{{Vietri}
  et~al.}{2018}]{2018A&A...617A..81V}
{Vietri} G.,  et~al., 2018, \mn@doi [\aap] {10.1051/0004-6361/201732335}, \href
  {https://ui.adsabs.harvard.edu/abs/2018A&A...617A..81V} {617, A81}

\bibitem[\protect\citeauthoryear{{Violino} et~al.,}{{Violino}
  et~al.}{2016}]{2016MNRAS.457.1371V}
{Violino} G.,  et~al., 2016, \mn@doi [\mnras] {10.1093/mnras/stv2937}, \href
  {https://ui.adsabs.harvard.edu/abs/2016MNRAS.457.1371V} {457, 1371}

\bibitem[\protect\citeauthoryear{{Weymann}, {Morris}, {Foltz}  \&
  {Hewett}}{{Weymann} et~al.}{1991}]{1991ApJ...373...23W}
{Weymann} R.~J.,  {Morris} S.~L.,  {Foltz} C.~B.,   {Hewett} P.~C.,  1991,
  \mn@doi [\apj] {10.1086/170020}, \href
  {https://ui.adsabs.harvard.edu/abs/1991ApJ...373...23W} {373, 23}

\bibitem[\protect\citeauthoryear{{Wolf} et~al.,}{{Wolf}
  et~al.}{2018}]{2018PASA...35...10W}
{Wolf} C.,  et~al., 2018, \mn@doi [\pasa] {10.1017/pasa.2018.5}, \href
  {https://ui.adsabs.harvard.edu/abs/2018PASA...35...10W} {35, e010}

\bibitem[\protect\citeauthoryear{Wolf et~al.,}{Wolf
  et~al.}{2019}]{10.1093/mnras/stz2955}
Wolf C.,  et~al., 2019, \mn@doi [Monthly Notices of the Royal Astronomical
  Society] {10.1093/mnras/stz2955}, 491, 1970

\makeatother
\end{thebibliography}




\begin{figure*}
    \textbf{J0514$-$3854 ($\boldsymbol{\zem=1.775}$, $\boldsymbol{\alpha=-2.15}$)}\\
	\includegraphics[height=0.215\textheight,left]{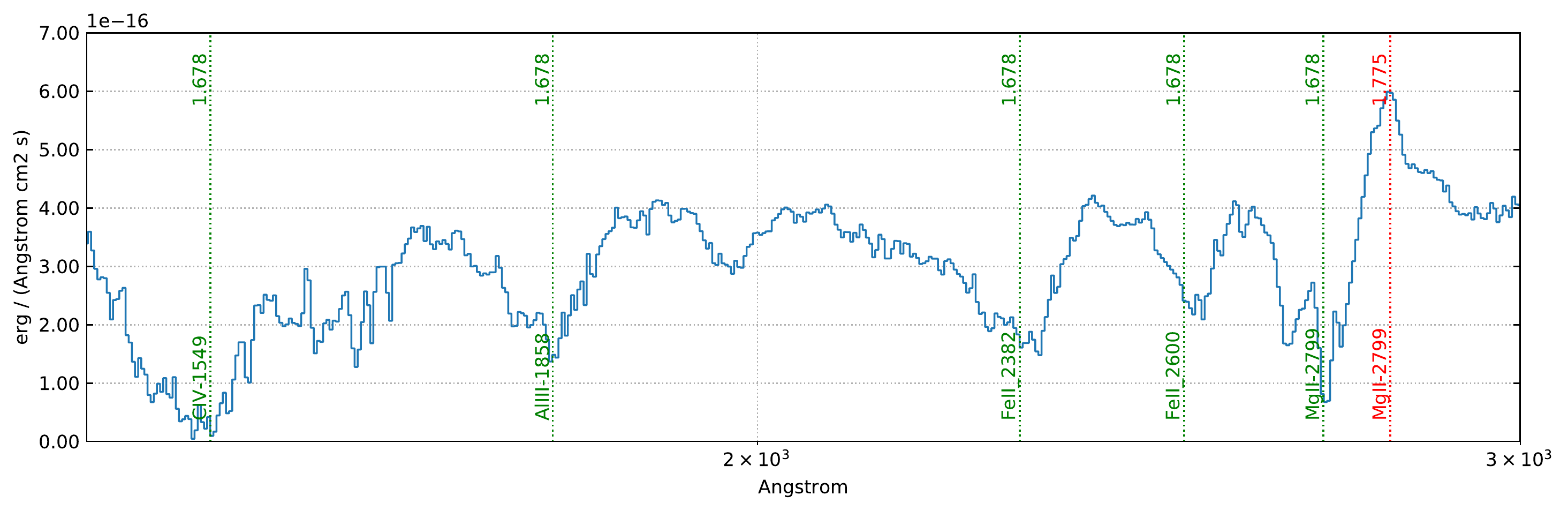}\\
	\includegraphics[height=0.215\textheight,left]{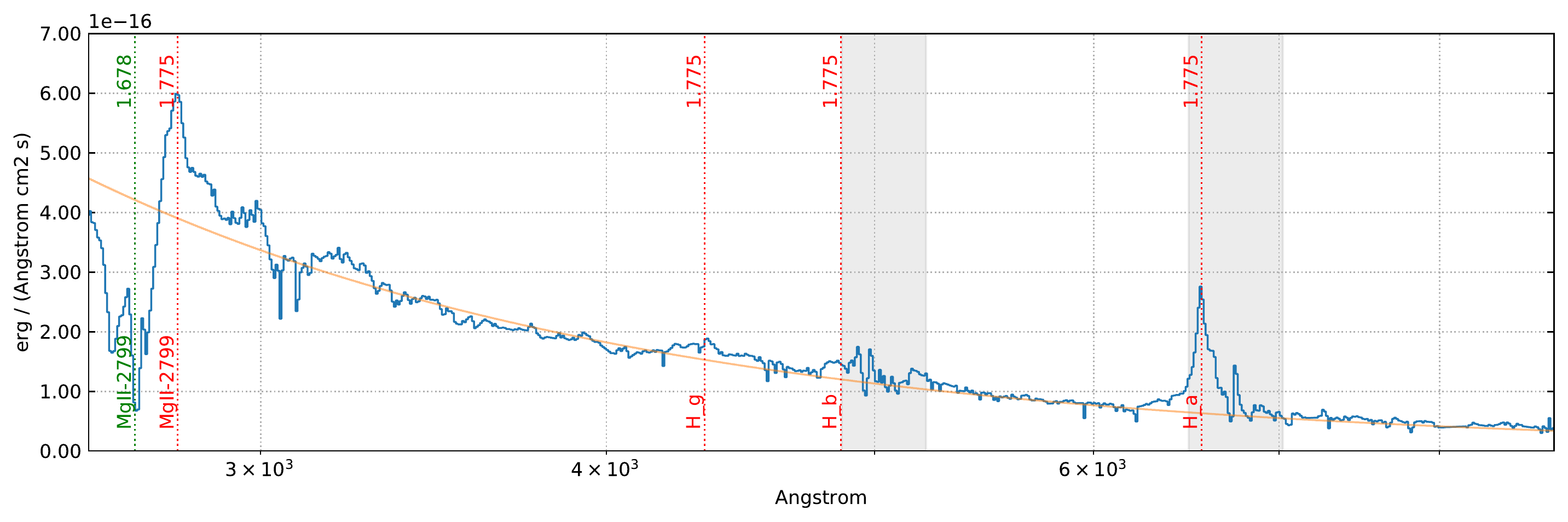}\\
	\vspace{0.5cm}
    \textbf{J1215$-$2129 ($\boldsymbol{\zem=1.464}$, $\boldsymbol{\alpha=-1.98}$)}\\
	\includegraphics[height=0.215\textheight,left]{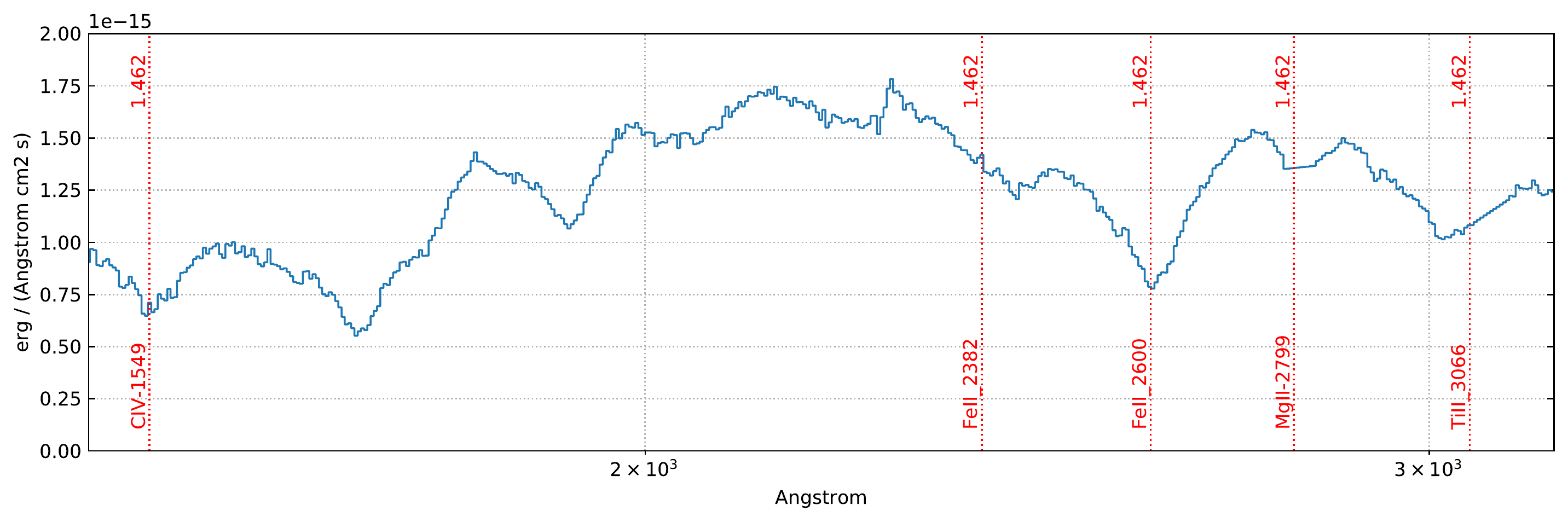}\\
	\includegraphics[height=0.215\textheight,left]{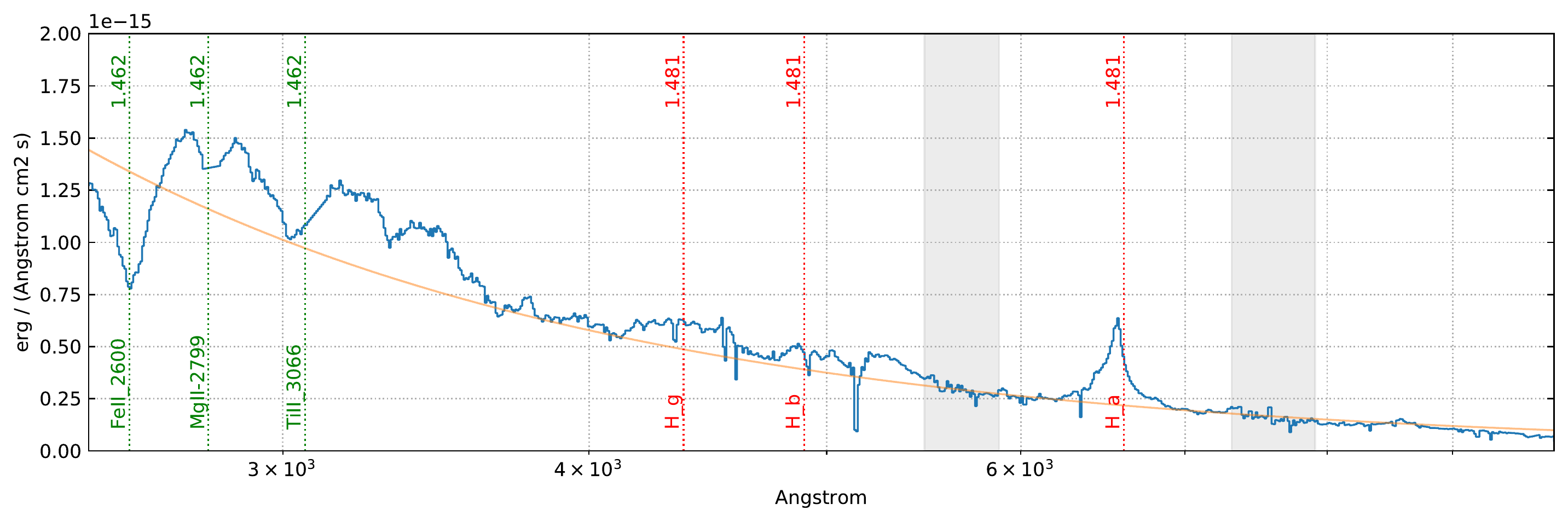}\\
    \caption{Spectra of FeLoBALQs from our sample. Wavelengths are rest-frame. Notable emission and absorption features are highlighted by dotted bars (red and green, respectively). Regions affected by strong telluric absorption are shaded. Each spectrum is split into two chunks for better visualization; in the red chunk, the best-fitting power-law continuum (orange line, see \autoref{sec:cont}) is superimposed to the extracted flux density (blue line). Some absorption lines may be omitted in the red chunk for better clarity.}
    \label{fig:felobal}
\end{figure*}

\begin{figure*}
    \centering
    \textbf{J1318$-$0245 ($\boldsymbol{\zem=1.404}$, $\boldsymbol{\alpha=-2.06}$)}\\
	\includegraphics[height=0.215\textheight,left]{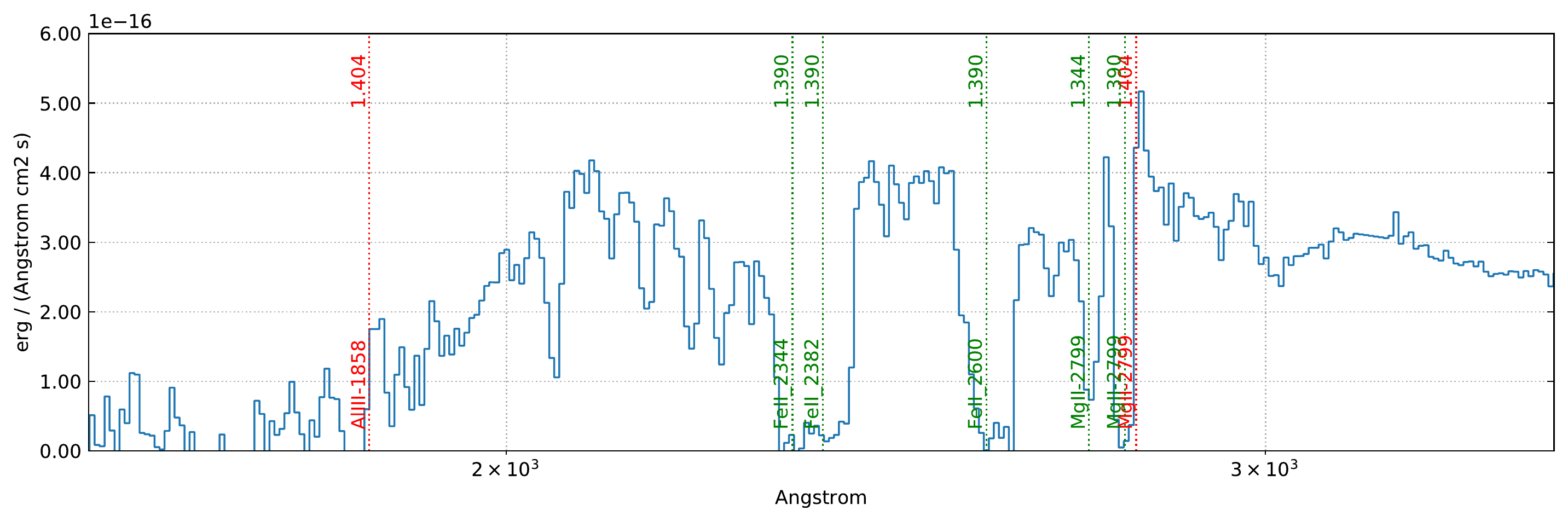}\\
	\includegraphics[height=0.215\textheight,left]{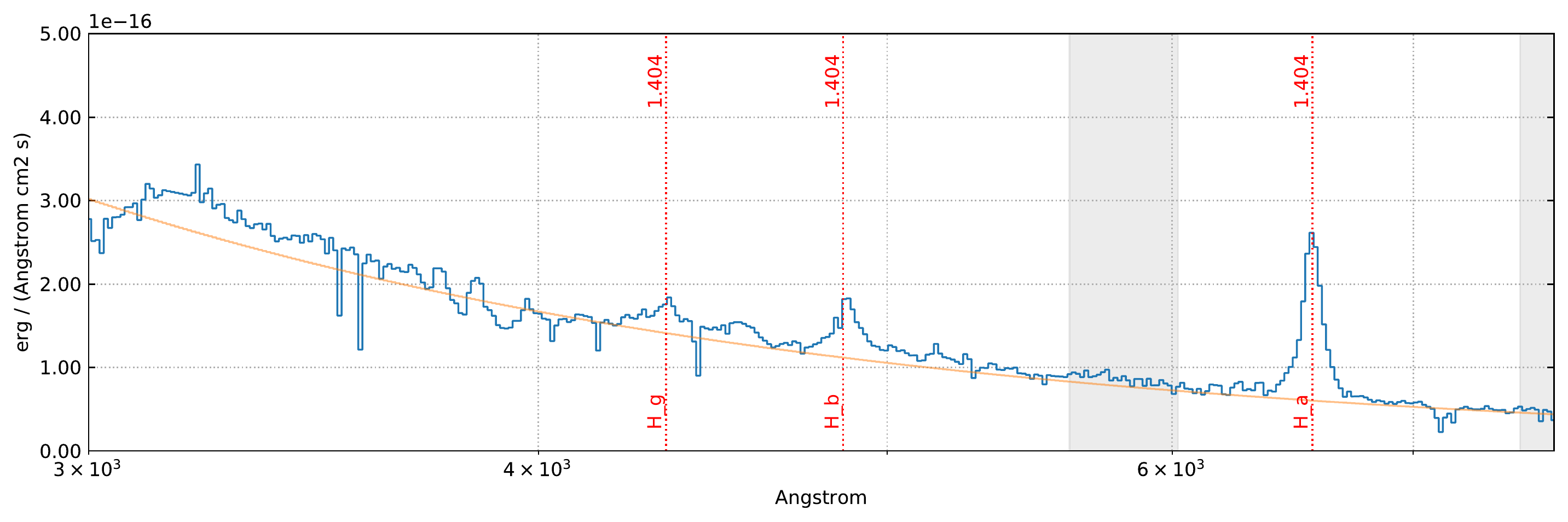}\\
	\vspace{0.5cm}
    \textbf{J1503$-$0451 ($\boldsymbol{\zem=0.929}$, $\boldsymbol{\alpha=-1.96}$)}\\
	\includegraphics[height=0.215\textheight,left]{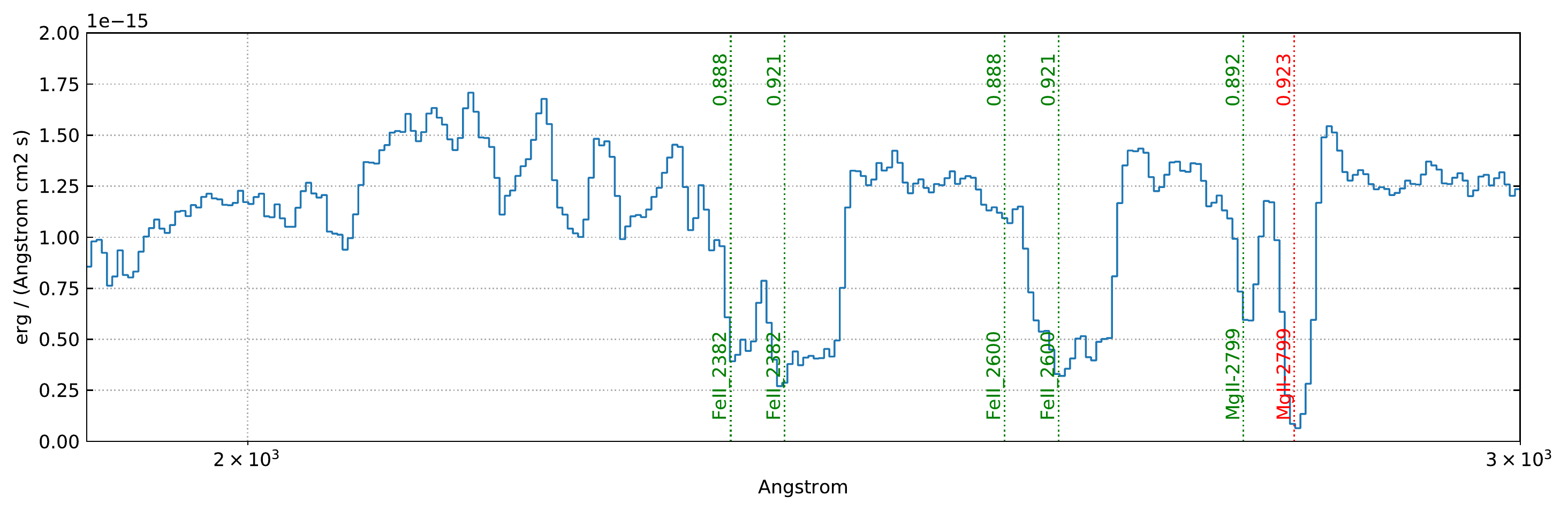}\\
	\includegraphics[height=0.215\textheight,left]{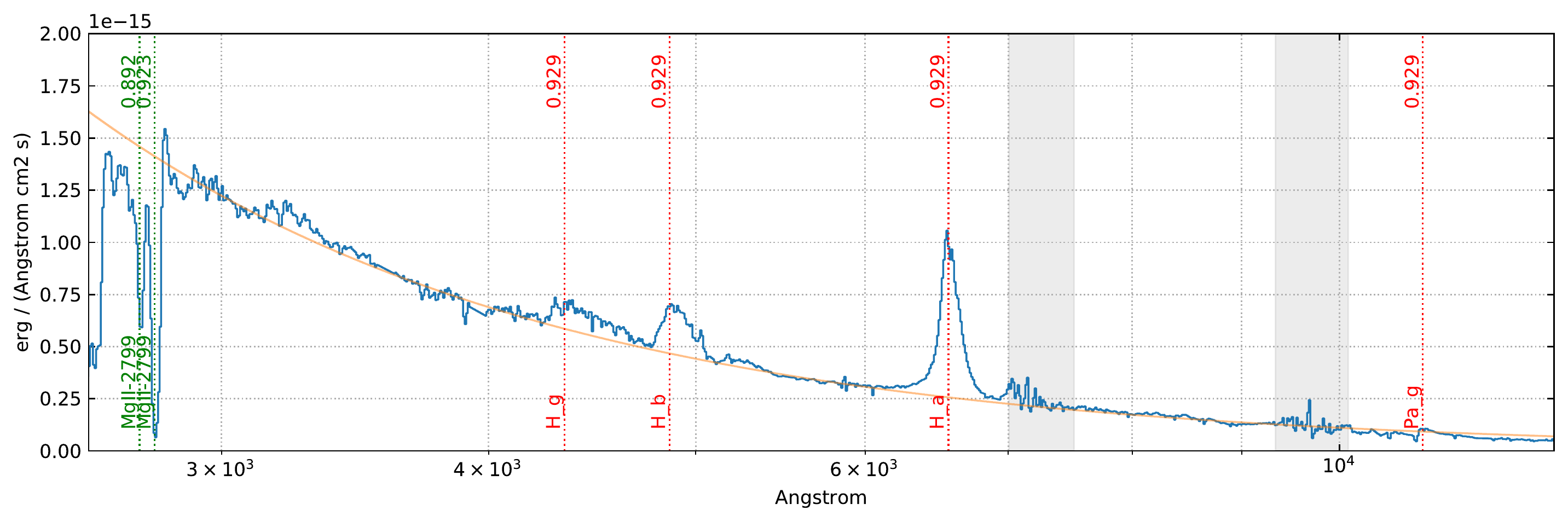}\\
	\contcaption{}
\end{figure*}

\begin{figure*}
    \centering
	\textbf{J2012$-$1802 ($\boldsymbol{\zem=1.275}$, $\boldsymbol{\alpha=-1.91}$)}\\
	\includegraphics[height=0.215\textheight,left]{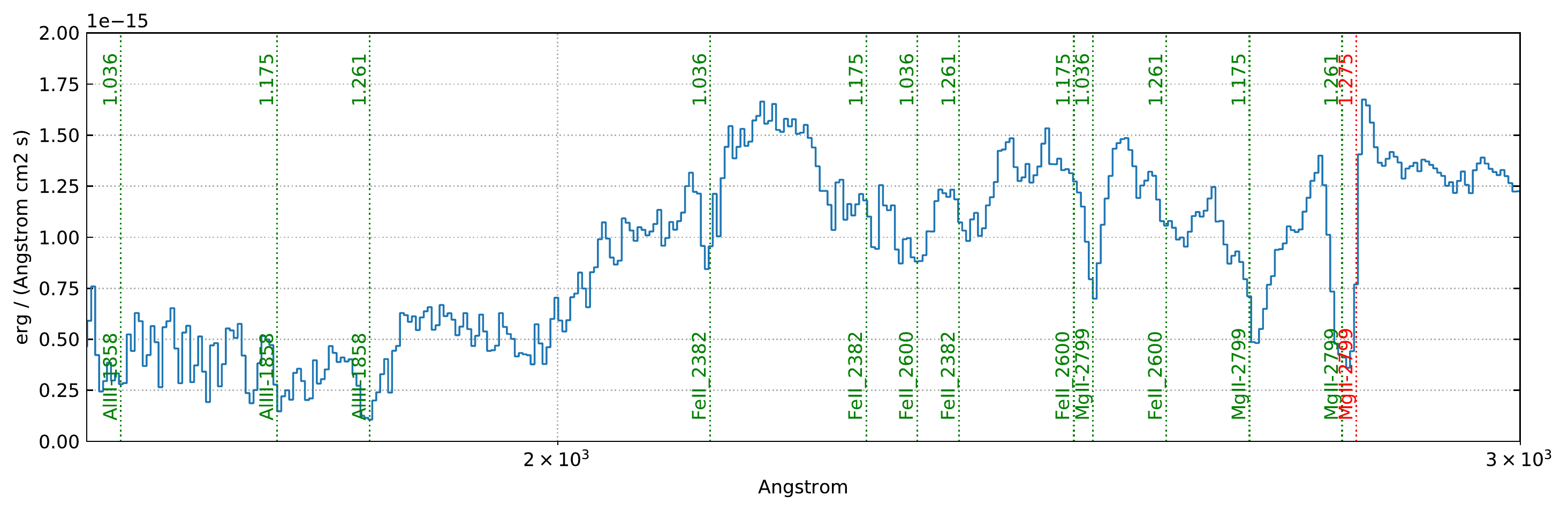}\\
	\includegraphics[height=0.215\textheight,left]{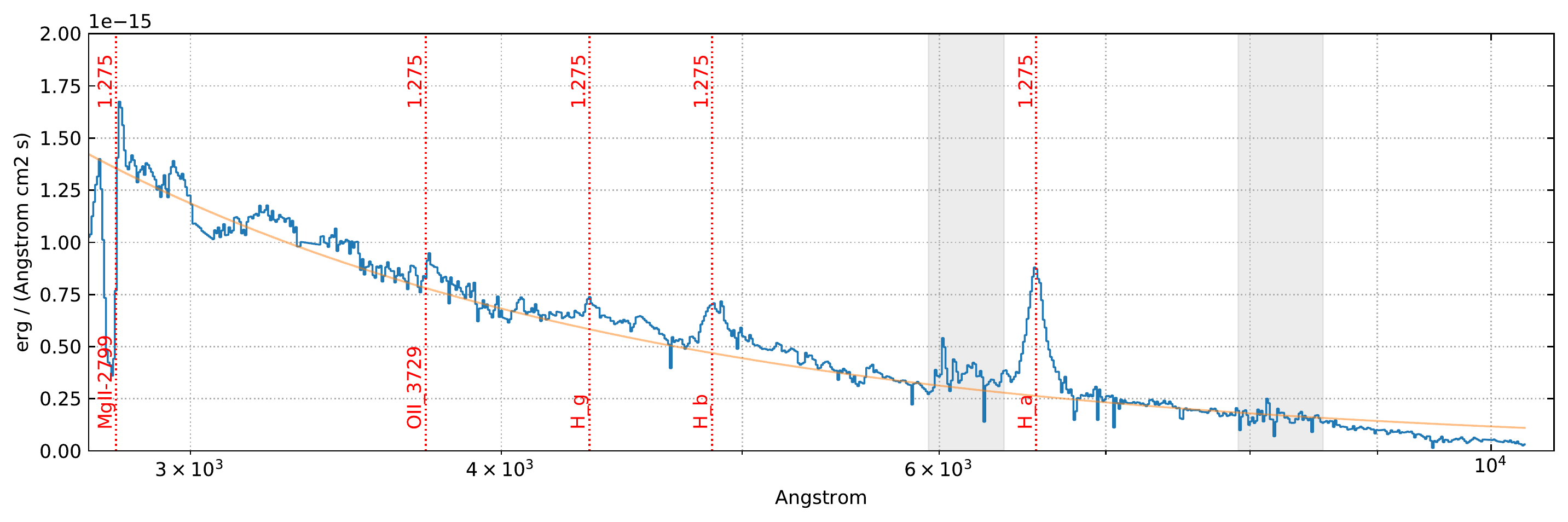}\\
	\vspace{0.5cm}
    \textbf{J2018$-$4546 ($\boldsymbol{\zem=1.352}$, $\boldsymbol{\alpha=-1.85}$)}\\
	\includegraphics[height=0.215\textheight,left]{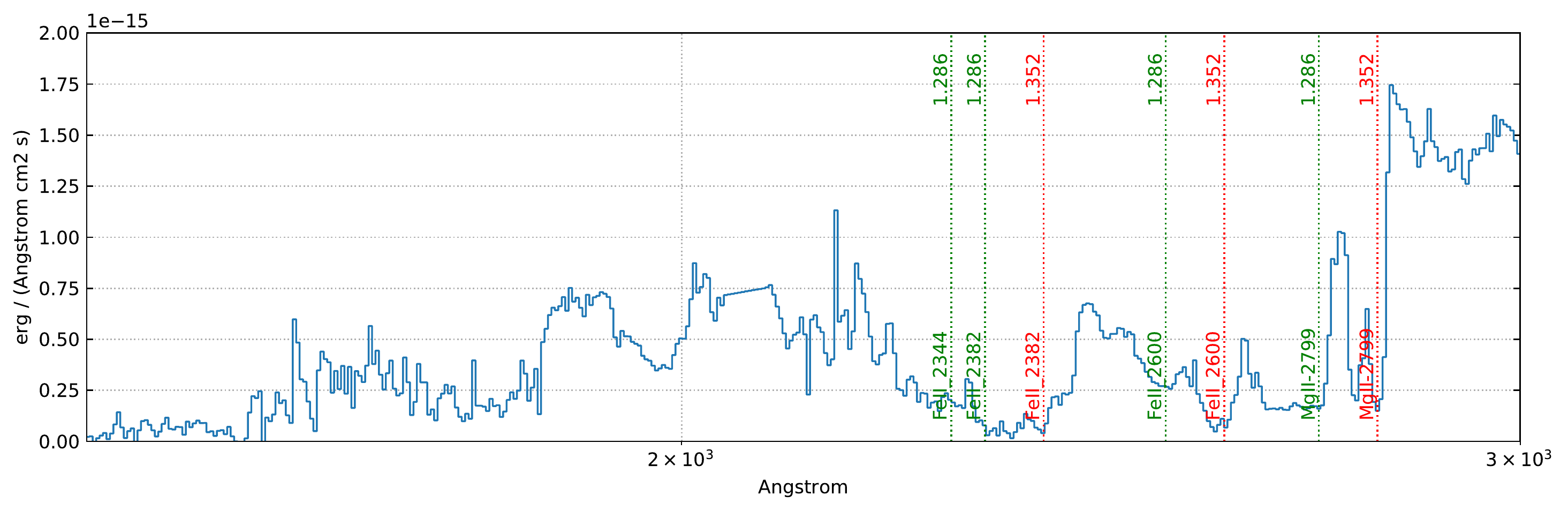}\\
	\includegraphics[height=0.215\textheight,left]{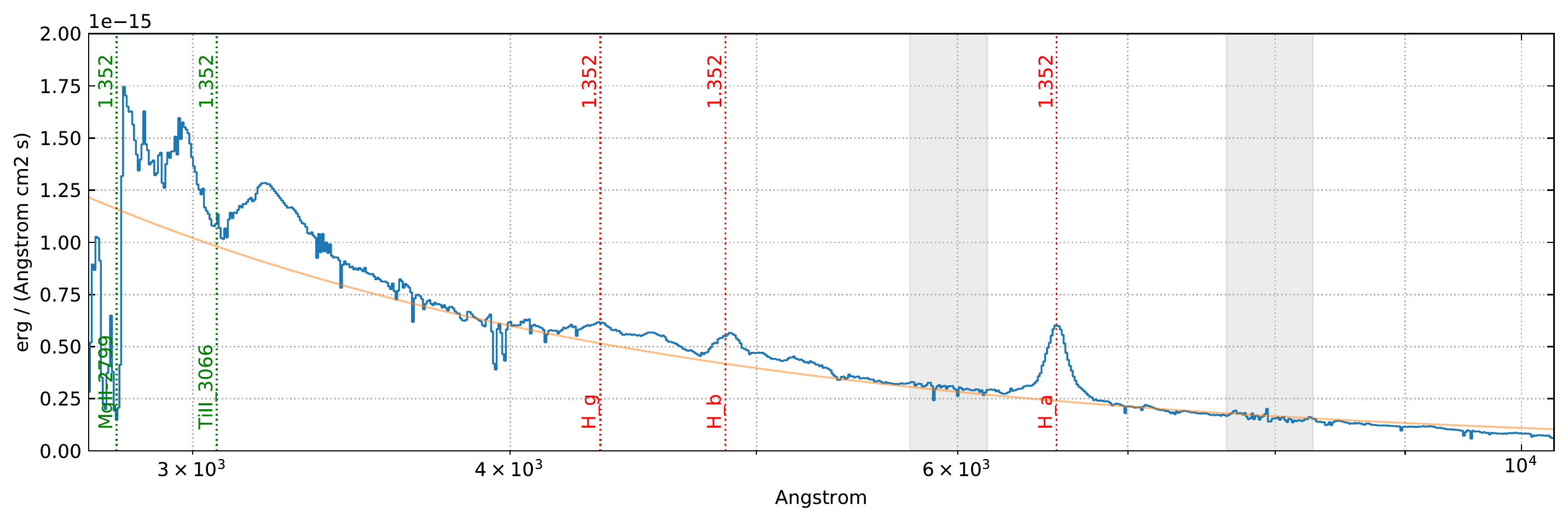}\\
	\contcaption{}
\end{figure*}

\begin{figure*}
    \centering
    \textbf{J2105$-$4104 ($\boldsymbol{\zem=2.247}$, $\boldsymbol{\alpha=-2.07}$)}\\
	\includegraphics[height=0.215\textheight,left]{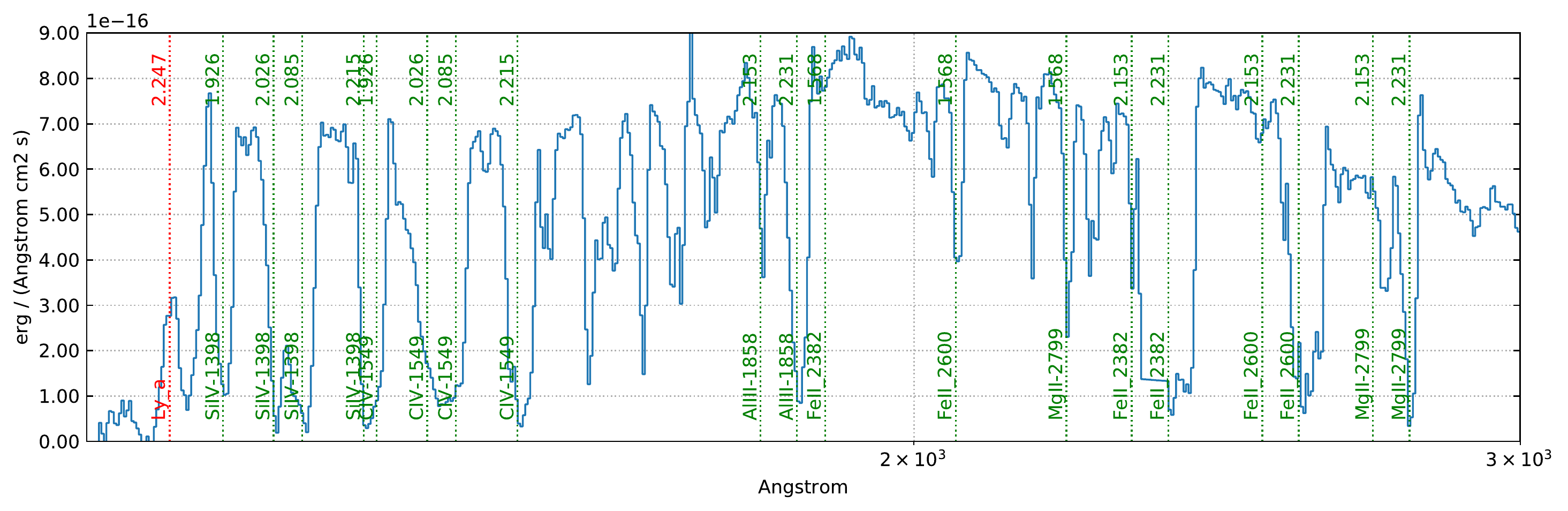}\\
	\includegraphics[height=0.215\textheight,left]{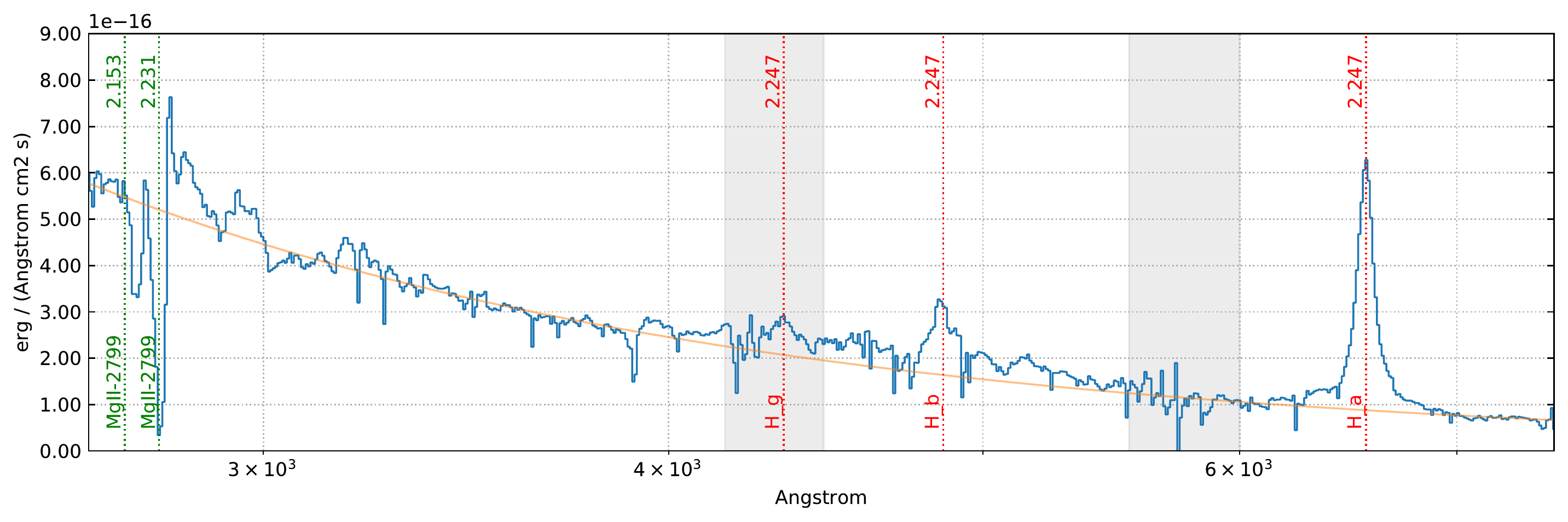}\\
    \vspace{0.5cm}
    \textbf{J2154$-$0514 ($\boldsymbol{\zem=1.629}$, $\boldsymbol{\alpha=-1.95}$)}\\
	\includegraphics[height=0.215\textheight,left]{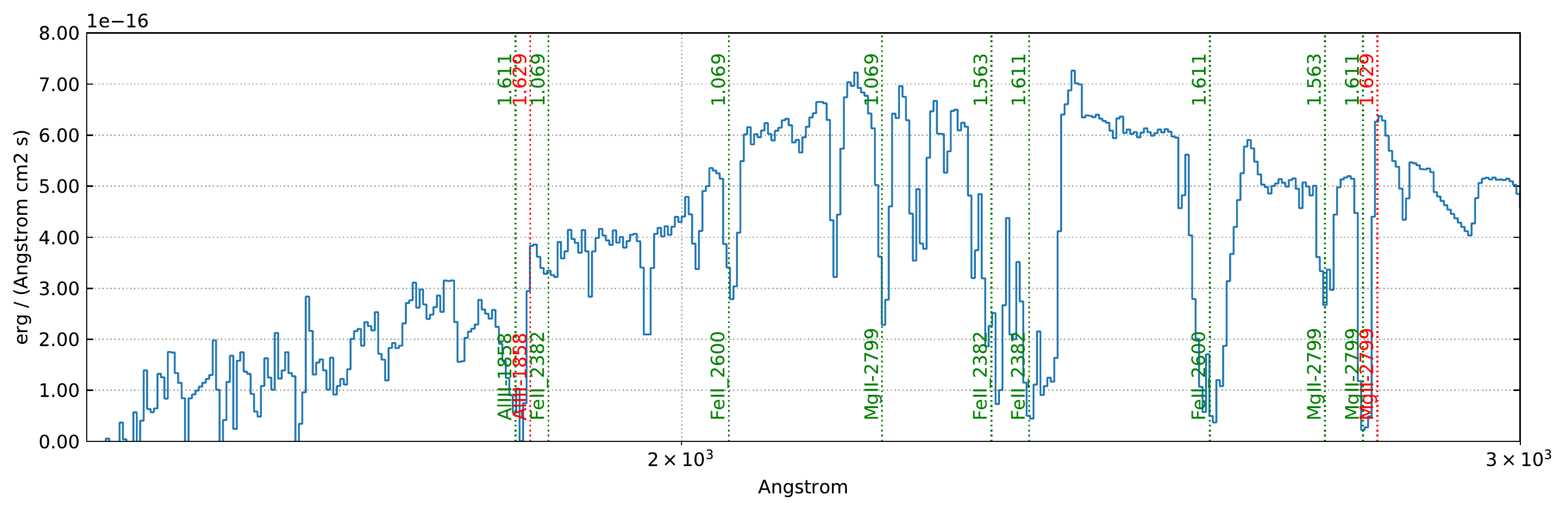}\\
	\includegraphics[height=0.215\textheight,left]{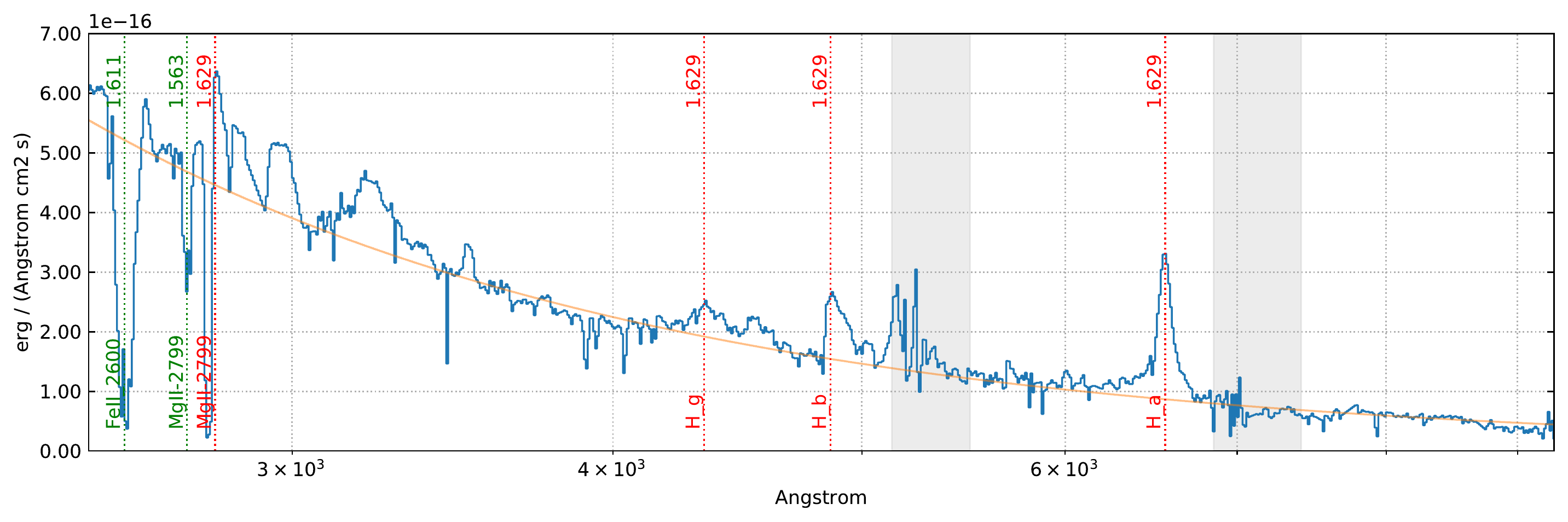}\\
    \contcaption{}
\end{figure*}

\begin{figure*}
    \centering
    \textbf{J0008$-$5058 ($\boldsymbol{\zem=2.041}$, $\boldsymbol{\alpha=-1.85}$)}\\
	\includegraphics[height=0.215\textheight,left]{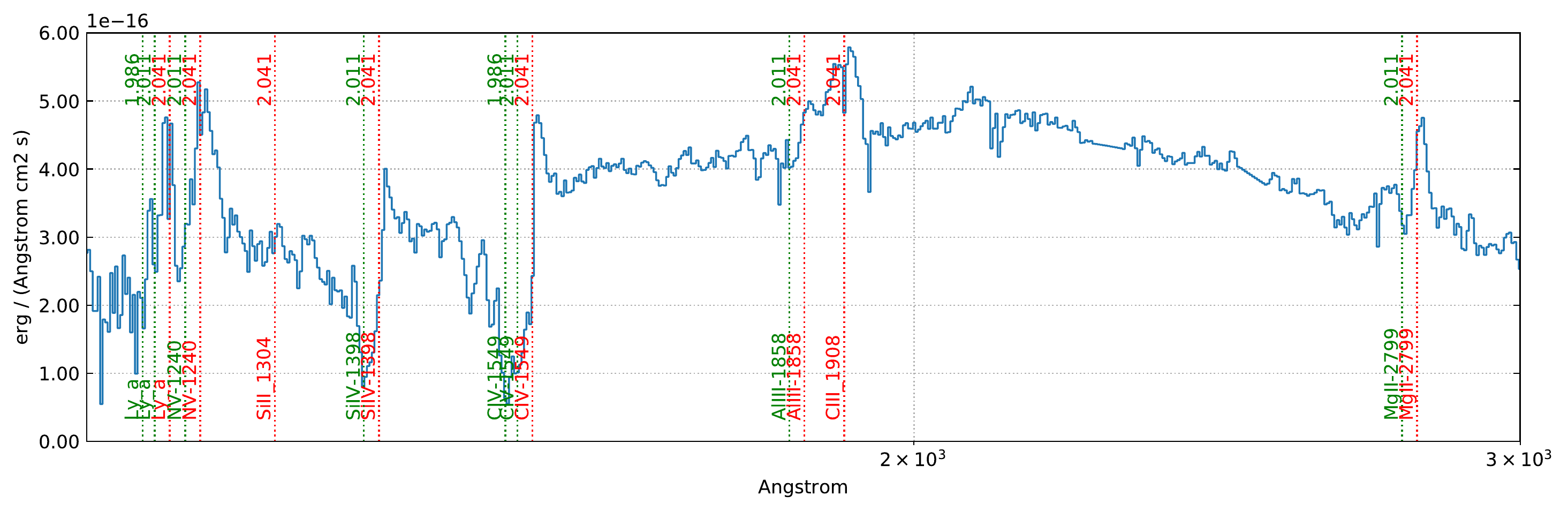}\\
	\includegraphics[height=0.215\textheight,left]{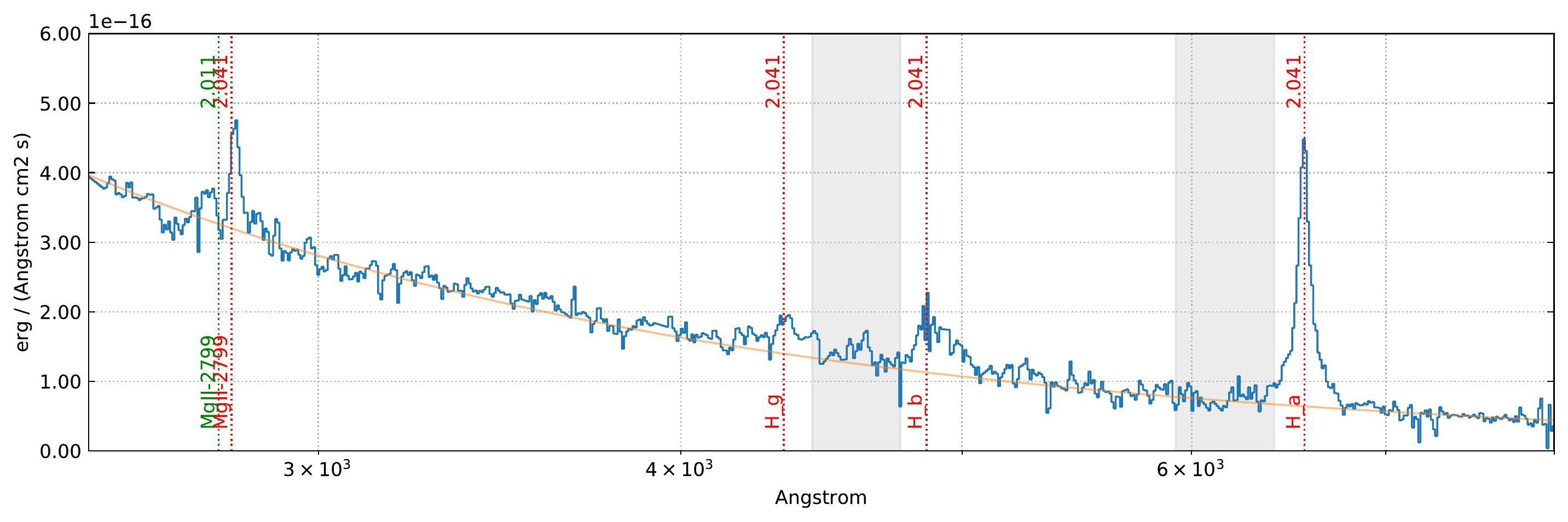}\\
	\vspace{0.5cm}
    \textbf{J0010$-$3201 ($\boldsymbol{\zem=2.379}$, $\boldsymbol{\alpha=-2.07}$)}\\
	\includegraphics[height=0.215\textheight,left]{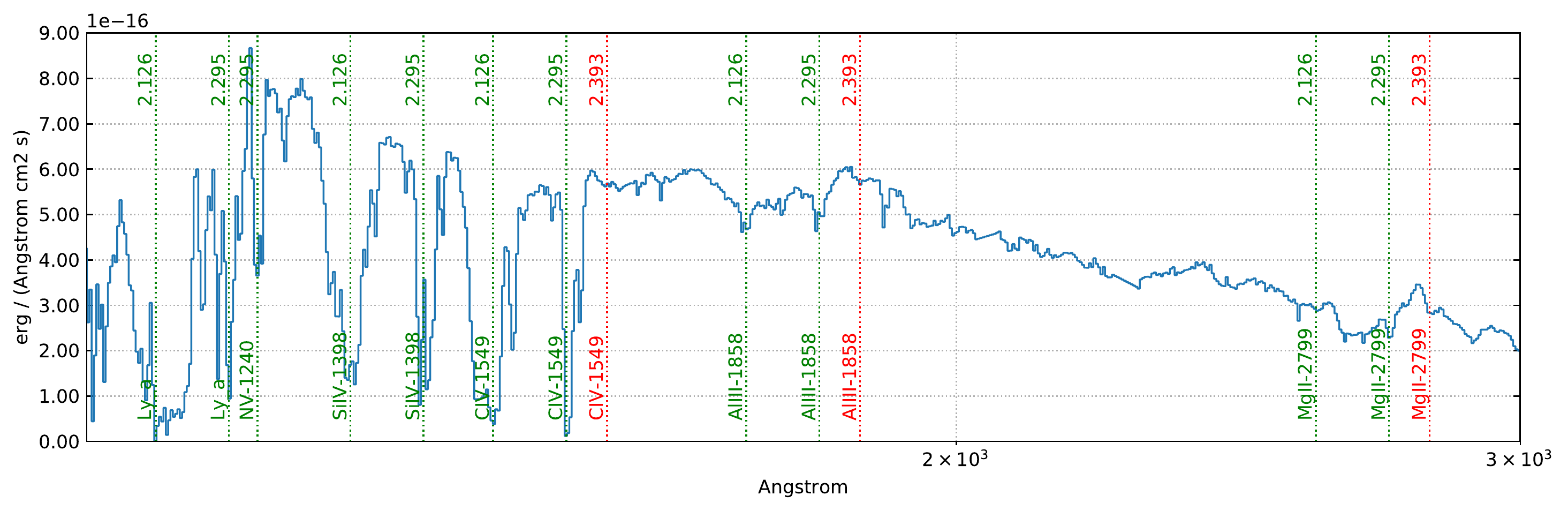}\\
	\includegraphics[height=0.215\textheight,left]{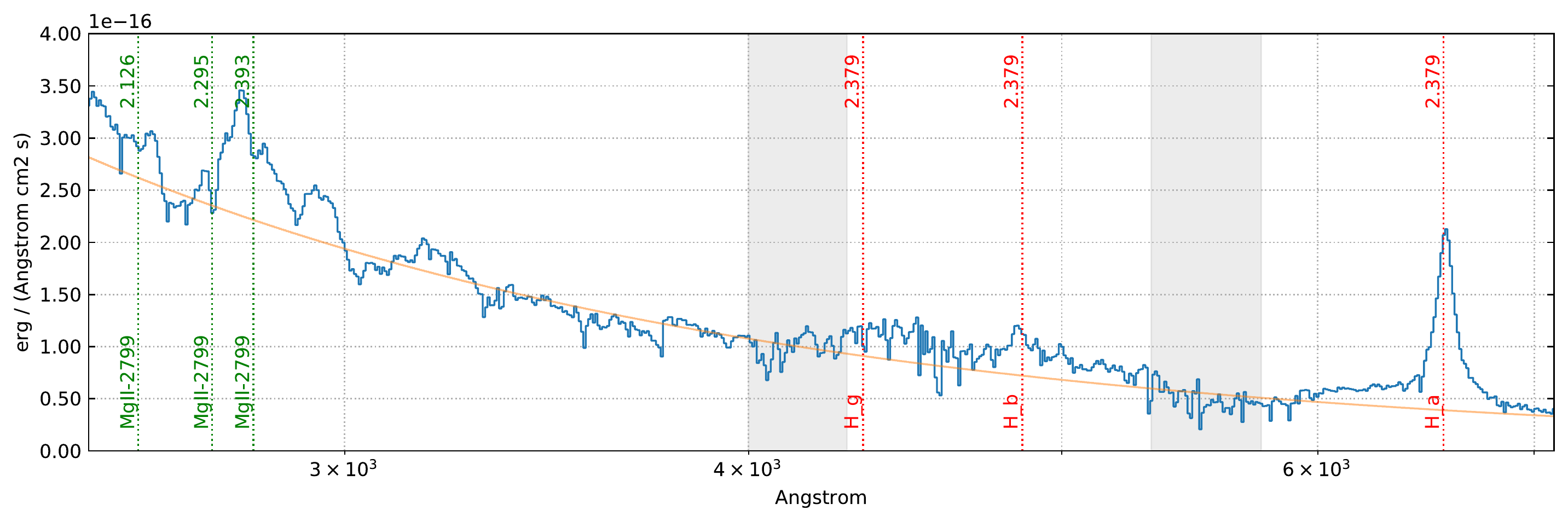}\\
    \caption{Spectra of other QUIP and non-QUIP QSOs from our sample. The legend is the same as \autoref{fig:felobal}.}
    \label{fig:non_felobal}
\end{figure*}

\begin{figure*}
    \centering
    \textbf{J0140$-$2531 ($\boldsymbol{\zem=2.947}$)}\\
	\includegraphics[height=0.215\textheight,left]{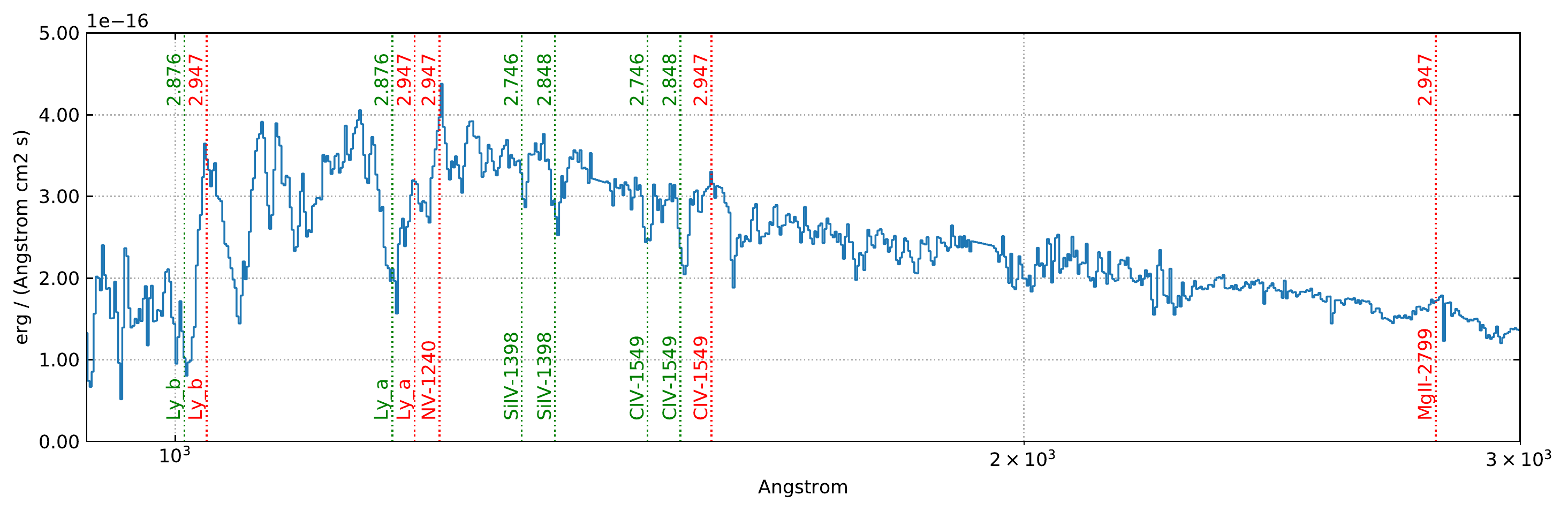}\\
	\includegraphics[height=0.215\textheight,left]{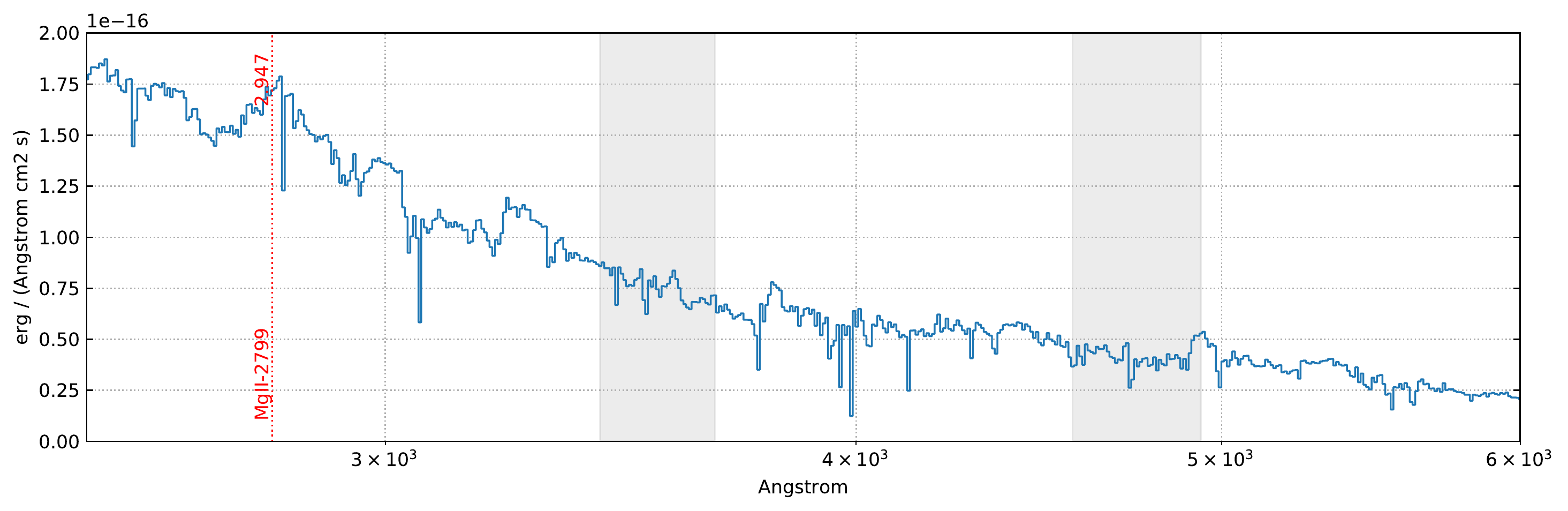}\\
	\vspace{0.5cm}
    \textbf{J0407$-$6245 ($\boldsymbol{\zem=1.289}$, $\boldsymbol{\alpha=-1.96}$)}\\
	\includegraphics[height=0.215\textheight,left]{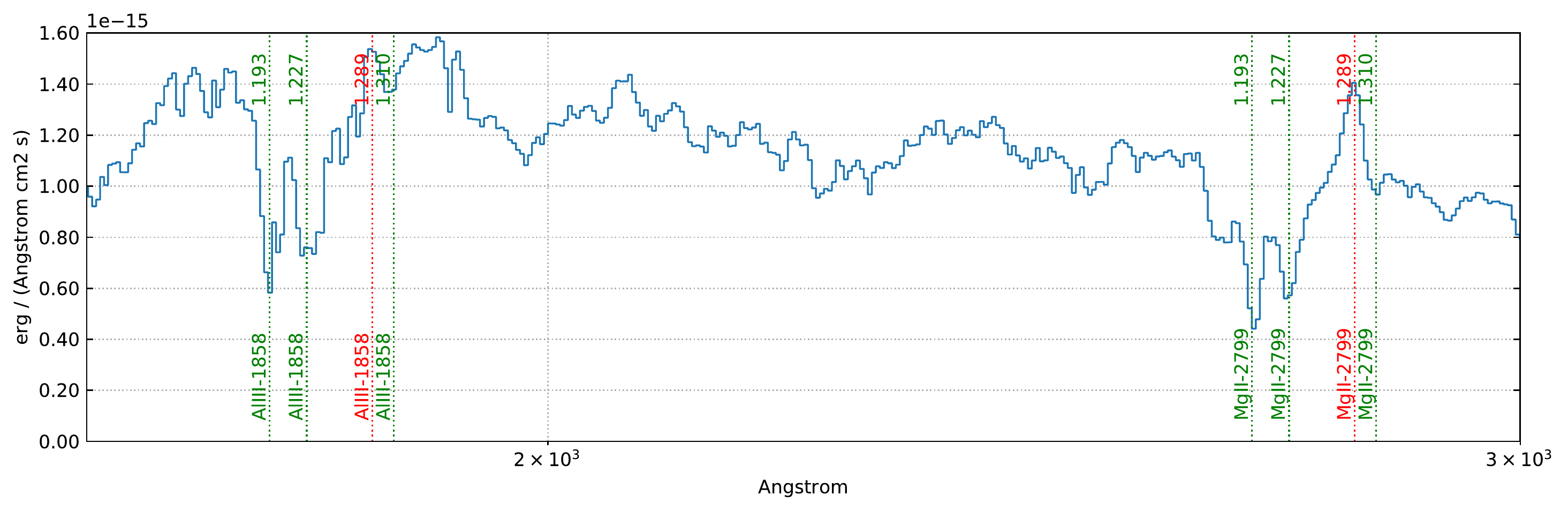}\\
	\includegraphics[height=0.215\textheight,left]{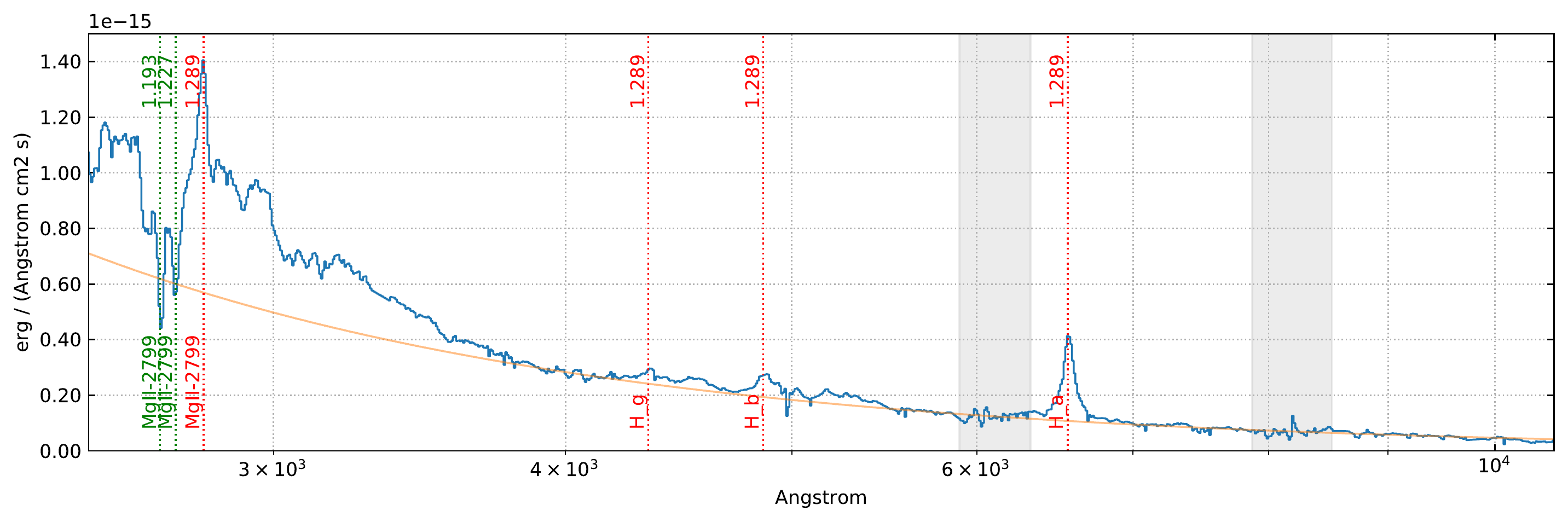}\\
	\contcaption{}
\end{figure*}

\begin{figure*}
    \centering
    \textbf{J2134$-$7243 ($\boldsymbol{\zem=2.178}$, $\boldsymbol{\alpha=-1.96}$)}\\
	\includegraphics[height=0.215\textheight,left]{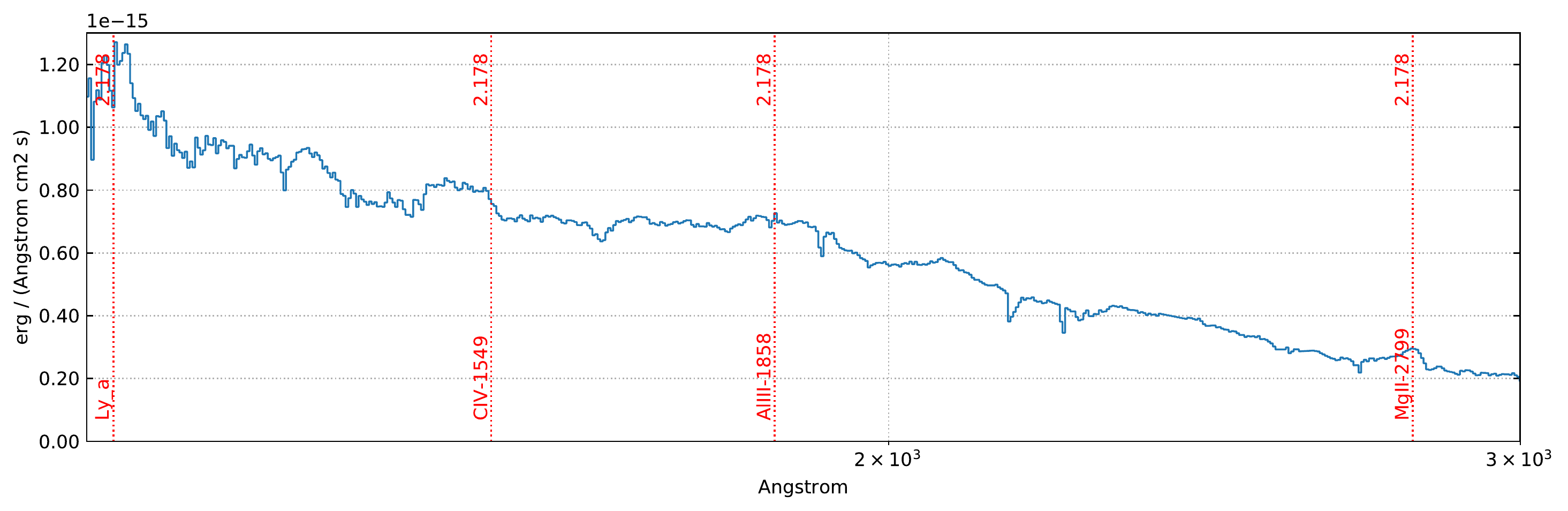}\\
	\includegraphics[height=0.215\textheight,left]{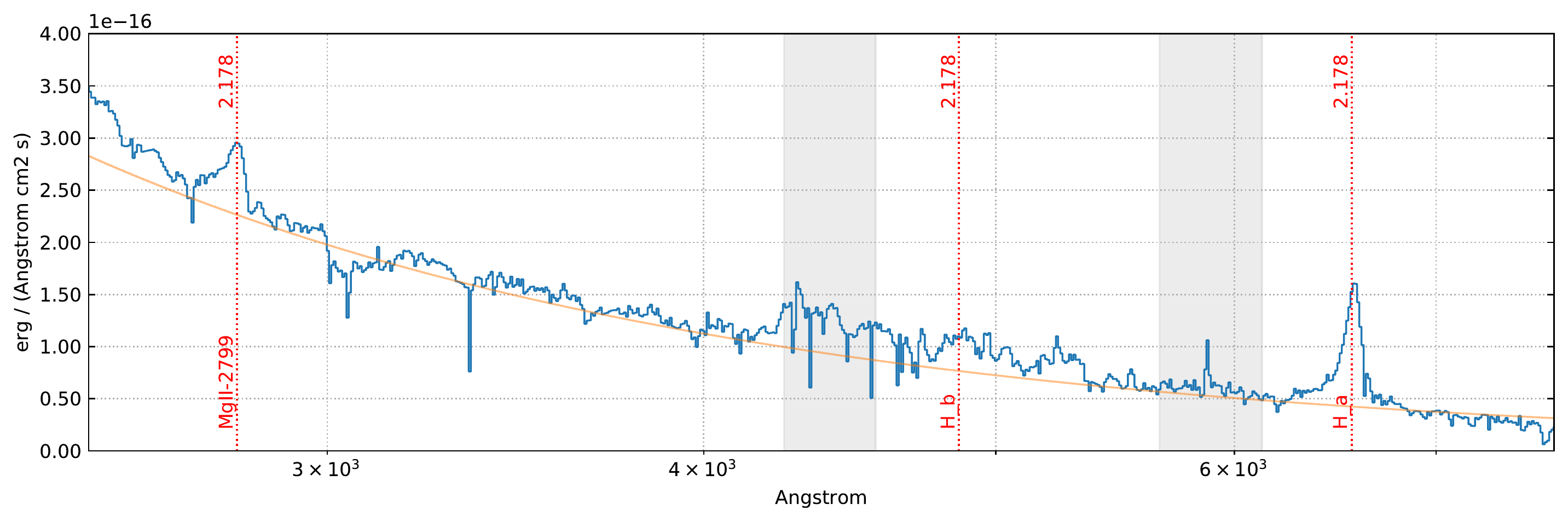}\\
	\vspace{0.5cm}
    \textbf{J2157$-$3602 ($\boldsymbol{\zem=4.665}$, $\boldsymbol{\alpha=-1.42}$)}\\
	\includegraphics[height=0.215\textheight,left]{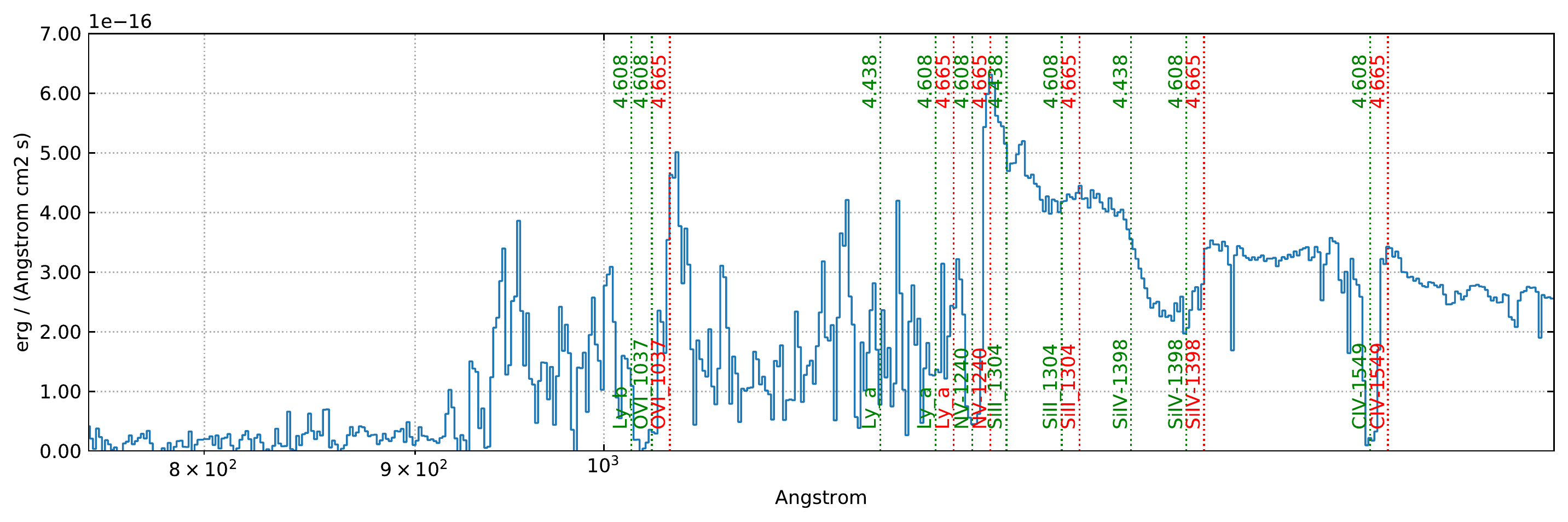}\\
	\includegraphics[height=0.215\textheight,left]{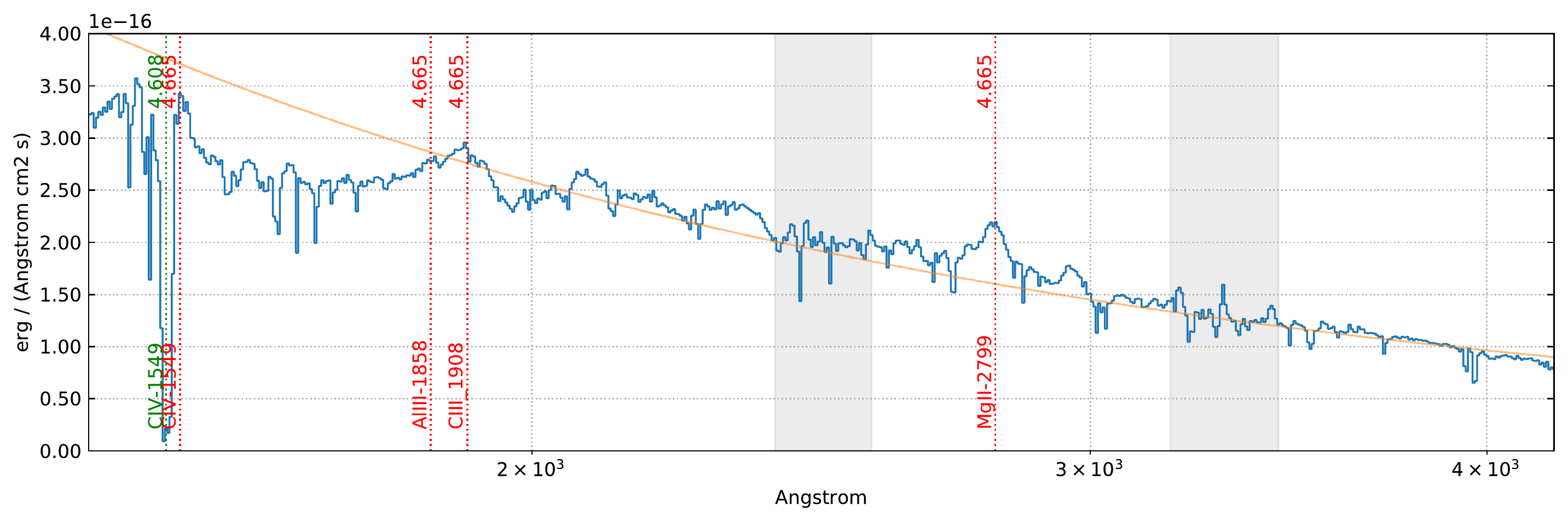}\\
	\contcaption{}
\end{figure*}

\begin{figure*}
    \centering
    \textbf{J2222$-$4146 ($\boldsymbol{\zem=2.192}$, $\boldsymbol{\alpha=-2.06}$)}\\
	\includegraphics[height=0.215\textheight,left]{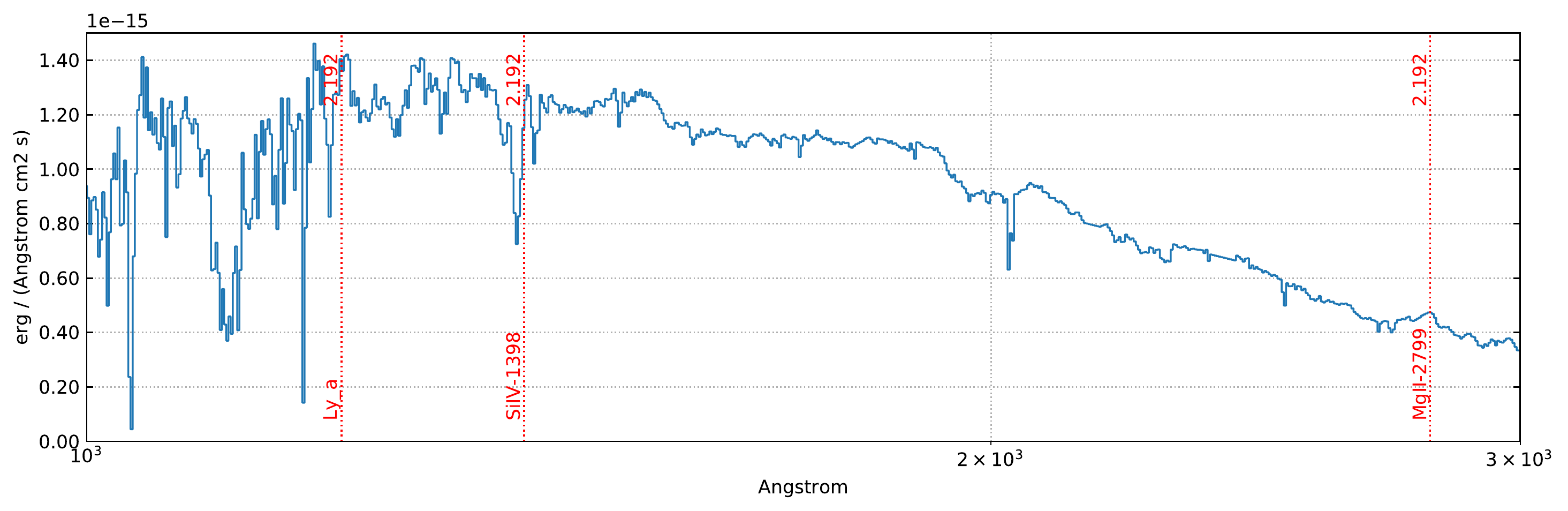}\\
	\includegraphics[height=0.215\textheight,left]{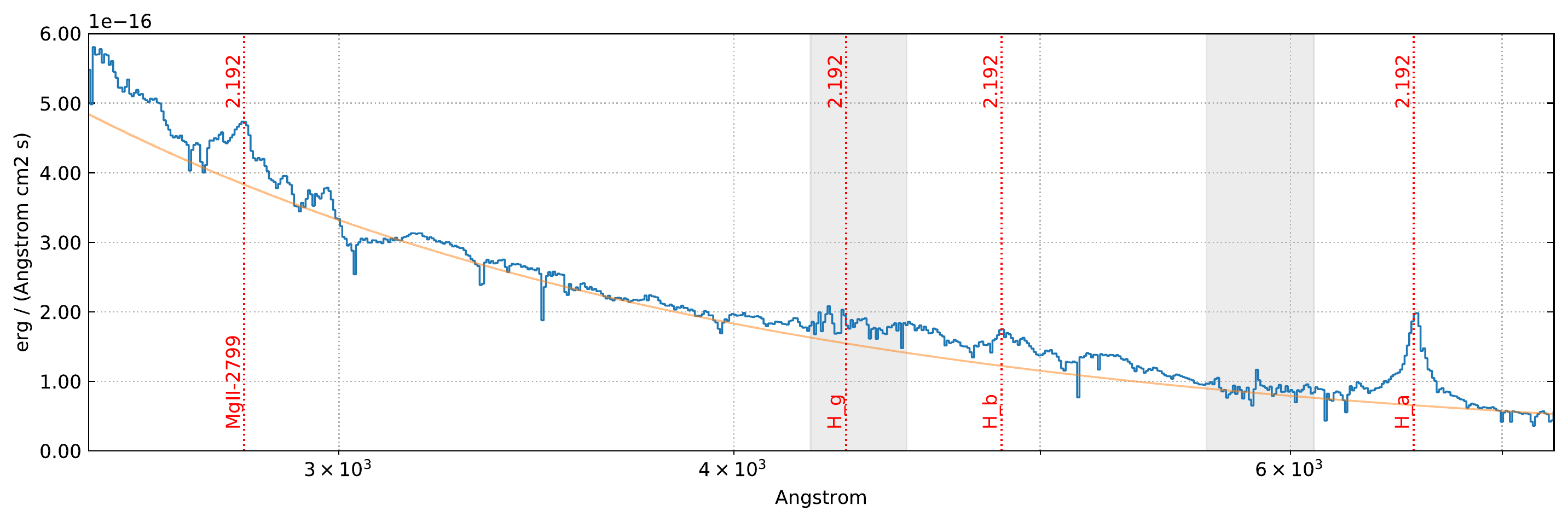}\\
	\vspace{0.5cm}
    \textbf{J2255$-$5404 ($\boldsymbol{\zem=2.255}$, $\boldsymbol{\alpha=-2.19}$)}\\
	\includegraphics[height=0.215\textheight,left]{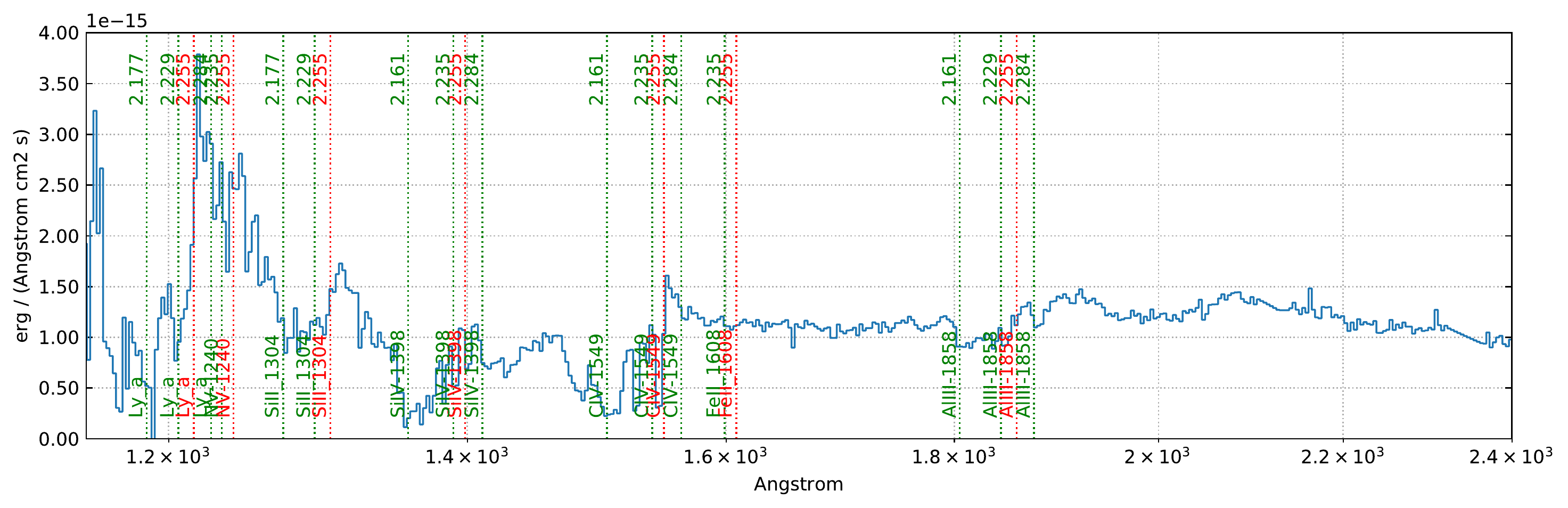}\\
	\includegraphics[height=0.215\textheight,left]{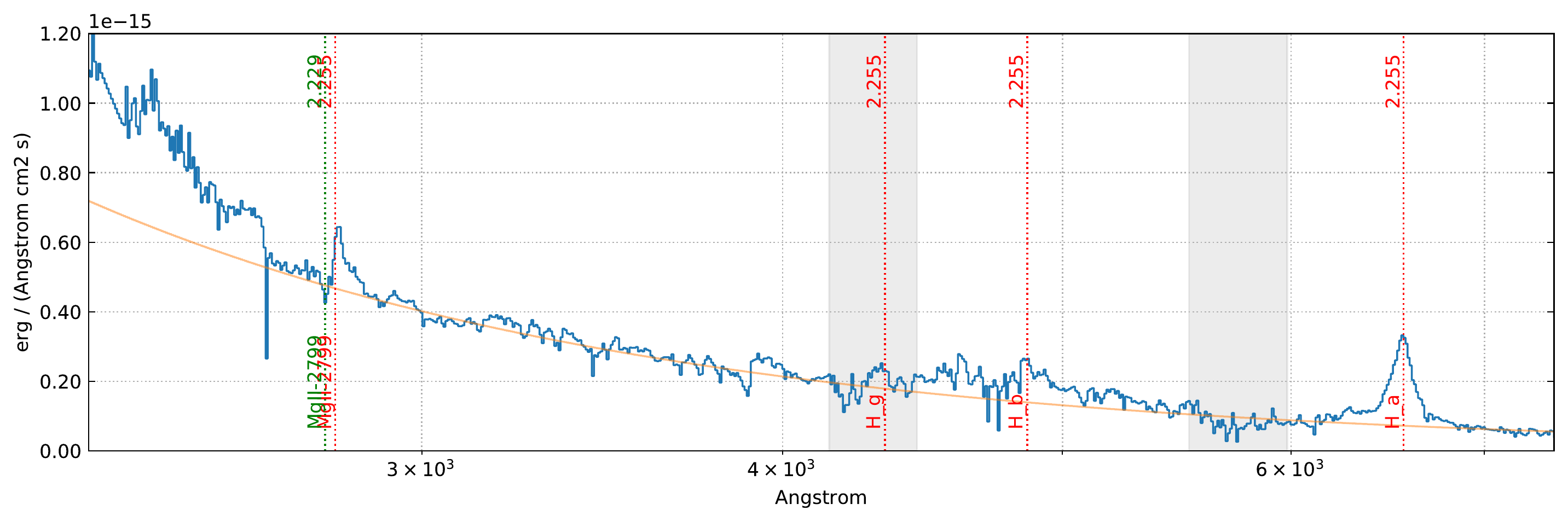}\\
	\contcaption{}
\end{figure*}

\begin{figure*}
    \centering
    \textbf{J2319$-$7322 ($\boldsymbol{\zem=2.612}$, $\boldsymbol{\alpha=-1.88}$)}\\
	\includegraphics[height=0.215\textheight,left]{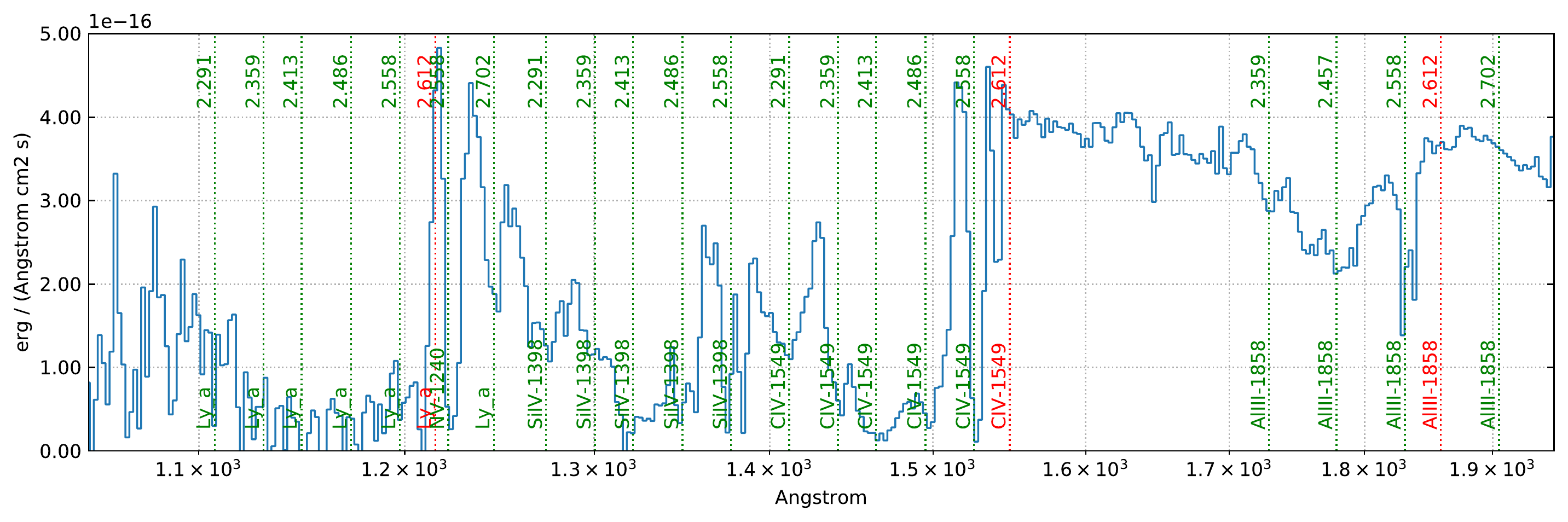}\\
	\includegraphics[height=0.215\textheight,left]{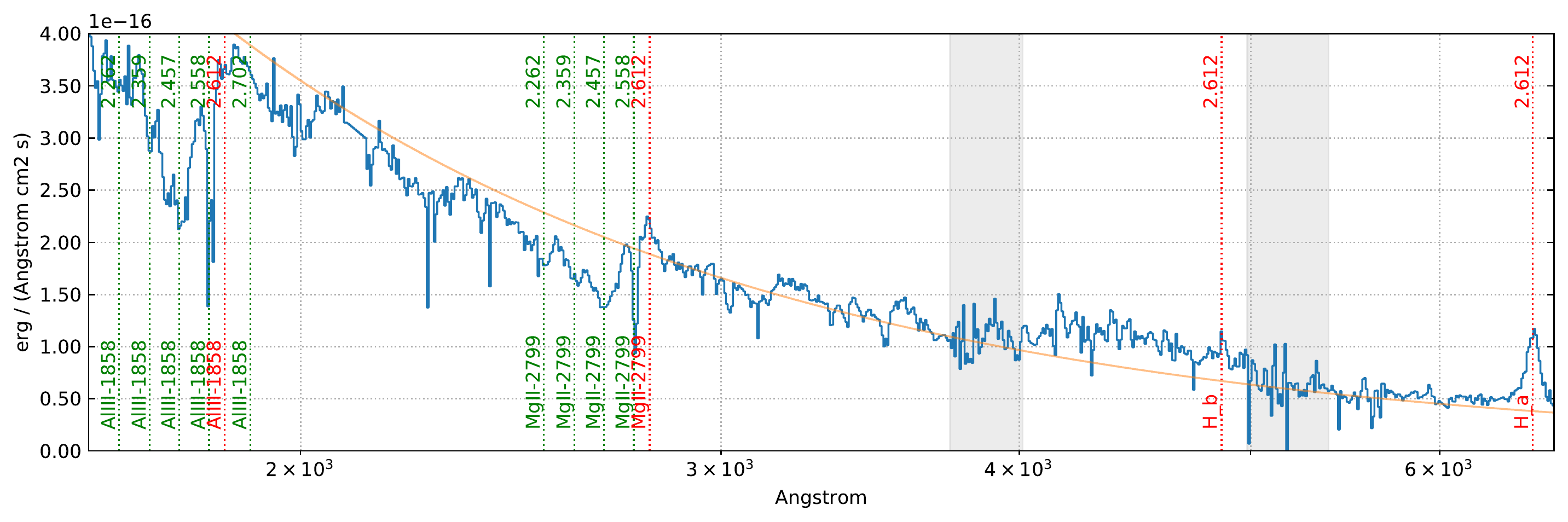}\\
	\vspace{0.5cm}
    \textbf{J2355$-$5253 ($\boldsymbol{\zem=2.363}$, $\boldsymbol{\alpha=-1.73}$)}\\
	\includegraphics[height=0.215\textheight,left]{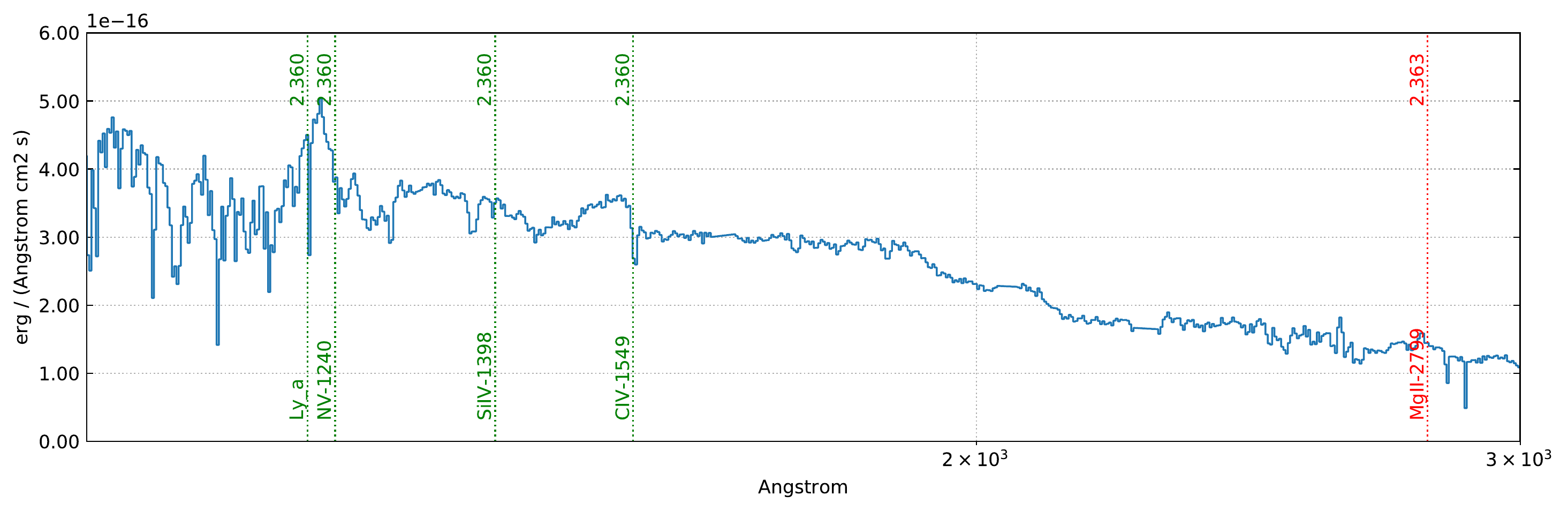}\\
	\includegraphics[height=0.215\textheight,left]{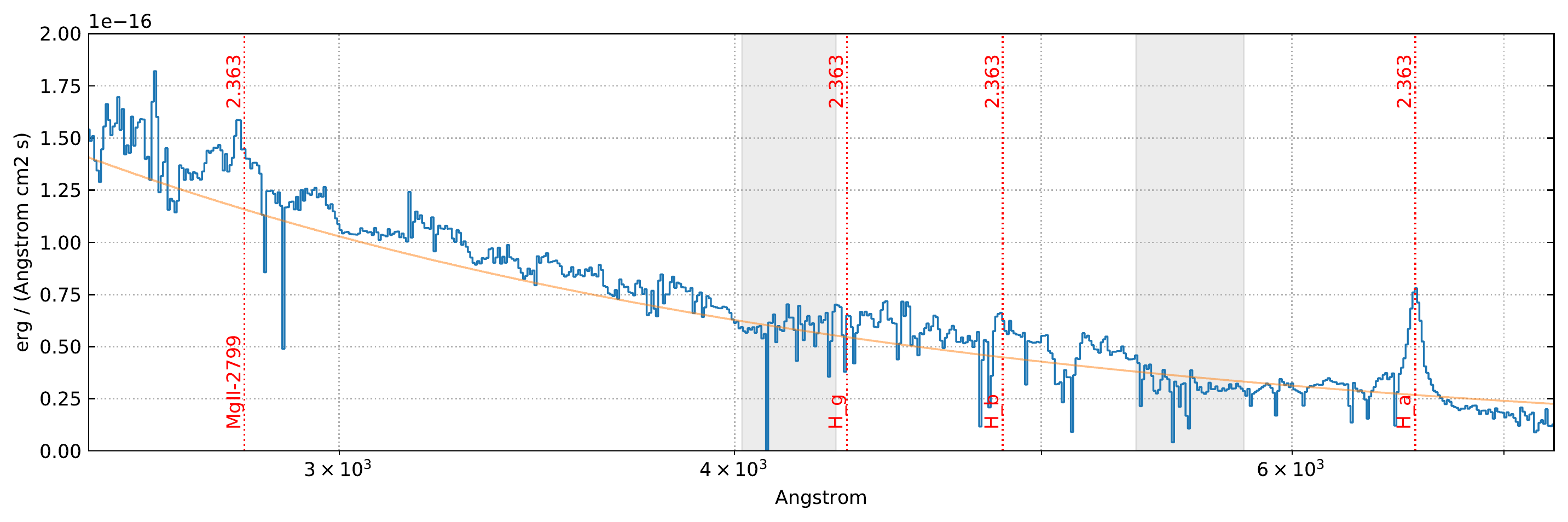}\\
	\contcaption{}
\end{figure*}


\bsp	
\label{lastpage}
\end{document}